\pdfoutput=1

\documentclass[%
aps,
pra,
twocolumn,
superscriptaddress,
amsfonts,amssymb,amsmath,
nofootinbib,
]{revtex4-2}

\usepackage[utf8]{inputenc}
\usepackage[T1]{fontenc}
\usepackage{lmodern}

\usepackage{graphicx}
\graphicspath{{figures/}}
\usepackage{dcolumn}
\usepackage{bm}

\usepackage{hyperref}
\hypersetup{
	pdftitle={Analog cosmological reheating in an ultracold Bose gas},
	pdfauthor={Aleksandr Chatrchyan, Kevin T. Geier, Markus K. Oberthaler, Jürgen Berges, Philipp Hauke},
}

\usepackage[english]{babel}
\usepackage[autostyle=true]{csquotes}
\usepackage[dvipsnames]{xcolor}
\usepackage{siunitx}
\usepackage{mathtools}
\usepackage{dsfont}
\usepackage{braket}
\usepackage[
caption=false,
subrefformat=simple,
labelformat=simple,
]{subfig}

\usepackage[
capitalise,
]{cleveref} 
\crefname{section}{Sec.}{Secs.}

\usepackage[acronym]{glossaries}
\glsdisablehyper
\newacronym{bec}{BEC}{Bose\textendash Einstein condensate}
\newacronym{1d}{$1$D}{one-dimensional}
\newacronym{2d}{$2$D}{two-dimensional}
\newacronym{3d}{$3$D}{three-dimensional}
\newacronym{ntfp}{NTFP}{non-thermal fixed point}
\newacronym{frw}{FRW}{Friedmann\textendash Robertson\textendash Walker}
\newacronym{gpe}{GPE}{Gross\textendash Pitaevskii equation}
\newacronym{tof}{TOF}{time-of-flight}
\newacronym{qcd}{QCD}{quantum chromodynamics}
\newacronym{uv}{UV}{ultraviolet}
\newacronym{2pi}{2PI}{two-particle irreducible}

\newcommand{\mean}[1]{\overline{#1}}
\newcommand{\vect}[1]{\bm{#1}}
\newcommand{\diff}{\mathop{}\!\mathrm{d}}
\newcommand{\etothepowerof}[1]{\mathrm{e}^{#1}}
\newcommand{\kb}{k_{\mathrm{B}}}
\newcommand{\commutator}[2]{\big[#1, #2\big]}

\DeclareMathOperator*{\argmax}{arg\,max}

\DeclareMathOperator{\real}{Re}
\DeclareMathOperator{\imag}{Im}

\newcommand{\initialturbulencetime}{t_{\mathrm{i}}}
\newcommand{\driventurbulencetime}{t_{\mathrm{d}}}
\newcommand{\reheatingtime}{t_{\mathrm{f}}}
\newcommand{\reheatingtemp}{T_{\mathrm{f}}}
\newcommand{\characteristicmomentum}{\bar{p}}
\newcommand{\initialcharmom}{p_{\mathrm{i}}}

\newcommand{\finalcharmom}{p_{\mathrm{f}}}
\newcommand{\alphadriven}{\alpha_{\mathrm{driven}}}
\newcommand{\alphafree}{\alpha_{\mathrm{free}}}
\newcommand{\betadriven}{\beta_{\mathrm{driven}}}
\newcommand{\betafree}{\beta_{\mathrm{free}}}
\newcommand{\timeunit}{t_0}
\newcommand{\lengthunit}{x_0}
\newcommand{\momentum}{p}

\newcommand{\cosmictime}{\tau}
\newcommand{\conformaltime}{\eta}
\newcommand{\labtime}{t}
\newcommand{\inflaton}{\phi}
\newcommand{\cosmicinflaton}{\inflaton}
\newcommand{\conformalinflaton}{\Phi}
\newcommand{\conformalinflatonmomentum}{\Pi}
\newcommand{\condensate}{\Psi}
\newcommand{\bosefield}{\hat{\condensate}}
\newcommand{\cosmiccondensate}{\psi}
\newcommand{\cosmicbosefield}{\hat{\cosmiccondensate}}
\newcommand{\labbosefield}{\bosefield}
\newcommand{\density}{n}
\newcommand{\phase}{\theta}

\begin{document}

\title{Analog cosmological reheating in an ultracold Bose gas}

\author{Aleksandr Chatrchyan}
\thanks{These authors contributed equally to this work.\\E-mail A.C.~at: \href{mailto:chatrchyan@thphys.uni-heidelberg.de}{chatrchyan@thphys.uni-heidelberg.de}\\E-mail K.T.G.~at: \href{mailto:geier@thphys.uni-heidelberg.de}{geier@thphys.uni-heidelberg.de}.}
\affiliation{Institute for Theoretical Physics, Ruprecht-Karls-Universität Heidelberg, Philosophenweg 16, 69120 Heidelberg, Germany}

\author{Kevin T. Geier}
\thanks{These authors contributed equally to this work.\\E-mail A.C.~at: \href{mailto:chatrchyan@thphys.uni-heidelberg.de}{chatrchyan@thphys.uni-heidelberg.de}\\E-mail K.T.G.~at: \href{mailto:geier@thphys.uni-heidelberg.de}{geier@thphys.uni-heidelberg.de}.}
\affiliation{Institute for Theoretical Physics, Ruprecht-Karls-Universität Heidelberg, Philosophenweg 16, 69120 Heidelberg, Germany}
\affiliation{INO-CNR BEC Center and Department of Physics, University of Trento, Via Sommarive 14, 38123 Povo (TN), Italy}
\affiliation{Kirchhoff Institute for Physics, Ruprecht-Karls-Universität Heidelberg, Im Neuenheimer Feld 227, 69120 Heidelberg, Germany}

\author{Markus K. Oberthaler}
\affiliation{Kirchhoff Institute for Physics, Ruprecht-Karls-Universität Heidelberg, Im Neuenheimer Feld 227, 69120 Heidelberg, Germany}

\author{Jürgen Berges}
\affiliation{Institute for Theoretical Physics, Ruprecht-Karls-Universität Heidelberg, Philosophenweg 16, 69120 Heidelberg, Germany}

\author{Philipp Hauke}
\affiliation{INO-CNR BEC Center and Department of Physics, University of Trento, Via Sommarive 14, 38123 Povo (TN), Italy}
\affiliation{Institute for Theoretical Physics, Ruprecht-Karls-Universität Heidelberg, Philosophenweg 16, 69120 Heidelberg, Germany}
\affiliation{Kirchhoff Institute for Physics, Ruprecht-Karls-Universität Heidelberg, Im Neuenheimer Feld 227, 69120 Heidelberg, Germany}

\date{\today}

\begin{abstract}
Cosmological reheating describes the transition of the post-inflationary universe to a hot and thermal state. In order to shed light on the underlying dynamics of this process, we propose to quantum-simulate the reheating-like dynamics of a generic cosmological single-field model in an ultracold Bose gas. In our setup, the excitations on top of an atomic Bose--Einstein condensate play the role of the particles produced by the decaying inflaton field after inflation. Expanding spacetime as well as the background oscillating inflaton field are mimicked in the non-relativistic limit by a time dependence of the atomic interactions, which can be tuned experimentally via Feshbach resonances. As we illustrate by means of classical-statistical simulations for the case of two spatial dimensions, the dynamics of the atomic system exhibits the characteristic stages of far-from-equilibrium reheating, including the amplification of fluctuations via parametric instabilities and the subsequent turbulent transport of energy towards higher momenta. The transport is governed by a non-thermal fixed point showing universal self-similar time evolution as well as a transient regime of prescaling with time-dependent scaling exponents. While the classical-statistical simulations can capture only the earlier stages of the dynamics for weak couplings, the proposed experiment has the potential of exploring the evolution up to late times even beyond the weak coupling regime.
\end{abstract}


\maketitle

\tableofcontents

\newpage
\section{\label{sec:introduction}Introduction}

Cosmological inflation is a well-established paradigm that solves the puzzles of flatness, homogeneity, and isotropy of the universe~\cite{Guth:1980zm, Starobinsky:1980te}, and explains the generation of density perturbations for structure formation~\cite{Starobinsky:1982ee}. According to this paradigm, the early universe underwent a stage of accelerated expansion which, in a typical scenario, is driven by a scalar field, the inflaton. Inflation is followed by a reheating phase~\cite{Kofman:1994rk}, during which the inflaton decays, e.g., into degrees of freedom of the standard model of particle physics. The heating process can start with a preheating stage of rapid particle production far from equilibrium, and finally completes after the particles, due to their interactions, approach thermal equilibrium at the reheating temperature~\cite{Kofman:1997yn, Amin:2014eta}. Over the recent years, it has been realized that many salient features of the dynamics become universal in the far-from-equilibrium regime, across physical systems at vastly different scales~\cite{Micha:2004bv, Berges:2014bba, Nowak:2012gd, Orioli:2015dxa, Prufer:2018hto, Erne:2018gmz}.  This universality makes it particularly appealing to gain insight into the non-equilibrium processes during reheating by use of table-top experiments, such as ultracold atomic gases. Due to their high degree of controllability, these systems are ideally suited to realize and study processes relevant for the early universe that are otherwise challenging to access~\cite{Fedichev:2003bv, Fischer:2004bf, Uhlmann:2005hf, Neuenhahn:2012dz, neuenhahn2012localized, Neuenhahn:2012dz,Posazhennikova:2016nut,Zache:2017dnz, Robertson:2018gwi, Wittemer:2019agm}.

\begin{figure}
	\includegraphics[width=\columnwidth]{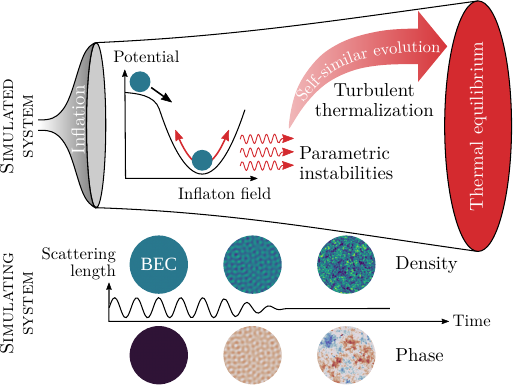}%
	\caption{\label{fig:reheating}Schematic illustration of post-inflationary reheating dynamics in the early universe and the simulation of an analogous process in an ultracold Bose gas. We consider a scenario where a single-component homogeneous \enquote{inflaton} field oscillates around the minimum of its potential, producing particles via parametric instabilities (\enquote{preheating}). Later, the system enters a turbulent state where energy is transported towards higher momenta in a self-similar way as the universe approaches thermal equilibrium (\enquote{reheating}). In the simulation, expanding spacetime as well as the oscillating inflaton field are mimicked in the non-relativistic limit by modulating the scattering length of a \acrfull{bec}, whose excitations play the role of particles produced by the decaying inflaton.}
\end{figure}

Here, we propose a quantum simulation of reheating-like dynamics in a single-component Bose gas, leveraging the ability to modulate the atomic interaction in time by means of Feshbach resonances~\cite{Chin2010}. The description of an ultracold Bose gas formally coincides with the non-relativistic limit of a scalar field in an expanding \gls{frw} spacetime, where the expansion is encoded in the time dependence of the interaction. This simplifies the simulation of such a system compared to approaches with explicit expansion~\cite{Eckel:2017uqx} by lifting the restrictions associated with the expansion of the trap. Remarkably, for the particular case of two spatial dimensions, where the Bose gas exhibits scale invariance~\cite{Gritsev:2009gf, Cha:2016esj, saint2019dynamical}, the dynamics becomes independent of the expansion, such that arbitrary expanding backgrounds may be studied by post-processing the data of a single experiment. It is our aim to reproduce the post-inflationary dynamics of generic single-field models~\cite{Micha:2004bv} in an ultracold Bose gas. Relativistic effects, such as particle production from the coherently oscillating inflaton field during the preheating stage, are absent in the atomic system, but preheating-like dynamics can be introduced also in the non-relativistic theory by a sinusoidal modulation of the coupling, mimicking an oscillating background field. As we demonstrate by means of classical-statistical (or truncated Wigner) simulations~\cite{Khlebnikov:1996mc,Berges:2007ym,Sinatra2002,Blakie2008}, the proposed setup exhibits the characteristic stages of reheating dynamics~\cite{Amin:2014eta}, including the preheating stage of parametric amplification of quantum fluctuations as well as the self-similar transport of energy towards higher momenta driving turbulent thermalization~\cite{Micha:2004bv}.
In this respect, our setup provides an analog to cosmological reheating, as illustrated schematically in \cref{fig:reheating}.

Analog models of gravity build on emergent curved spacetimes for low-energy excitations on top of \glspl{bec} or other systems~\cite{Barcelo2011}.
These can be generated, for instance, by modulating trapping potentials or interactions, allowing one to simulate cosmological particle production during inflation~\cite{Fedichev:2003bv,Fischer:2004bf,Jain:2007gg}.
The mapping to an emergent expanding spacetime is, however, valid only for linear excitations, and incorporating non-linear effects into this framework, which are crucial for post-inflationary reheating dynamics and responsible for thermalization, can be challenging.
Nonetheless, several experimental setups have been proposed for simulating post-inflationary dynamics in Bose gases~\cite{neuenhahn2012localized,Neuenhahn:2012dz,Posazhennikova:2016nut,Zache:2017dnz, Robertson:2018gwi}. Besides these theoretical proposals, preheating-like dynamics has recently been observed in experiments with a ring-shaped condensate after rapidly expanding the trap~\cite{Eckel:2017uqx}. In our setup, a preheating-like parametric amplification of sound waves on top of the \gls{bec} is induced by modulating the atomic interaction, which mimics the oscillations of the inflaton field around the minimum of its potential~\cite{Kofman:1997yn}. Parametric instabilities in Bose gases have been studied extensively in the literature~\cite{ Staliunas2002,Engels:2007zz,PhysRevA.81.053627,PhysRevA.84.013618,Jaskula:2012ab,PhysRevX.9.011052} and the concept has evolved into a promising tool for state preparation~\cite{Zhang2020}. Going beyond the well-understood linear regime of parametric resonance, our main focus is geared towards the non-linear stages of the dynamics, involving a secondary amplification of excitations~\cite{Kofman:1997yn,Berges:2002cz,Zache:2017dnz} that herald the development of turbulence~\cite{Micha:2004bv}.

The preheating-like instabilities drive the system far away from equilibrium into the vicinity of a non-thermal fixed point, where the dynamics becomes universal and self-similar~\cite{Berges:2008sr}. A non-thermal fixed point characterizes the far-from-equilibrium behavior of an entire universality class of physical systems as diverse as the inflaton in the early universe, ultracold atomic gases in the laboratory, or even the quark-gluon plasma explored in heavy-ion collisions~\cite{Micha:2004bv, Berges:2014bba, Orioli:2015dxa, Schmied:2018mte}.
Such universal dynamics far from equilibrium has been observed experimentally in spinor and tunnel-coupled Bose gases in form of inverse cascades transporting conserved quantities towards lower momenta~\cite{Prufer:2018hto, Erne:2018gmz}. Bidirectional cascades involving additionally a self-similar transport of energy towards higher momenta have been studied numerically in \gls{1d} spinor Bose gases~\cite{Schmied:2018osf} and have recently been observed experimentally in an isolated \gls{3d} Bose gas following a cooling quench~\cite{Glidden:2020qmu}. Our protocol of parametrically exciting a pure single-component \gls{bec} constitutes a complementary path of entering the regime of self-similar dynamics far from equilibrium and allows one to study the direct energy cascade in both driven and isolated systems.

We distinguish a regime of driven turbulence in the presence of continuous parametric driving and a regime of free turbulence after the modulation of the interaction is switched off, corresponding to the decay of the inflaton.
Recently, a new type of prescaling phenomenon~\cite{Mazeliauskas:2018yef,Schmied2019} has been established, which is closely related to the onset of a hydrodynamic behavior far from equilibrium~\cite{Mazeliauskas:2018yef}. In this regime, the dynamics is governed by the fixed-point scaling function and time-dependent scaling exponents, which slowly relax to their universal values. Remarkably, our numerical results indicate that such a behavior emerges during the transition from driven to free turbulence, which opens the door for an experimental observation of this phenomenon in cold Bose gases.

At late times, when the occupancies at characteristic momenta become of order unity, the system is expected to thermalize~\cite{Micha:2004bv}. This final stage of the dynamics is dominated by quantum fluctuations and therefore cannot be captured by means of classical-statistical simulations~\cite{Berges:2007ym,Berges2014}. In contrast, it can be accessed by experimental studies in cold atomic systems, which are quantum-mechanical by nature.

For reasons of consistency with standard cosmology, the inflaton must necessarily decay into more familiar forms of matter and gauge fields, e.g., those that constitute the standard model of particle physics, before the universe thermalizes at the reheating temperature. In contrast, reheating in our setup involves the decay of the condensate into its own fluctuations only, as in the relativistic $\lambda \varphi^4$ model that has been studied in Ref.~\cite{Micha:2004bv}. Despite this simplicity, these models capture many characteristic features of the reheating process, which can be directly probed in our proposed experiment.

This paper is organized as follows. In \cref{sec:expansion}, we present our approach of simulating expansion in a Bose gas via a time-dependent interaction, establishing an analogy to the inflaton field in the early universe after inflation. We proceed in \cref{sec:preheating} (\enquote{analog preheating}) with discussing preheating-like dynamics, where particle production is mimicked by parametrically exciting Bogoliubov quasi-particles via a modulation of the interaction. Though our approach works in principle for any dimensionality, we focus here on the case of a spatially \gls{2d} Bose gas. We explain and interpret the momentum spectrum resulting from these instabilities, emphasizing the important role of non-linear effects leading to the onset of turbulent dynamics. In \cref{sec:reheating} (\enquote{analog reheating}), we analyze the universal self-similar time evolution for both the driven and the free direct cascade, comparing the numerically extracted scaling exponents with predictions from kinetic theory. We also show that the system exhibits prescaling during the transition from driven to free turbulence. Moreover, we discuss how expansion affects the dynamics and can, under certain conditions, even prevent the system from thermalizing. In \cref{sec:experiment}, we discuss details relevant for an experimental implementation, before we conclude in \cref{sec:conclusion}.

\section{\label{sec:expansion}Expanding spacetime in Bose gases}

The inflaton is commonly described by a relativistic real scalar field $\inflaton(x)$ in curved spacetime with the action~\cite{Kolb:1990vq}
\begin{equation}
\label{eq:inflatonaction}
S = \int \diff^{d+1}x \sqrt{-\mathfrak{g}} \left[ \frac{1}{2} g^{\mu \nu} \partial_\mu \inflaton \partial_\nu \inflaton - V(\inflaton) \right],
\end{equation}
where $V$ is the potential, $g_{\mu \nu}$ is the metric tensor, $\mathfrak{g}$ its determinant, and $d$ the number of spatial dimensions.
The expansion of the universe is well described with the help of the flat \gls{frw} metric~\cite{Kolb:1990vq}. It corresponds to a homogeneous and isotropic universe, in accordance with the cosmological principle, which is Minkowskian at each time slice. The \gls{frw} metric has the form 
\begin{equation}
\label{eq:metric}
\diff s^2 = g_{\mu \nu} \diff x^\mu \diff x^\nu =  c^2 \diff \cosmictime^2 - a^2(\cosmictime) \diff \vect{x}^2,
\end{equation}
where $c$ is the speed of light and $\cosmictime$ denotes the cosmic time. The cosmic scale factor $a(\cosmictime)$ grows with time in an expanding universe and relates the comoving distance~$\vect{x}$ to the proper distance~${\vect{r}(\cosmictime) = a(\cosmictime) \vect{x}}$.

Our goal is to observe reheating-like dynamics in an ultracold Bose gas in analogy to reheating in the early universe.
A Bose gas is described by a non-relativistic complex field operator $\bosefield$ subject to the Hamiltonian~\cite{Pitaevskii2016}
\begin{equation}
\label{eq:hamiltonian}
\hat{H} = \int \diff^d x \left[ \bosefield^\dagger \left( -\frac{\hbar^2 \nabla^2}{2m} + U \right) \bosefield + \frac{g}{2} \bosefield^\dagger \bosefield^\dagger \bosefield \bosefield \right],
\end{equation}
where $\hbar$ is the reduced Planck constant, $m$ the atomic mass, $U$ an external trapping potential, and $g$ the quartic coupling which determines the strength of the atomic interactions via the $s$-wave scattering length.
In what follows, we briefly discuss possible strategies of how to realize an expanding spacetime according to \cref{eq:metric} in an ultracold Bose gas, before we detail the approach chosen in this work.

The most straightforward approach to incorporate expansion in a trapped Bose gas is to physically expand the trap geometry~\cite{Uhlmann:2005hf, Eckel:2017uqx, Cha:2016esj}. While this approach is simple and direct in principle, there are practical limitations such as the restriction to short times or small expansion velocities due to a finite optical imaging system.

Alternatively, one can keep the trap geometry fixed, associating physical distances with the co-moving distances of an expanding system.
In this way, however, the redshift of momenta as well as dilution of the system due to the expansion have to be explicitly accounted for in form of a modified kinetic term and a non-unitary evolution, as discussed in the subsequent subsection (see also Ref.~\cite{Suarez:2015fga}).
Besides possible technical difficulties in engineering such non-Hermitian Hamiltonians, this formulation has the additional drawback that it implies a decreasing atomic density, which diminishes the experimental signal with progressing laboratory time.

We follow here a third approach, wherein the expansion of spacetime is encoded solely in the time dependence of the atomic interaction~\cite{Jain:2007gg,Fedichev:2003bv,Cha:2016esj}. For experiments with ultracold atoms where a broad Feshbach resonance is available to tune the interactions, this third approach may be the preferred way of studying a broad range of expansion scenarios. In what follows, we will discuss our approach of incorporating expansion. In \cref{sec:preheating,sec:reheating}, we will then illustrate the dynamics of preheating and reheating that becomes accessible in this setup.

\subsection{Relativistic versus non-relativistic dynamics in expanding spacetime}

We model the inflaton as a real scalar field, whose dynamics is described by the action~\eqref{eq:inflatonaction} with a potential ${V(\inflaton)=m^2 c^2 \inflaton^2 / 2 \hbar^2 + \lambda \inflaton^4/4!}$, where $m$ is the mass and $\lambda$ a quartic coupling.

The classical equations of motion for the inflaton field can be obtained from \cref{eq:inflatonaction} using the principle of least action. For the metric defined in \cref{eq:metric} they read
\begin{equation}
\label{eq:eomrel}
\frac{1}{c^2} \ddot{\inflaton} + \frac{1}{c^2} d H \dot{\inflaton} - \frac{\nabla^2}{a^2} \inflaton + \frac{m^2 c^2}{\hbar ^2} \inflaton + \frac{\lambda}{6} \inflaton^3 = 0,
\end{equation}
where ${H = \dot{a} / a}$ is the Hubble parameter and the dot denotes the derivative with respect to the cosmic time $\cosmictime$. This equation is of a Klein\textendash Gordon type with a modified spatial derivative term and an additional friction term, which express the redshift of momenta in the co-moving frame and the dilution of the field due to the expansion, respectively~\cite{Kolb:1990vq}.

The analogous equations for the non-relativistic system have a similar structure.
To see this, we factor out the fast oscillations due to the mass term in \cref{eq:eomrel} from the canonically quantized inflaton field~$\hat{\cosmicinflaton}(\tau, \vect{x})$, defining a slowly varying complex field~$\cosmicbosefield(\tau, \vect{x})$ via the relation
\begin{equation}
\label{eq:slowlyvarying}
	\hat{\cosmicinflaton}  = \frac{\hbar}{\sqrt{2 m c}} \left[ \cosmicbosefield \etothepowerof{-i mc^2 \cosmictime / \hbar} + \mathrm{h.c.} \right] ,
\end{equation}
where $\mathrm{h.c.}$ denotes the Hermitian conjugate.
As shown in \cref{app:nonrelativisticlimit}, provided typical momenta, expansion velocities, and field values are small, this new field evolves according to the equation~\cite{Suarez:2015fga}
\begin{equation}
\label{eq:eomnonherm}
i \hbar \dot \cosmicbosefield = \left( -\frac{\hbar^2}{2m} \frac{\nabla^2}{a^2} - i\hbar \frac{d}{2} H + g \cosmicbosefield^\dagger \cosmicbosefield \right) \cosmicbosefield ,
\end{equation}
with coupling~$g= {\lambda \hbar^4 }/{ 8 m^2 c}$.
\Cref{eq:eomnonherm} is reminiscent of the Heisenberg equations of motion generated by the Hamiltonian of an ultracold Bose gas~\eqref{eq:hamiltonian} in absence of an external trapping potential. The kinetic term is proportional to $a^{-2}$, which describes the redshift of momenta in the co-moving frame, and the dilution of the system due to the expansion is expressed by a non-Hermitian term causing the norm of the field to decay.

In analogy to the transformation to conformal variables known in quantum field theory on curved spacetime~\cite{mukhanov2007introduction}, we now introduce a new time variable $\labtime$ via the relation ${\diff \labtime = \diff \cosmictime / a^2}$, which we refer to as the laboratory time, and the rescaled field operator ${\labbosefield  = \cosmicbosefield a^{d/2}}$.
\Cref{eq:eomnonherm} then takes the standard form of the equations of motion for a non-relativistic bosonic field generated by the Hamiltonian~\eqref{eq:hamiltonian},
\begin{equation}
\label{eq:eomnonrel}
	i\hbar \frac{\partial \labbosefield}{\partial \labtime}= \left( -\frac{\hbar^2 \nabla^2}{2m} + g_{\mathrm{eff}}(\labtime) \labbosefield^\dagger \labbosefield \right) \labbosefield ,
\end{equation}
with the time-dependent effective coupling
\begin{equation}
\label{eq:coupling}
g_{\mathrm{eff}}(\labtime) = g a^{2-d}(\labtime).
\end{equation}
The transformation back to the original variables is performed in post-processing.

\Cref{eq:eomnonrel,eq:coupling} describe a Bose gas in an expanding spacetime, where the expansion enters only in the time dependence of the interaction term via the scale factor~$a(t)$. We stress that the correspondence to \cref{eq:eomrel} is restricted to the non-relativistic limit and therefore not capable of capturing relativistic effects, such as the resonant amplification of fluctuations during the preheating stage. As discussed in the next section, we can mimic an analogous process in the non-relativistic system by periodically modulating the interaction. Despite these restrictions, it is worth emphasizing that the relation between \cref{eq:eomrel,eq:eomnonrel} holds on the non-linear level, i.e., it does not rely on any linearization. This is in contrast to the analog gravity framework, where a mapping to \cref{eq:eomrel} is achieved by engineering an emergent expanding spacetime for linear excitations on top of the condensate~\cite{Jain:2007gg}. Such a mapping, however, typically breaks down on the non-linear level. Thus, our approach based on \cref{eq:eomnonrel} is well suited for simulating reheating-like dynamics, where non-linear effects are essential.

\subsection{\label{sec:special2d}The special case of expansion in 2D}
For $d = 2$, the effective coupling in \cref{eq:coupling} becomes independent of the scale factor $a(t)$. This is a consequence of a dynamical symmetry in \gls{2d} Bose gases known as scale invariance~\cite{Gritsev:2009gf}. While quantum anomalies in strongly interacting systems can lead to violations of scale invariance~\cite{Olshanii2010,Holten2018}, it has been well tested experimentally in the weakly interacting regime~\cite{Hung2011,saint2019dynamical}.
Thus, if scale invariance holds, the equations of motion are the same as in the case of a static spacetime. The nature of the expansion, encoded in the scale factor $a(t)$, then enters the transformation only back to the original temporal and spatial coordinates as well as field variables.
This makes the simulation in two spatial dimensions particularly efficient, since the evolution of a single experiment can conveniently be mapped to arbitrary expanding spacetimes in a post-processing step. For this reason, we will focus on weakly interacting \gls{2d} Bose gases in the remainder of this work. Although our universe is clearly not \gls{2d}, this geometry captures most of the essential physics of reheating dynamics, as we demonstrate in the following sections.

\section{\label{sec:preheating}Analog preheating}

Having discussed the connection between a scalar field in an expanding spacetime and an ultracold Bose gas, we are now in the position to address the question of how an analog of the preheating stage of post-inflationary dynamics can be simulated in a Bose gas.

\subsection{Preheating in the early universe}

In this subsection, we briefly review the basic mechanism of preheating in the early universe.
After inflation, the energy budget of the universe is assumed to be stored predominantly in the homogeneous inflaton field, which oscillates around the minimum of its potential. The stage of preheating consists of an explosive, non-perturbative particle creation from the decaying inflaton.

A common mechanism for preheating is via the parametric amplification of quantum fluctuations in the presence of the effective potential induced by the inflaton~\cite{Kofman:1997yn, Berges:2002cz}. This effect can most conveniently be understood by linearizing the fluctuations of the inflaton field around its homogeneous background, ${\inflaton(\cosmictime, \vect{x}) = \inflaton_0(\cosmictime) + \delta \inflaton(\cosmictime, \vect{x})}$, and inserting it in \cref{eq:eomrel}. The resulting equation describes damped oscillations of the background field,
\begin{equation}
\label{eq:eomrelhomog}
\frac{1}{c^2} \ddot{\inflaton}_0 + \frac{1}{c^2} d H \dot{\inflaton}_0 + \frac{m^2 c^2}{\hbar ^2} \inflaton_0 + \frac{\lambda}{6} \inflaton_0^3 = 0 .
\end{equation}
The equations of motion for the fluctuations, dropping all terms of quadratic or higher order in $\delta \inflaton$, read in Fourier space
\begin{equation}
\label{eq:eomrelfluct}
\delta \ddot{\inflaton}_{\vect{\momentum}} + d H \delta \dot{\inflaton}_{\vect{\momentum}} + \omega_{\vect{\momentum}}^2(\inflaton, \cosmictime) \delta \inflaton_{\vect{\momentum}} = 0
\end{equation}
with ${\omega_{\vect{\momentum}}^2(\phi_0, \cosmictime) / c^2 = \vect{\momentum}^2 / \hbar^2 a^2(\cosmictime) + m^2 c^2 / \hbar^2 + \lambda \inflaton_0^2 / 2}$ and the Fourier transform ${\delta \inflaton_{\vect{\momentum}}(\cosmictime) = \int \diff^d x \, \delta \inflaton(\cosmictime, \vect{x}) \etothepowerof{-i \vect{\momentum} \vect{x} / \hbar}}$. \Cref{eq:eomrelfluct} describes a collection of parametric oscillators for each momentum mode~$\vect{\momentum}$, driven by the background oscillations of $\inflaton_0(\cosmictime)$. Modes within certain instability bands satisfy resonance conditions, which leads to an exponential growth of occupancies corresponding to particle production~\cite{Greene:1998pb,Berges:2002cz}. The growth of fluctuations eventually invalidates the linearized approach of \cref{eq:eomrelfluct}.
Note that in more realistic models, couplings of the inflaton to other bosonic fields can lead, in a similar way, to a production of those degrees of freedom via parametric resonances as well~\cite{Kofman:1997yn}. Here, we focus on the simplest scenario corresponding to a decay of the inflaton solely into its own quanta of excitation.

\subsection{\label{sec:analogpreheatingbosegas}Analog preheating in Bose gases}

Our aim is to observe an analog of preheating dynamics in a cold-atom Bose gas. Motivated by the discussions in the previous sections, we mimic the state of the universe after inflation by a homogeneous \gls{bec}. (Possible implications of the presence of an external trapping potential are discussed in \cref{sec:experiment}.)
It is tempting to search for a direct analog of the inflaton's oscillations, which could trigger parametric instabilities in the condensate. However, this effect is not present for a single-component Bose gas in its ground state.

In fact, the absence of parametric resonance is a consequence of the non-relativistic limit considered in \cref{sec:expansion}.
To re-introduce parametric instabilities in the non-relativistic model, we add a periodic modulation of the coupling with frequency $\omega$ as $g \to g (1 + \sin \omega \cosmictime)$. As discussed in \cref{app:analogpreheating}, this procedure does not map one-to-one to the parametric resonance scenario described by \cref{eq:eomrelfluct}. For instance, the modulation frequency $\omega$ is non-relativistic in the simulation, while according to \cref{eq:eomrelhomog}, the inflaton oscillates at relativistic frequencies ${\omega_{\rm rel} \simeq m c^2 / \hbar}$ on the scale given by its rest mass. Furthermore, there is no dynamical mechanism of back reaction of the fluctuations on the modulation frequency or amplitude, as contained in the full model~\eqref{eq:eomrel}.

Despite these differences, modulating the interaction as described above suffices to generate the desired parametric resonance phenomena in the Bose gas~\cite{Staliunas2002,Engels:2007zz,PhysRevA.81.053627,PhysRevA.84.013618,Jaskula:2012ab,PhysRevX.9.011052,Zhang2020}, analogous to the ones taking place in the preheating stage of most inflationary models~\cite{Kofman:1997yn,Berges:2002cz}. This is demonstrated in detail in the subsequent subsections.

\subsection{\label{sec:preheatingnumerical}Numerical study of preheating dynamics}

We illustrate our implementation of preheating dynamics by means of numerical simulations based on \cref{eq:eomnonrel}, which corresponds to a static ultracold Bose gas with a time-dependent interaction.
We consider here a Bose gas in two dimensions, where the expansion does not explicitly enter the equations of motion, as discussed in \cref{sec:special2d}. (Nonetheless, we keep the number of spatial dimensions~$d$ in all formulas general.)
\Cref{eq:eomnonrel} is then formally equivalent to the equation of motion generated by the Hamiltonian~\eqref{eq:hamiltonian}, and the back transformation to the original time and field variables can be performed in post-processing. The latter is illustrated at the example of a matter-dominated expansion in \cref{sec:thermalizationvsexpansion}.

Following the argumentation in \cref{sec:analogpreheatingbosegas}, we induce parametric instabilities by harmonically modulating the coupling as
\begin{equation}
\label{eq:modulation}
g(\labtime) = g_0 \left( 1 + r \sin(\omega \labtime) \right) ,
\end{equation}
where $g_0$ is a positive offset value, $r$ is the amplitude of the modulation and $\omega$ is its frequency. Note that here we have used the laboratory time~$\labtime$ in the argument of the sine function instead of the cosmic time~$\cosmictime$, which, when expressed in terms of the laboratory time~$\labtime$, results in oscillations with increasing frequency. This substitution is based on the simplifying assumption that the expansion is insignificant during the duration of the modulation, in which case the relation between both time variables becomes linear and the corresponding proportionality constant can be absorbed in the modulation frequency.

In contrast to the post-inflationary setting, where the produced particles back-react on the inflaton, the modulation of the coupling is imposed externally in the simulation. Switching off the modulation thus corresponds to the decay of the inflaton, which constitutes a free parameter in the model.

A particularly useful observable for the study of parametric resonance and turbulent thermalization is the single-particle momentum distribution
\begin{equation}
\label{eq:spmd}
f(t, \vect{\momentum}) = \frac{1}{V} \braket{ \bosefield_{\vect{\momentum}}^\dagger(t) \bosefield_{\vect{\momentum}}(t) } ,
\end{equation}
where $V$ is the volume and ${\bosefield_{\vect{\momentum}}(t) = \int \diff^d x \, \bosefield(t, \vect{x}) \etothepowerof{-i \vect{\momentum} \vect{x} / \hbar}}$ denotes the Fourier transform of the Bose field $\bosefield(t, \vect{x})$.
Importantly, this quantity can be experimentally accessed in time-of-flight measurements (see \cref{sec:experiment} for further discussions).

\begin{figure}
	\includegraphics[width=\columnwidth]{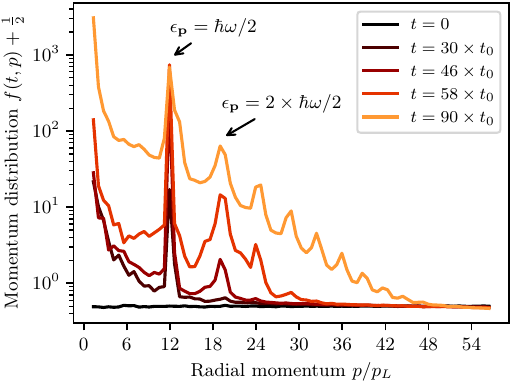}
	\caption{\label{fig:spmdpreheating} Radially averaged momentum distribution $f(t, \momentum)$ as a function of the radial momentum~${\momentum = |\vect{p}|}$ at different times $t$, demonstrating preheating dynamics. The coupling is modulated with relative strength ${r = \num{0.25}}$ at a frequency~$\omega$, chosen such that the resonance condition ${\epsilon_{\momentum_{\rm res}} = \hbar \omega / 2}$ for the momentum ${\momentum_{\rm res} = 12 \times p_L}$ with ${p_L = 2 \pi \hbar / L}$ is fulfilled [see discussion below \cref{eq:parametricoscillator}]. At early times ${t \lesssim 30 \times \timeunit}$, a single narrow resonance can be observed around $\momentum_{\rm res}$. At later times, a broad resonance band emerges with peaks at higher harmonics of the modulation frequency. These secondary resonances are due to non-linear interactions, as discussed in \cref{sec:secondaries}. See the video accompanying the arXiv version of this paper for a qualitative illustration of the dynamics of a single realization~\protect\cite{SuppMatArXiv}.}
\end{figure}

We evaluate the quantum expectation value in \cref{eq:spmd} by means of classical-statistical (or truncated Wigner) simulations~\cite{Khlebnikov:1996mc,Berges:2007ym,Sinatra2002,Blakie2008}. This method takes into account quantum fluctuations by stochastically sampling classical field configurations from the Wigner distribution of the initial state. Each realization is propagated deterministically according to the \gls{gpe} and quantum mechanical observables are obtained as statistical averages over multiple realizations. Here, the initial state is taken to be a spatially homogeneous \gls{bec}, and for each realization, all non-zero momentum modes are populated with vacuum noise corresponding to an average occupancy of half a particle per mode. This mimics quantum fluctuations acting as a seed for parametric instabilities. It is important to note that this approach goes beyond a mean-field description, which fails to capture parametric resonance since the occupancies of all excited modes are exactly zero. Further details about the simulation method can be found in \cref{app:numerical}.

In what follows, we express length and time in units of the characteristic scales ${\lengthunit = \hbar / \sqrt{m n_0 g_0}}$ and ${\timeunit = \hbar / n_0 g_0}$, respectively, where $n_0 = N / V$ is the homogeneous particle density in a system of $N$ particles in a volume $V$. Momenta are given either in units of the lowest non-zero momentum~${p_L = 2 \pi \hbar / L}$ or in units of the characteristic momentum~${p_{\xi} = 2 \pi \hbar / \xi}$ corresponding to the healing length ${\xi = \hbar / \sqrt{2 m n_0 g_0}}$.
In a quasi-\gls{2d} Bose gas, the interaction strength $g_0 = \tilde{g} \hbar^2 / m$ is characterized by the dimensionless parameter $\tilde{g} = \sqrt{8 \pi} a_s / a_{\mathrm{HO}}$, where $a_s$ is the $s$-wave scattering length and $a_{\mathrm{HO}}$ is the oscillator length of the harmonic potential in the strongly confined direction~\cite{Hadzibabic2011}. If not stated otherwise, we consider a uniform quasi-\gls{2d} system of $N = \num{e6}$ weakly interacting particles with coupling $\tilde{g} = \num{2.5e-3}$ in a square box with periodic boundary conditions. This choice of parameters fixes the box length as $L / x_0 = 50$.
Moreover, in this isotropic setting, the momentum distribution depends only on the magnitude $\momentum = |\vect{\momentum}|$ of the momentum, $f(t, \vect{\momentum}) \equiv f(t, \momentum)$.

\Cref{fig:spmdpreheating} shows the radially averaged momentum distribution of a parametrically excited system described by \cref{eq:eomnonrel}. The interaction is modulated according to \cref{eq:modulation} with $r = \num{0.25}$, and the modulation frequency $\omega$ is chosen as the resonance frequency of the momentum $|\momentum_{\rm res}| = 12 \times p_L$, as discussed below \cref{eq:parametricoscillator} in the next subsection. In this simulation, the noise cutoff has been chosen as $p_{\Lambda} = p_{\xi} \approx \num{70.7} \times p_L$. To gain a qualitative impression of the induced preheating dynamics, see the video accompanying the arXiv version of this paper, where we show the evolution of both the density and the phase of a single realization~\cite{SuppMatArXiv}.

At early times, $t \lesssim 30 \times \timeunit$, in \cref{fig:spmdpreheating}, we observe a single narrow resonance around the resonance momentum $\momentum_{\rm res}$ satisfying the resonance condition ${\epsilon_{\momentum_{\rm res}} = \hbar \omega / 2}$ (cf.~\cref{sec:lininst}). Due to particle number conservation, the growing occupancy of the resonant momentum causes the condensate to decay, mimicking particle production from the decaying inflaton in the early universe. At later times, secondary resonances at higher harmonics of the modulation frequency appear as a result of non-linear interactions among the produced quasi-particles.

In addition to the narrow resonance peaks, a transient growth of fluctuations at low momenta occurs at early times before the primary peak becomes visible. This growth can be interpreted as a consequence of the fact that the sampled initial state corresponds to the ground state of an ideal Bose gas. At $t = 0$, the system is effectively quenched to a finite interaction, producing a power law in the momentum distribution proportional to $p^{-1}$ at low momenta (cf.~\cref{app:truncatedwigner}). This early-time behavior has, however, no influence on the preheating dynamics we are interested in here.

In the following subsections, we provide analytical insights into both the linear regime of parametric resonance and the non-linear regime of secondary resonances.

\subsection{The linear regime of parametric resonance}
\label{sec:lininst}

Parametric instabilities play a crucial role in many modern experiments with \glspl{bec}~\cite{Staliunas2002,Engels:2007zz,PhysRevA.81.053627,Jaskula:2012ab,PhysRevX.9.011052,Zhang2020} and can conveniently be understood by adopting a semi-classical point of view. To this end, we consider the \gls{gpe} for the condensate wave function $\condensate(t, \vect{x})$,
\begin{equation}
\label{eq:gpe}
i \hbar \partial_t \condensate = \left( - \frac{\hbar^2 \nabla^2}{2m} + g(t) \left| \condensate \right|^2 \right) \condensate ,
\end{equation}
which is formally obtained as the classical equation of motion after replacing the Bose field operator in \cref{eq:eomnonrel} by its expectation value, ${\bosefield(t, \vect{x}) \to \condensate(t, \vect{x}) = \braket{\bosefield(t, \vect{x})}}$.
It is instructive to work in the Madelung representation, $\condensate(t, \vect{x}) = \sqrt{\density(t, \vect{x})} \exp \left( i \phase(t, \vect{x}) \right)$, which allows us to express the \gls{gpe}~\eqref{eq:gpe} in form of the hydrodynamic equations~\cite{Pitaevskii2016}
\begin{subequations}
\label{eq:gpehydro}
\begin{align}
\partial_t \density &= -\frac{\hbar}{m} \nabla \left( \density \nabla \phase \right) , \\
\hbar \partial_t \phase &= -g(t) \density + \frac{\hbar^2}{2m} \left[ \frac{\nabla^2 \sqrt{\density}}{\sqrt{\density}} - ( \nabla \phase )^2 \right] .
\end{align}
\end{subequations}
To obtain some intuition about the early stages of the evolution, we express both the density and the phase in terms of a homogeneous background with fluctuations on top of it, ${\density(t, \vect{x}) = \density_0(t) + \density_1(t, \vect{x})}$ and ${\phase(t, \vect{x}) = \phase_0(t) + \phase_1(t, \vect{x})}$. Linearizing \cref{eq:gpehydro} with respect to the fluctuations, yields the equations ${\partial_t \density_0 = 0}$ and ${\partial_t \phase_0 = - g(t) n_0 / \hbar}$ for the background condensate. For the fluctuations, we obtain
\begin{subequations}
	\label{eq:gpehydrolinear}
	\begin{align}
	\partial_t \density_1 &= -\frac{\hbar n_0}{m} \nabla^2 \phase_1 , \\
	\hbar \partial_t \phase_1 &= -g(t) \density_1 + \frac{\hbar^2}{4mn_0} \nabla^2 \density_1 .
	\end{align}
\end{subequations}
Taking the time derivative and inserting the resulting equations into each other, fluctuations of the density and the phase decouple to linear order. Transforming to momentum space, ${\density_{1 \, \vect{\momentum}}(t) = \int \diff^d x \, \density_1(t, \vect{x}) \etothepowerof{-i \vect{\momentum} \vect{x} / \hbar}}$ and similarly for the phase, the linearized equations can be expressed as
\begin{subequations}
\label{eq:parametricoscillator}
\begin{gather}
\label{eq:parametricoscillatordensity}
\partial_t^2 \density_{1 \, \vect{\momentum}} + \omega^2_{\vect{\momentum}}(t) \density_{1 \, \vect{\momentum}} = 0 , \\
\label{eq:parametricoscillatorphase}
\partial_t^2 \phase_{1 \, \vect{\momentum}} + \frac{2 n_0 \partial_t g(t)}{2 n_0 g(t) + \epsilon_{\vect{\momentum}, 0}} \partial_t \phase_{1 \, \vect{\momentum}} +  \omega^2_{\vect{\momentum}}(t) \phase_{1 \, \vect{\momentum}} = 0 ,
\end{gather}
\end{subequations}
where ${\hbar^2 \omega^2_{\vect{\momentum}}(t) = \epsilon_{\vect{\momentum}, 0} ( \epsilon_{\vect{\momentum}, 0} + 2 n_0 g(t) )}$ is a time-dependent form of the famous Bogoliubov dispersion relation \cite{Bogolyubov:1947zz}, and ${\epsilon_{\vect{\momentum}, 0} = \vect{\momentum}^2 / 2m}$ denotes the dispersion relation of a free particle.

\Cref{eq:parametricoscillator} describes a collection of parametric oscillators for each momentum mode $\vect{\momentum}$, which are undamped for the density and damped for the phase.
As discussed in \cref{app:mathieu}, these are special cases of Mathieu's equation~\cite{McLachlan1951}, which admits oscillatory solutions with exponentially growing amplitudes $\sim \etothepowerof{\zeta_{\vect{\momentum}} t}$, describing parametric resonance~\cite{Greene:1998pb,Berges:2002cz}. The resonance condition for the momentum mode $\vect{\momentum}_{\mathrm{res}}$ reads $\epsilon_{\vect{\momentum}_{\mathrm{res}}} = \hbar \omega / 2$, where $\epsilon_{\vect{\momentum}} = \sqrt{\epsilon_{\vect{\momentum}, 0} (\epsilon_{\vect{\momentum}, 0} + 2 n_0 g_0)}$ denotes the Bogoliubov dispersion relation and $\omega$ is the modulation frequency defined in \cref{eq:modulation}. That is, resonance occurs for those momentum modes $\vect{\momentum}_{\mathrm{res}}$ whose energy equals half a quantum of energy $\hbar \omega / 2$ injected in the system through the modulation.

In fact there is an entire range of modes around $\vect{\momentum}_{\mathrm{res}}$ which experience a positive growth rate, and the width of this instability band increases with the modulation amplitude $r$. To leading order in perturbation theory~\cite{Nayfeh1973} (see \cref{app:mathieu}), the growth rate of the resonant momentum mode is given by
\begin{equation}
\label{eq:growthrateres}
\zeta_{\vect{\momentum}_{\mathrm{res}}} = r \omega \left( \frac{n_0 g_0}{\hbar \omega} \right)^2 \left( \sqrt{1 + \left( \frac{\hbar \omega}{2 n_0 g_0} \right)^2} - 1 \right) .
\end{equation}
For $\hbar \omega \ll n_0 g_0$, this rate simplifies to $\zeta_{\vect{\momentum}_{\mathrm{res}}} \approx r \omega / 8$. In this regime, the Bogoliubov dispersion becomes linear, $\epsilon_{\vect{\momentum}} \approx c_{\mathrm{s}} |\vect{\momentum}| / \hbar$ with the speed of sound $c_{\mathrm{s}} = \sqrt{n_0 g_0 / m}$, and the produced quasi-particles have the character of sound waves. In the opposite limit, $\hbar \omega \gg n_0 g_0$, particles with quadratic dispersion $\epsilon_{\vect{\momentum}} \approx n_0 g_0 + \vect{\momentum}^2 / 2 m$ are produced.

In the simulation presented in \cref{fig:spmdpreheating}, parametric resonance is clearly visible as a pronounced peak at the momentum satisfying the resonance condition.
Likewise, the excitation of a single dominant wave length in the linear stage of the dynamics is qualitatively confirmed in the evolution of both the density and the phase of a single realization, as can be seen in the video accompanying the arXiv version of this paper~\cite{SuppMatArXiv} as well as in the snapshots\footnote{Since $\density_1$ and $\phase_1$ are conjugate variables, they are phase shifted such that the density fluctuations reach their maximum when the phase fluctuations cross zero, and vice versa. To create a better visual impression, the central snapshot of the phase in \protect\cref{fig:reheating} has been shifted forward in time by a quarter of an oscillation period until the phase fluctuations reach their next maximum.} shown in the central panels in the lower part of \cref{fig:reheating}.
It is worthwhile emphasizing that parametric instabilities can be triggered only if the initial occupancy is non-zero. This seed is not contained in the mean field analysis presented in this subsection, but is added in the simulation in form of vacuum noise according to the truncated Wigner prescription.

The linearized equations~\eqref{eq:gpehydrolinear} are helpful to get an intuitive analytical understanding for the early stages of the dynamics and describe the emergence of the primary resonant peak in \cref{fig:spmdpreheating}. However, as a result of the exponential growth of occupancies, this approach fails to describe the later stages where non-linear effects play a fundamental role. These non-linearities are taken into account by our numerical simulations, which are based on the full \gls{gpe}~\eqref{eq:gpe}, and include secondary excitations outside the resonance band, as shown in \cref{fig:spmdpreheating}. These will be discussed further in the subsequent section.

\subsection{\label{sec:secondaries}Secondary instabilities}

In this section we discuss the first non-linear corrections, which lead to secondary instabilities~\cite{Berges:2002cz,Zache:2017dnz}. Within our semi-classical picture, these can be understood by considering the hydrodynamic equations of motion~\eqref{eq:gpehydro} for the fluctuations to quadratic order, i.e., including terms $\mathcal{O}(\phase_1^2)$, $\mathcal{O}(\phase_1 (n_1/n_0))$ and $\mathcal{O}((\density_1/\density_0)^2)$. This leads to
\begin{subequations}
	\begin{align}
	\partial_t \density_1 = &-\frac{\hbar \density_0}{m} \nabla^2 \phase_1 - \frac{\hbar}{m} \left(\density_1 \nabla^2 \phase_1 + \nabla \density_1 \nabla \phase_1\right) , \\
	\hbar \partial_t \phase_1 = &-g \density_1 +\frac{\hbar^2}{4mn_0} \nabla^2 \density_1 \nonumber \\
	&- \frac{\hbar^2}{2m} \left[ \left( \nabla \phase_1 \right)^2 + \frac{\left( \nabla \density_1 \right)^2}{4\density_0^2} + \frac{\density_1 \nabla^2 \density_1}{2\density_0^2}  \right] .
	\end{align}
\end{subequations}
As can be seen, density and phase fluctuations no longer decouple to this order.

Taking the time derivative of the above equation for the density fluctuations and inserting the expressions for the first-order time derivatives of the fluctuations, the result can be written in momentum space as
\begin{equation}
\label{eq:momint}
\begin{gathered}
\partial_t^2 \density_{1\, \vect{p}} +\omega_{\vect{p}}^2(t) \density_{1\, \vect{p}}
= \frac{\density_0}{2m^2\hbar^2} \int_{\vect{q}} \phase_{1\, \vect{p-q}}  \phase_{1\, \vect{q}} u(\vect{p}, \vect{q}) \\
+\int_{\vect{q}} \density_{1\, \vect{p-q}}  n_{1\, \vect{q}} \left[  \frac{v(\vect{p}, \vect{q})}{8m^2\hbar^2 \density_0}- \frac{g(t)}{m\hbar^2} \vect{pq} \right]
\end{gathered}
\end{equation}
with
$u(\vect{p}, \vect{q}) = \vect{p}^2(\vect{pq} - \vect{q}^2) + 2(\vect{pq})(\vect{p}-\vect{q})^2$,
$v(\vect{p}, \vect{q}) = \vect{p}^2 (2\vect{p}^2 - 3\vect{pq}+\vect{q}^2) - 2 (\vect{pq})\vect{q}^2$ and $\int_{\vect{q}} = \int \diff^d q / (2 \pi \hbar)^d$. In deriving \cref{eq:momint}, we have neglected all cubic terms in the fluctuations.

The above momentum integrals are dominated by the contribution from the exponentially growing unstable modes $\vect{q}$, for which ${|\vect{p}-\vect{q}| \approx |\vect{q}| \approx p_{\mathrm{res}}}$. For modes that are stable on the linear level, the integrals therefore act as source terms and \cref{eq:momint} describes forced harmonic oscillators with an exponentially growing force. This effect is analyzed in more detail in \cref{app:secondaries}. There, we show that the forcing leads to an exponential growth of the momentum modes with $p \lesssim 2 p_\mathrm{res}$ proportional to $\etothepowerof{2\zeta_{\vect{\momentum}_{\mathrm{res}}} t}$, with a growth rate $\zeta_{\vect{\momentum}_{\mathrm{res}}}$ given by \cref{eq:growthrateres}. Modes, for which in addition $\epsilon_{\vect{p}} \approx 2\epsilon_{\vect{p}_\mathrm{res}}$ holds, experience a resonant amplification and are strongest enhanced.

These features of secondary instabilities are captured by our numerical simulations. In particular, one can observe both the narrow peak and the broad band in the distribution function in \cref{fig:spmdpreheating} between $t = 46 \times \timeunit$ and $t = 58 \times \timeunit$. Similar peaks at higher multiples of the resonance frequency appear at later times due to higher-order corrections to \cref{eq:gpehydro}.

The perturbative analysis breaks down when ${n_1 / n_0 \approx 1}$, i.e., when the number of excited atoms becomes comparable to the number of condensate atoms. At this point, the exponential growth stops and turbulent dynamics sets in, which can be observed qualitatively in the video accompanying the arXiv version of this paper~\cite{SuppMatArXiv}. A typical snapshot of the density and the phase of a single realization after the onset of turbulence is shown in the right panels in the lower part of \cref{fig:reheating}. In the next section, we present a quantitative analysis of turbulent dynamics in momentum space.

In \cref{app:instabilitiesquantum}, we discuss the theory of parametric instabilities and secondary excitations from the perspective of quantum equations of motion for correlation functions, which validates the use of the classical-statistical approximation.

\subsection{Analysis of growth rates}

\begin{figure}
	\includegraphics[width=\columnwidth]{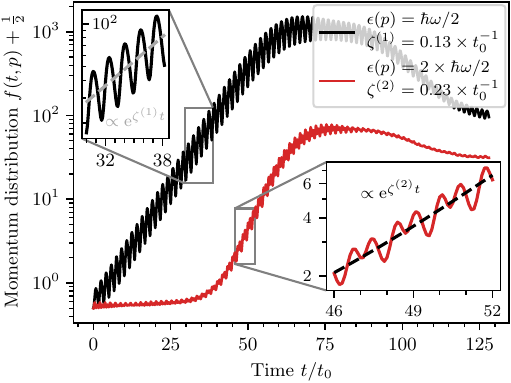}
	\caption{\label{fig:growthrates}Radially averaged momentum distribution $f(t, p)$ as a function of time $t$ showing the exponential growth of the primary and secondary resonances corresponding to the annotated peaks in \cref{fig:spmdpreheating}. We have extracted the growth rates by fitting a straight line to the quantity $\ln f(t, p)$, as shown in the insets. The resulting growth rate of the primary resonance ${\zeta_{\mathrm{num}}^{(1)} = \num{0.13} \times \timeunit^{-1}}$ agrees well with the analytical prediction ${2 \zeta_{\mathrm{pert}}^{(1)} = \num{0.15} \times \timeunit^{-1}}$ obtained from \cref{eq:growthrateres}. The growth of the secondary instability starts later, but its rate ${\zeta_{\mathrm{num}}^{(2)}= \num{0.23} \times \timeunit^{-1}}$ is approximately twice as large as the one of the primary resonance, as expected from our discussion in \cref{sec:secondaries}. The exponential growth stops when the number of excited atoms becomes comparable to the number of condensate atoms.}
\end{figure}

\Cref{fig:growthrates} depicts the time evolution of the occupancies corresponding to the primary and secondary resonances annotated in \cref{fig:spmdpreheating}. The growth of the secondary resonance marks the onset of the non-linear regime where the quasi-particles produced in the primary resonance start to interact, and the linear picture presented in \cref{sec:lininst} breaks down.

Before comparing the numerically extracted growth rates to the analytical predictions presented in the previous sections, we first relate the momentum distribution $f(t, \vect{\momentum})$ to the hydrodynamic density and phase variables. On the mean-field level, linear fluctuations of the condensate wave function, expressed as ${\condensate(t, \vect{x}) = \condensate_0(t) + \condensate_1(t, \vect{x})}$, are related to linear density and phase fluctuations via ${\condensate_1(t, \vect{x}) / \condensate_0(t)} = { \density_1(t, \vect{x}) / 2 n_0 + i \phase_1(t, \vect{x}) }$.
The momentum distribution on the linear level then corresponds to
\begin{equation}
\label{eq:spmdlinear}
\left| \condensate_{1\, \vect{\momentum}} \right|^2 = n_0 \left[ \frac{n_{1\, \vect{\momentum}}^2}{(2 n_0)^2} + \phase_{1\, \vect{\momentum}}^2 \right] .
\end{equation}
Therefore, a parametric resonance where the density and phase fluctuations grow as ${\density_{1\, \vect{\momentum}} \sim \phase_{1\, \vect{\momentum}} \sim \etothepowerof{\zeta_{\vect{\momentum}} t}}$ results in a growth of the momentum distribution as ${f(t, \vect{\momentum}) \sim \etothepowerof{2 \zeta_{\vect{\momentum}} t}}$.

The oscillations of the occupancies in \cref{fig:growthrates} can be understood from the linearized parametric oscillator equations~\eqref{eq:parametricoscillator}. Recall that the latter admit oscillatory solutions with exponentially growing amplitudes. However, being conjugate variables, the oscillations of $n_1$ and $\theta_1$ are shifted in phase by approximately $\pi / 2$. According to \cref{eq:spmdlinear}, the momentum distribution thus corresponds to the sum of two phase-shifted oscillating functions with slightly different initial amplitudes. This results in the residual oscillations on top of the exponential growth observed in \cref{fig:growthrates}. We have checked that the oscillation frequency of the primary resonance agrees with the modulation frequency, while the oscillations of the secondary resonance additionally contain frequency components corresponding to twice the modulation frequency, reflecting the interactions between the resonantly produced quasi-particles.

We have extracted the growth rate of the primary resonance ${\zeta_{\mathrm{num}}^{(1)} = \num{0.13} \times \timeunit^{-1}}$ by fitting an exponential function to the numerical data, as shown in the insets of \cref{fig:growthrates}. The result is close to the analytical prediction~\eqref{eq:growthrateres} obtained from perturbation theory, ${2 \zeta_{\mathrm{pert}}^{(1)} = \num{0.15} \times \timeunit^{-1}}$. The secondary resonance at $2\epsilon_{\vect{\momentum}_\mathrm{res}}$ in \cref{fig:growthrates} grows at the rate ${\zeta_{\mathrm{num}}^{(2)}= \num{0.23} \times \timeunit^{-1}}$, which is indeed approximately twice the growth rate of the primary resonance.
The secondary resonance quickly saturates, indicating the breakdown of the perturbative analysis presented in the previous section. This marks the transition into a highly non-linear regime leading to the formation of turbulence, which is the topic of the next section.

\section{\label{sec:reheating}Analog reheating}

Having studied the main features of preheating-like dynamics along with a possible implementation in an ultracold Bose gas, we now turn to the reheating process and the question of how the system finally approaches thermal equilibrium.

Particle spectra formed during preheating are highly non-thermal with large occupation numbers at low momenta.
As a consequence, the system enters a turbulent state, characterized by a local transport of conserved quantities in momentum space. Typically, an inverse cascade transports particles towards lower momenta, while a direct cascade transports energy towards higher momenta, constituting a key process in the context of turbulent thermalization~\cite{Micha:2004bv}. At early times, this transport is driven, i.e., the oscillating inflaton acts as a source injecting energy into the system at resonant momenta. Eventually, the inflaton decays, marking the transition from driven to free turbulence. The system remains in the turbulent state for a long time, until the occupancy of characteristic momenta eventually becomes comparable to the vacuum expectation value given by the \enquote{quantum half}. In this final stage, which is dominated by quantum fluctuations, the system relaxes to thermal equilibrium, completing the reheating process~\cite{Micha:2004bv, Amin:2014eta}.

Turbulent dynamics is accompanied by the emergence of self-similarity and universality. This is reflected by a power-law behavior of the momentum distribution within a certain inertial range of momenta.
One well-known example within the theory of weak wave turbulence is the prediction of a stationary direct cascade with a universal power-law distribution ${f(p) \propto p^{- d}}$~\cite{Zakharov1992}.
More generally, self-similarity in far-from-equilibrium systems can become manifest in their time evolution. In such a scenario the evolution of the distribution function can be expressed as
\begin{equation}
\label{eq:selfsimilar}
f(t, \vect{p}) = s^{\alpha} f_{\mathrm{S}}(s^\beta \vect{p} ) ,
\end{equation}
where $s = t / t_{\mathrm{ref}}$ and $t_{\mathrm{ref}}$ is an arbitrary reference time.
The scaling hypothesis~\eqref{eq:selfsimilar} constitutes a significant reduction of complexity as it allows to describe a relevant part of the system's dynamics by simply rescaling a single-variable scaling function $f_{\mathrm{S}}$ as determined by the scaling exponents $\alpha$ and $\beta$.
Remarkably, in many far-from-equilibrium scenarios, both the exponents as well as the scaling form of the distribution are universal, which means they are insensitive to microscopic details as well as initial conditions, and depend only on a few macroscopic system parameters like dimensionality, symmetry or the number of field components~\cite{Micha:2004bv, Berges:2014bba, Orioli:2015dxa, Berges:2017ldx, Schmied:2018mte}. As a result, if universality holds, physical systems with vastly differing energy scales can behave quantitatively the same. This makes ultracold Bose gases a particularly promising target for simulating universal aspects of far-from-equilibrium dynamics like that of the inflaton in the early universe. Universal self-similar time evolution reflects the system being in the vicinity of a non-thermal fixed point, which acts as an attractor on the way towards thermal equilibrium~\cite{Berges:2008sr, Berges:2008wm, Scheppach:2009wu}.

In this section, we demonstrate by means of classical-statistical simulations that the salient features of these phenomena can be observed in our implementation of reheating dynamics. We first study the regimes of driven and free turbulence separately, before considering a transient prescaling regime, where the universal shape of the scaling function is maintained, but the scaling exponents change over time. From our numerical simulations, we extract the scaling exponents as well as the scaling form of the distribution and compare the results with analytical predictions from kinetic theory. Finally, we discuss the relaxation to thermal equilibrium, which is, however, not captured by our classical-statistical simulations as it is dominated by quantum fluctuations. We conclude this section with a discussion of an example of how expansion may prevent the system from thermalizing, building a bridge to \cref{sec:expansion}.

\subsection{Driven versus free turbulence}

In order to drive the system into a turbulent state, we follow the protocol presented in \cref{sec:preheatingnumerical} of parametrically exciting a homogeneous \gls{bec}. Here, our focus lies on the later stages of the non-linear dynamics after the proliferation of secondary instabilities when a smooth distribution in form of a power law has formed.
We distinguish the regime of driven turbulence, realized by continuously modulating the interaction according to \cref{eq:modulation}, and free turbulence, developed if the modulation is switched off shortly after the primary resonance has saturated. In our analogy to reheating in the early universe, the former case corresponds to the situation where the inflaton possesses enough energy to drive turbulence for a long time, while in the latter case, the inflaton runs out of energy rather quickly at around the same time when turbulence sets in.

To maximize the inertial range where self-similar scaling can be observed, it is desirable to inject energy at momentum scales close to the lowest momenta supported by the system where occupancies can become large. To this end, the interaction is modulated with a relative amplitude $r = 1$ at frequency $\omega$ chosen such that the resonant momentum becomes $p_{\rm res} = 3 \times 2 \pi \hbar / L$ (cf.~\cref{sec:lininst}).

\begin{figure*}
	\subfloat[][\label{fig:driventurbulence}Direct cascade with continuously modulated interaction.]{\includegraphics[width=\columnwidth]{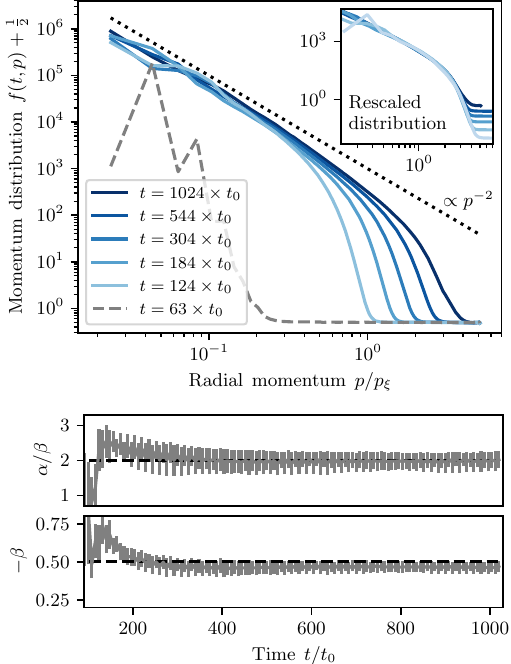}}\hspace{18pt}\subfloat[][\label{fig:freeturbulence}Direct cascade after switching off the modulation.]{\includegraphics[width=\columnwidth]{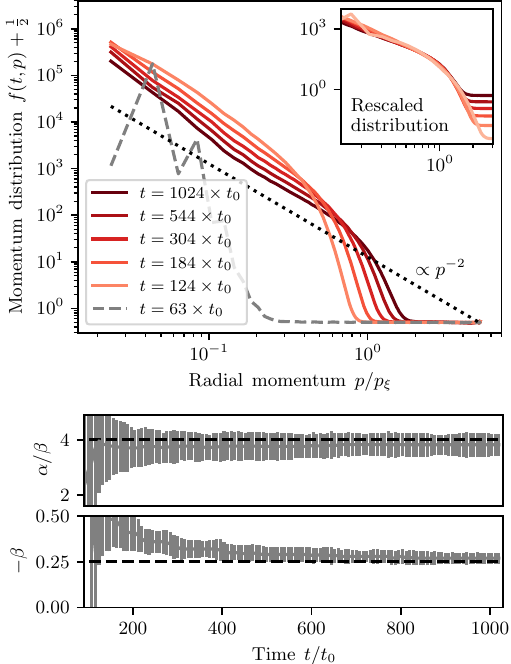}}
	\caption{\label{fig:turbulence}Self-similar time evolution of the momentum distribution in form of a direct energy cascade for driven (a) and free turbulence (b). Energy is injected at low momenta by modulating the scattering length according to \cref{eq:modulation} with a relative amplitude ${r = 1}$ at a frequency $\omega$ chosen such that ${\hbar \omega / 2 = \epsilon_{p_{\rm res}}}$ with ${p_{\rm res} = 3 \times 2 \pi \hbar / L}$. In the case of continuous modulation (a), a stationary distribution with a power law close to $p^{-2}$ develops, whose front is evolving self-similarly. If the driving is switched off once the primary resonance has saturated, corresponding here to ${t = 80 \times \timeunit}$ (b), energy is propagated in a self-similar way to higher momenta, but the distribution at a given momentum decreases with time, reflecting energy conservation. A power law proportional to $p^{-2}$ is shown in form of a dotted line as a guide to the eye. The insets show the distributions rescaled according to \cref{eq:selfsimilar} using the numerically extracted scaling exponents displayed below the respective distributions.}
\end{figure*}

The left and right panels of \cref{fig:turbulence} show a comparison of the direct cascades emerging in the regimes of driven and free turbulence, respectively. All parameters for both simulations are identical, with the exception that in \cref{fig:driventurbulence}, the interaction is modulated continuously, while in \cref{fig:freeturbulence}, the modulation is switched off smoothly\footnote{We have checked that the differences between switching off the modulation suddenly or smoothly within a few modulation periods are insignificant.} within two modulation periods ${2 \pi / \omega \approx \num{8.2} \times t_0}$ at time ${t = 80 \times \timeunit}$, roughly corresponding to the time when the primary resonance saturates. In the latter case, energy is conserved already at the onset of turbulence and the driven regime is skipped. In both scenarios, the momentum distribution takes a scaling form corresponding to a power law close to ${f(t, p) \propto p^{-2}}$, which is indicated by a dotted line as a guide to the eye. A power law proportional to ${p}^{-d}$ is characteristic for weak wave turbulence~\cite{dyachenko1992optical, Zakharov1992, Nowak:2011sk} and has been observed experimentally in Ref.~\cite{navon2016emergence}.

Moreover, as it can be seen in \cref{fig:turbulence}, the distributions in the two regimes exhibit self-similar time evolution in different ways.
In the case of driven turbulence, \cref{fig:driventurbulence}, the front of the cascade evolves self-similarly, leaving behind a stationary distribution.
Stationary turbulence arises in the theory of weak wave turbulence as a stationary solution of the scattering integral. However, such a configuration necessarily requires the existence of at least a source and a sink for energy to be injected and dissipated, respectively~\cite{Zakharov1992}. In the present case, the energy source is provided by the modulation of the interaction. We have verified that the rate of energy injection into the system is approximately constant in this stationary regime, as expected for driven turbulence~\cite{Micha:2004bv}.
Since energy is transported locally in momentum space, unoccupied higher momentum modes play the role of an energy sink~\cite{Micha:2004bv}. This allows for the build-up of a stationary distribution, although the model lacks a mechanism of dissipation.
By contrast, the distribution in the case of free turbulence, \cref{fig:freeturbulence}, is not stationary, but decreases as a function of time for a given momentum, reflecting energy conservation.

To quantify the self-similar time evolution, we have extracted the scaling exponents $\alpha$ and $\beta$ defined in \cref{eq:selfsimilar} from our numerical data at different times using the maximum likelihood technique described in \cref{app:extractscalingexponents}. In the turbulent stage of the dynamics, the exponents are approximately constant at late times, as shown in the lower part of \cref{fig:turbulence}. The observed slow relaxation of the exponents at early times in \cref{fig:freeturbulence} can be interpreted as prescaling, as discussed in the subsequent subsection.
At the latest simulated times, the exponents take the values
\begin{align}
\label{eq:exponentsdriven}
\alpha_{\mathrm{driven}} = -0.95 \pm 0.17 , \quad \beta_{\mathrm{driven}} = -0.46 \pm 0.06 ,
\end{align}
for driven turbulence, and
\begin{align}
\label{eq:exponentsfree}
\alpha_{\mathrm{free}} = -1.05 \pm 0.11 , \quad \beta_{\mathrm{free}} = -0.27 \pm 0.03 ,
\end{align}
for free turbulence, respectively. Rescaling the distribution according to \cref{eq:selfsimilar} with the extracted exponents, as described in \cref{app:extractscalingexponents}, all data points collapse to a single universal scaling function $f_{\mathrm{S}}(p)$, as shown in the insets of \cref{fig:turbulence}. The exponent $\beta$ describes the speed of energy propagation towards higher momenta, which is higher for the driven cascade than for the free cascade. Analytical predictions for this exponent from kinetic theory as well as the relations between $\alpha$ and $\beta$ following from conservation laws will be discussed in \cref{sec:kinetictheory}.

The above values of the exponents are insensitive to the details of how the far-from-equilibrium state is approached. In particular, we have observed the same exponents starting from an initial state with a highly occupied narrow window of momenta on top of a condensate background. Such an initial state is similar to the state of the system at the end of the preheating stage, when a certain momentum mode is overpopulated as a consequence of parametric resonance. 

The slow power-law dynamics of the direct cascade can be challenging to capture with classical-statistical simulations for experimentally realistic configurations, since this method is known to be prone to instabilities caused by the vacuum noise~\cite{Blakie2008}. These instabilities manifest in a decay of the \enquote{quantum half} and the generation of spurious quantum pressure, resulting in an unphysical dependence of the results on the \gls{uv} cutoff~\cite{Berges2014}. Classical statistical simulations are therefore restricted to large occupancies and weak couplings, where these inevitable deficiencies are mitigated via a separation of scales.
To simulate reheating dynamics in the turbulent regime, we have therefore increased the particle number and reduced the coupling such that the validity of the classical-statistical approximation can be ensured. This is discussed in detail in \cref{app:validity}, where we also assess the range of accessible coupling strengths for our setup. In \cref{sec:experiment}, we discuss how our numerical results relate to realistic experimental conditions.

\subsection{\label{sec:prescaling}Prescaling}

 \begin{figure*}
	\includegraphics[width=\textwidth]{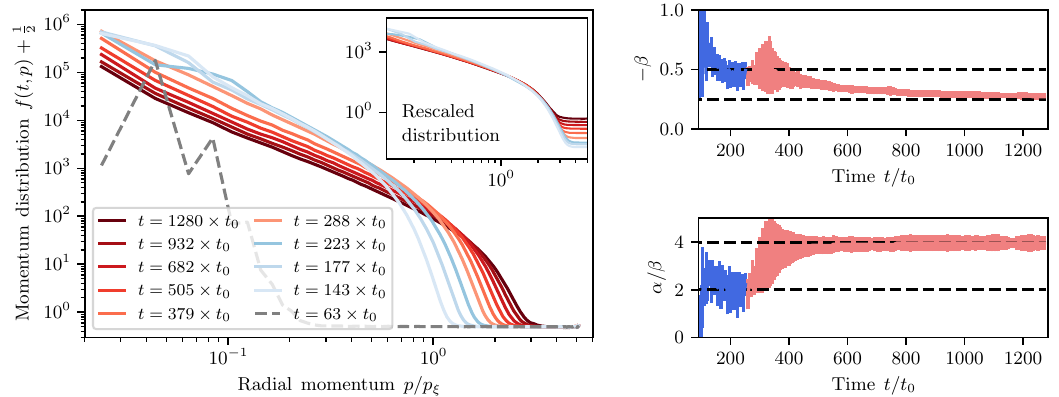}
	\caption{\label{fig:prescaling}Prescaling at the transition from driven to free turbulence. The simulation parameters are identical to those in \cref{fig:turbulence}, but the modulation is switched off suddenly at a later time $t = 256 \times \timeunit$. Before this time (blue curves), turbulence is driven and both the momentum distribution as well as the scaling exponents $\alpha$ and $\beta$ are the same as in \cref{fig:driventurbulence}. After switching off the modulation (red curves), the ratio $\alpha / \beta$ quickly changes to the one expected for free turbulence, reflecting energy conservation. The exponent $\beta$ gradually changes towards the value obtained for free turbulence in \cref{fig:freeturbulence}, reducing the speed of energy transport in the cascade. Although the scaling exponents still change in time, the distribution has already attained its universal scaling form. This important hallmark of prescaling is indicated in the inset, where all data points collapse to a single curve after rescaling according to \cref{eq:selfsimilar} with the extracted time-dependent scaling exponents $\alpha(t)$ and $\beta(t)$, as described in \cref{app:extractscalingexponents}.}
\end{figure*}

So far, we have investigated the two limiting cases where turbulence is either driven or free. Now, we address the transient regime corresponding to the somewhat more realistic situation where, in the beginning, turbulence is driven by the inflaton oscillations, but at some point goes over to free turbulence when the inflaton has decayed.

This scenario is illustrated in \cref{fig:prescaling}. Up to the time ${t = 256 \times \timeunit}$, the direct cascade is driven (blue curves), as in \cref{fig:driventurbulence}. At this time, the modulation is switched off, mimicking the decay of the inflaton. We then observe a slowing down of the speed of energy propagation and the distribution decreases in time for a given momentum (red curves), reminiscent of the direct cascade of free turbulence shown in \cref{fig:freeturbulence}.

On the right-hand side of \cref{fig:prescaling}, the numerically extracted scaling exponents are shown as a function of time. After switching off the modulation, the exponent $\beta$ slowly evolves from a value close to the one reported in \cref{eq:exponentsdriven} for driven turbulence to a value close to the one reported in \cref{eq:exponentsfree} for free turbulence. By contrast, the ratio $\alpha / \beta$ changes rather quickly between these two regimes. This behavior is expected since this ratio is fixed by energy conservation, which is enforced instantaneously after switching off the modulation, cf.~\cref{sec:kinetictheory}. Surprisingly, the self-similar scaling form of the distribution is approximately preserved during the transition, which is clearly visible in the inset of \cref{fig:prescaling}, where the distributions, rescaled according to \cref{eq:selfsimilar} with the time-dependent scaling exponents, fall on top of each other.

Recently, such a situation, where the system's dynamics is governed by a universal scaling function much before the corresponding exponents have attained their universal values, has been studied in the context of heavy-ion collisions~\cite{Mazeliauskas:2018yef}. This phenomenon, termed prescaling, is closely related to the emergence of a far-from-equilibrium hydrodynamic behavior, as it allows to describe the dynamics in terms of few slowly changing parameters~\cite{Mazeliauskas:2018yef}.
A different type of prescaling, where certain correlation functions already scale with their universal exponents at early times while others do so only at much later times, has been studied numerically in three-component Bose gases~\cite{Schmied2019}.
Our results indicate that prescaling, as defined in Ref.~\cite{Mazeliauskas:2018yef}, can be observed during the transition from driven to free turbulence\footnote{A much shorter and less pronounced stage of prescaling occurs also in the scenario shown in \cref{fig:freeturbulence}, where the modulation is switched off much earlier, at $t = 80 \times t_0$.}, opening up new paths to studying this phenomenon experimentally.

In \cref{app:methodofmoments}, we corroborate our analysis of prescaling using an alternative method based on the moments of the distribution, which is particularly suitable for extracting time-dependent scaling exponents~\cite{Mazeliauskas:2018yef}.

\subsection{\label{sec:kinetictheory}Kinetic description}

We now discuss how the values of the scaling exponents for the direct cascade presented in \cref{eq:exponentsdriven,eq:exponentsfree} can be understood from an analytical point of view.

One relation between the exponents $\alpha$ and $\beta$ follows from the scaling of the total energy. To see this, we assume a self-similar time evolution according to \cref{eq:selfsimilar} as well as a power-law scaling of the dispersion relation, $\epsilon_{\vect{p}} \propto |\vect{p}|^z$, where $z \approx 2$ in the particle regime of the Bogoliubov dispersion. Then, the total kinetic energy scales as
\begin{equation}
\label{eq:scalingenergy}
E(t) = V \int \frac{\diff^d p}{(2 \pi \hbar)^d} \, \epsilon_{\vect{p}} f(t, \vect{p})
= \left( \frac{t}{t_{\mathrm{ref}}} \right)^{\alpha-(d+z)\beta} E(t_{\mathrm{ref}}) .
\end{equation}
In the case of free turbulence, energy is conserved across the cascade, $E = \mathrm{const}$, while in the case of driven turbulence, it grows linearly with time, $E \propto t$. This implies the relation
\begin{equation}
\label{eq:scaling1}
\alpha = \gamma +(d + z) \beta ,
\end{equation}
where $\gamma = 1$ in the driven regime and $\gamma = 0$ in the absence of driving.

Another relation between the scaling exponents can be obtained with the help of perturbative%
\footnote{Large occupation numbers usually require going beyond perturbative kinetic theory for the description of the dynamics near a non-thermal fixed point. An example is the dynamics of the inverse cascade leading to Bose condensation~\cite{Orioli:2015dxa}. For the direct cascade, however, occupation numbers of characteristic momenta are typically lower (although still much larger than one), such that a perturbative analysis is applicable (see also~\cite{Micha:2004bv}).}
kinetic theory for Bose gases~\cite{dyachenko1992optical, Orioli:2015dxa}. The time evolution of the distribution function in this framework is written in the form of a Boltzmann equation (see, e.g., Ref.~\cite{Proukakis:1998zz}),
\begin{equation}
\label{eq:boltzmann}
\partial_t f(t,\vect{p}) = C_{2\leftrightarrow2}[f](t,\vect{p}) +C_{1\leftrightarrow2}[f](t,\vect{p}) ,
\end{equation}
where
\begin{equation}
\label{eq:222}
\begin{split}
C_{2\leftrightarrow2}[f] = 4 \pi g^2 &\int_{\vect{p_2}, \vect{p_3}, \vect{p_4}}
\delta(\vect{p} + \vect{p}_2 - \vect{p}_3 - \vect{p}_4) \\
&\times \delta (\epsilon_{\vect{p}} + \epsilon_{\vect{p}_2} - \epsilon_{\vect{p}_3} - \epsilon_{\vect{p}_4}) \\
&\times[(f_{\mathbf{p}}+1) (f_{\mathbf{p}_2}+1) f_{\mathbf{p}_3} f_{\mathbf{p}_4} \\
&\phantom{\times[}- f_{\mathbf{p}} f_{\mathbf{p}_2} (f_{\mathbf{p}_3}+1) (f_{\mathbf{p}_4}+1)  ]
\end{split}
\end{equation}
describes ${2 \leftrightarrow 2}$ collisions between non-condensate atoms and
\begin{equation}
\label{eq:122}
\begin{split}
C_{1\leftrightarrow2}[f] = 2 (2 \pi)^4 \hbar n_0 g^2 \int_{\vect{p_1}, \vect{p_2}, \vect{p_3}}
\delta(\vect{p}_1-\vect{p}_2-\vect{p}_3) \\
\times \delta (\epsilon_{0}+\epsilon_{\vect{p}_1}- \epsilon_{\vect{p}_2}-\epsilon_{\vect{p}_3}) \\
\times \left[ \delta(\vect{p}-\vect{p}_1) - \delta(\vect{p}-\vect{p}_2) - \delta(\vect{p}-\vect{p}_3) \right] \\
\times \left[(f_{\vect{p}_1}+1) f_{\vect{p}_2} f_{\vect{p}_3} - f_{\vect{p}_1} (f_{\vect{p}_2}+1) (f_{\vect{p}_3}+1) \right]
\end{split}
\end{equation}
arises in the presence of the condensate and describes effective ${1 \leftrightarrow 2}$ collisions between the non-condensate atoms. Here, $\epsilon_0$ and $\epsilon_{\vect{p}}$ are the energies of condensate and non-condensate atoms, respectively, and we have used the abbreviations ${f_{\vect{p}} = f(t, \vect{p})}$ and ${\int_{\vect{p}} = \int \diff^d p / (2 \pi \hbar)^d}$. In our case, $\epsilon_0$ can be neglected since we are interested in the direct cascade, which predominantly involves atoms that are much more energetic than condensate atoms.

In order to derive the desired relation between the scaling exponents, let us assume that the right-hand side of the Boltzmann equation~\eqref{eq:boltzmann} is dominated by one of the two scattering integrals, and denote the number of particles involved in the corresponding scattering processes by $l$, i.e., we have $l=4$ for $2\leftrightarrow2$ scattering and $l=3$ for $1\leftrightarrow2$ scattering. Comparing how the left- and right-hand sides of \cref{eq:boltzmann} scale with time, and assuming large occupation numbers, $f_{\mathbf{p}} \gg 1$, one arrives at~\cite{Orioli:2015dxa}
\begin{equation}
(l-2)\alpha = [(l-2)d - 2]\beta - 1 .
\label{eq:scaling2}
\end{equation}
Combining this result with \cref{eq:scaling1} yields
\begin{equation}
\label{eq:predictionbeta}
\beta = - \frac{-1-(l-2)\gamma}{(l-2)z+2} .
\end{equation}

For driven turbulence, $\gamma=1$, and a quadratic dispersion, $z=2$, both collision terms lead to $\alpha = -1$ and $\beta = -1/2$. These values are close to those in \cref{eq:exponentsdriven}, extracted from our simulation in the presence of a modulated coupling. For free turbulence, $\gamma =0$, one recovers the ratio $\alpha/\beta=4$ reflecting energy conservation, which is close to the numerical result in \cref{eq:exponentsfree}. The value of $\beta$ in this case is ${\beta = -1/6}$ for ${2\leftrightarrow2}$ scattering and ${\beta = -1/4}$ for ${2\leftrightarrow1}$ scattering. The latter is closer to \cref{eq:exponentsfree}, which is observed at late times in our simulations.
By contrast, in the experiment reported in Ref.~\cite{Glidden:2020qmu}, where, instead of parametrically driving a pure condensate out of equilibrium, a cooling quench has been applied to an initially uncondensed Bose gas, a scaling exponent closer to the $2 \leftrightarrow 2$ scattering solution, $\beta = - 1 / 6$, has been observed.
Deviations from the predicted scaling can in general be induced by a non-quadratic scaling of the dispersion relation~$\epsilon_{\mathbf{p}}$ or by the time dependence of the number of atoms in the condensate.

\subsection{\label{sec:thermalization}Thermalization}

As we have seen, on its way to thermal equilibrium the system takes a detour via a non-thermal fixed point, in the vicinity of which the dynamics is dominated by a turbulent transport of energy towards higher momenta. The self-similar dynamics of the direct cascade stops once the occupancy of the characteristic momentum dominating the kinetic energy budget becomes comparable to the expectation value of the vacuum noise given by the \enquote{quantum half}. At this point, quantum fluctuations become dominant over statistical fluctuations~\cite{Micha:2004bv} and the system is expected to relax to a Bose--Einstein distribution
\begin{equation}
f_{\mathrm{BE}}(\vect{p}) = \frac{1}{\exp(\epsilon_{\vect{p}} / \kb \reheatingtemp) - 1} ,
\end{equation}
where $\epsilon_{\vect{p}}$ is the dispersion relation, $\kb$ is the Boltzmann constant and $\reheatingtemp$ is the reheating temperature.

The dominance of quantum fluctuations makes this final stage of the dynamics inaccessible to classical-statistical simulations. In contrast to the expected relaxation to a Bose--Einstein distribution, at sufficiently late times in the numerical simulation, the cascade stops being self-similar, slows down, and approaches a classical equilibrium distribution~\cite{Berges:2014bba} with a temperature $\widetilde{T}(\Lambda)$ that depends on the \gls{uv} cutoff $\Lambda$ and is determined by the equipartition theorem, ${f_{\mathrm{cl, th}}(\vect{p})+1/2 \propto \kb \widetilde{T}(\Lambda)/\epsilon_{\vect{p}}}$. In particular, mode occupancies drop to unphysical values below the vacuum noise of $1/2$.

The late-time regime dominated by quantum fluctuations is beyond the regime of classical-statistical simulation methods. Here, we resort to analytical estimates for the reheating time and the reheating temperature in our system based on the self-similar time evolution. Following Ref.~\cite{Micha:2004bv}, we neglect the final stage of quantum relaxation to a Bose--Einstein distribution and consider thermalization as complete once the occupancy of the characteristic momentum dominating the kinetic energy budget becomes on the order of unity.
More precisely, we define the characteristic momentum $\characteristicmomentum(t)$ as the momentum that maximizes the integrand in the expression~\eqref{eq:scalingenergy} for the total kinetic energy. For an isotropic system with dispersion relation ${\epsilon_{\vect{p}} \propto p^z}$, where $p = |\vect{p}|$, the characteristic momentum is given by
${\characteristicmomentum(t) = \argmax_{p} p^{d + z - 1} f(t, p)}$.

If the scaling exponents as well as the momentum distribution at some reference time are known, the assumption of self-similar time evolution according to \cref{eq:selfsimilar} is sufficient to predict the time $\reheatingtime$ when the occupancy of the final characteristic momentum $\finalcharmom$ will reach unity. The time $\reheatingtime$ can be regarded as the best possible approximation to the reheating time obtainable from classical-statistical simulations.
However, \cref{eq:selfsimilar} is not directly applicable since the scaling exponents are not constant throughout the entire time evolution. In fact, they take different universal values in the regimes of driven and free turbulence, respectively, which are interpolated during a transient regime of prescaling.
Importantly, as shown in \cref{sec:prescaling}, the universal scaling form of the distribution is preserved during prescaling, allowing us to describe the full evolution solely in terms of time-dependent scaling exponents.

To properly account for time-dependent exponents $\alpha(t)$ and $\beta(t)$, we generalize the power-law expressions in \cref{eq:selfsimilar} as ${(t / t_{\rm ref})^\alpha \to \exp \int_{t_{\rm ref}}^t \alpha(t^\prime) \diff t^\prime / t^\prime}$, and similarly for $\beta$~\cite{Mazeliauskas:2018yef}. Note that the original expressions are recovered for constant $\alpha$ and $\beta$.
This way, the evolution of the initial characteristic momentum scale for turbulence~$\initialcharmom$ is given by
\begin{equation}
\label{eq:scalingcharacteristicmomentum}
\characteristicmomentum(t) = \exp \left[ - \int_{\initialturbulencetime}^t \beta(t^\prime) \frac{\diff t^\prime}{t^\prime} \right] \initialcharmom ,
\end{equation}
where $\initialturbulencetime$ denotes the time when turbulence sets in.
While the exponent $\beta$ determines the scaling of the characteristic momentum, the evolution of its occupancy is governed by the exponent $\alpha$,
\begin{equation}
\label{eq:scalingcharacteristicoccupancy}
f \left( t, \characteristicmomentum(t) \right) = \exp \left[ \int_{\initialturbulencetime}^t \alpha(t^\prime) \frac{\diff t^\prime}{t^\prime} \right] f(\initialturbulencetime, \initialcharmom) .
\end{equation}

By virtue of \cref{eq:scalingcharacteristicoccupancy}, it is in principle possible to compute the time $\reheatingtime$ by solving ${ f \left( \reheatingtime, \finalcharmom \right) = 1 }$, given full knowledge of the time dependence of the exponent $\alpha$. It is instructive, however, to rewrite this formula using some simplifying approximations. To this end, we assume that $\alpha$ is approximately constant during the driven regime, taking its universal value $\alphadriven$, and that, after switching off the modulation at time $\driventurbulencetime$, the relaxation to its universal value for free turbulence $\alphafree$ occurs fast compared to the overall reheating time scale.
We then recover power-law scaling for the occupancy of the final characteristic momentum,
\begin{equation}
\begin{split}
\label{eq:scalingcharacteristicoccupancysimplified}
f \left( \reheatingtime, \finalcharmom \right) \approx \left( \frac{\driventurbulencetime}{\initialturbulencetime} \right)^{\alphadriven} \left( \frac{\reheatingtime}{\driventurbulencetime} \right)^{\alphafree} f(\initialturbulencetime, \initialcharmom) ,
\end{split}
\end{equation}
and it reaches unity at the time
\begin{equation}
\label{eq:reheatingtimeestimate}
\reheatingtime = \driventurbulencetime \left( \frac{\driventurbulencetime}{\initialturbulencetime} \right)^{- \alphadriven / \alphafree} \left[ f(\initialturbulencetime, \initialcharmom) \right]^{- 1 / \alphafree} .
\end{equation}
Recall that a direct energy cascade is described by negative exponents $\alpha$ and $\beta$ with large initial occupancies~$f(\initialturbulencetime, \initialcharmom)$, which decrease in time according to \cref{eq:scalingcharacteristicoccupancysimplified}. Furthermore, the condition ${\initialturbulencetime \le \driventurbulencetime \le \reheatingtime}$ requires ${\driventurbulencetime \le \initialturbulencetime [ f(\initialturbulencetime, \initialcharmom) ]^{- 1 / \alphadriven}}$. If the latter inequality is satisfied as an equality, the occupancy of the characteristic momentum reaches unity already during driven turbulence before switching off the modulation, such that $\reheatingtime = \driventurbulencetime$. Here, we focus on the situation where $\driventurbulencetime < \reheatingtime$ such that the system spends a dominant part of its evolution in the regime of free turbulence.

Using an analogous line of arguments, we can estimate the final characteristic momentum scale from \cref{eq:scalingcharacteristicmomentum} as
\begin{equation}
\label{eq:finalcharacteristicmomentumestimate}
\begin{split}
\finalcharmom &\approx \left( \frac{\driventurbulencetime}{\initialturbulencetime} \right)^{-\betadriven} \left( \frac{\reheatingtime}{\driventurbulencetime} \right)^{-\betafree} \initialcharmom \\
&= \left( \frac{\driventurbulencetime}{\initialturbulencetime} \right)^{- \betadriven + \alphadriven / (d + z) } \left[ f(\initialturbulencetime, \initialcharmom) \right]^{1 / (d + z)} \initialcharmom ,
\end{split}
\end{equation}
where the second equality follows after inserting our estimate for the reheating time~\eqref{eq:reheatingtimeestimate} and substituting the relation~\eqref{eq:scaling1} between $\alphafree$ and $\betafree$ imposed by energy conservation. The use of the latter identity makes the estimate~\eqref{eq:finalcharacteristicmomentumestimate} of the final characteristic momentum independent of the scaling exponents in the regime of free turbulence, reflecting the fact that $\finalcharmom$ is closely connected to the total energy in the system~\cite{Micha:2004bv}, which is conserved after switching off the modulation. By contrast, the reheating time, according to \cref{eq:reheatingtimeestimate}, is sensitive to the values of the scaling exponents in both regimes, and, in particular, can be influenced by the non-universal behavior during prescaling.

Finally, we can obtain an order-of-magnitude estimate for the reheating temperature $\reheatingtemp$ in our system by identifying the latter with the typical kinetic energy of a particle with momentum $\finalcharmom$. For a dispersion relation of non-relativistic particle, this reads
\begin{equation}
\label{eq:reheatingtemperatureestimate}
\kb \reheatingtemp \sim \frac{\finalcharmom^2}{2 m} .
\end{equation}

When interpreting the estimates~\labelcref{eq:reheatingtimeestimate,eq:finalcharacteristicmomentumestimate,eq:reheatingtemperatureestimate} in the cosmological context, it is important to remember that our time variable~$\labtime$ denotes the laboratory time and should be transformed back to the cosmic time~$\cosmictime$ by integrating the relation ${\diff \labtime = \diff \cosmictime / a^2}$. Similarly, we need to take into account the redshift of momenta, ${p \to p / a}$, as expressed in the kinetic term of \cref{eq:eomnonherm}. For the estimate of the reheating temperature~\eqref{eq:reheatingtemperatureestimate}, this means ${\reheatingtemp \to \reheatingtemp / a^2(\reheatingtime)}$. This back transformation of variables is illustrated further in the subsequent subsection at a specific example.

Although \cref{eq:reheatingtimeestimate,eq:finalcharacteristicmomentumestimate,eq:reheatingtemperatureestimate} are useful to estimate asymptotic quantities without requiring to simulate the dynamics up to the point where the system thermalizes, the outlined argumentation is based on the strong assumption that the neglected final stage of the dynamics, which is dominated by quantum fluctuations, does not have a significant impact on these estimates. The latter remains to be checked against physical reality, which can be provided by comparison to an experiment.

\subsection{\label{sec:thermalizationvsexpansion}Thermalization versus expansion}

In an expanding spacetime, interaction rates decrease with the expansion. If the expansion is too rapid, it is therefore possible that the system is unable to ever reach thermal equilibrium. In a cosmological setting, this ability is usually quantified by comparing the typical interaction rate $\Gamma$ with the Hubble parameter ${H = \dot{a} / a}$. If $\Gamma \gg H$, thermalization is possible, whereas for $\Gamma \ll H$, the occupation numbers are expected to freeze without thermalizing as the mean free path of the particles exceeds the horizon size of the universe.

The full interpolation between both limiting cases, and the associated thermalization behavior, can be studied in our proposed setup. According to \cref{eq:eomnonrel}, for $d = 2$, the dynamics of a Bose gas in an expanding spacetime can be described in terms of a constant interaction~$g$. Consequently, the simulating experimental system will always thermalize eventually. However, this does not necessarily imply the same for the simulated expanding system, whose description is recovered only after a back transformation to the original spatial and temporal coordinates as well as field variables. A necessary condition for thermalization to occur in the simulated system is that the laboratory time $\labtime$ at which the cosmic time $\cosmictime$ becomes infinite is larger than the reheating time, ${\labtime_{\rm f} <  \labtime(\cosmictime = \infty)}$.

\begin{figure}
	\includegraphics[width=\columnwidth]{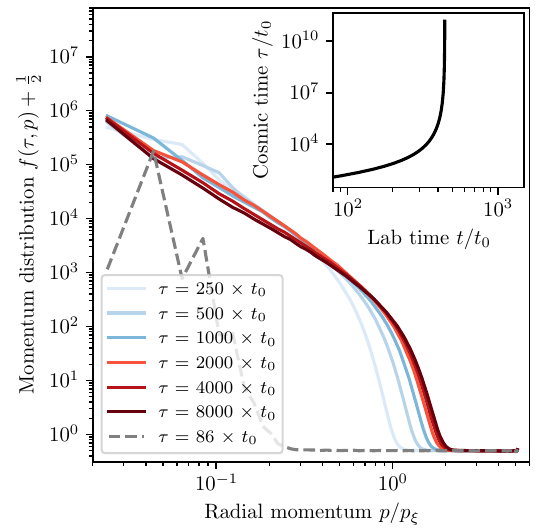}%
	\caption{\label{fig:spmdcosmic}Freeze-out of the momentum distribution of a system with power-law expansion according to \cref{powerexp} with ${H_0 = \num{0.0045} \times t_0^{-1}}$ and ${\nu=2/3}$. The numerical data are the same as in \cref{fig:prescaling}, but the laboratory time~$\labtime$ has been transformed back to the cosmic time~$\cosmictime$ by virtue of \cref{eq:physicalcomp}. As shown in the inset, the cosmic time diverges at the finite laboratory time ${t(\tau = \infty) \approx 444 \times t_0}$, which is, in particular, shorter than the reheating time of the associated simulating system. As a result, the evolution of the simulated expanding system slows down and eventually freezes before thermalization is complete.}
\end{figure}

An interesting feature of our transformation of the time coordinate is that in the case of a sufficiently rapid expansion, infinite cosmic time corresponds to a finite value of the laboratory time, i.e., ${\labtime(\cosmictime = \infty) < \infty}$. To demonstrate this, we consider a power-law expansion~\cite{Felder:2000hq},
\begin{equation}
\label{powerexp}
a(\cosmictime)=\left( 1 + \frac{H_0 \cosmictime}{\nu} \right)^{\nu},
\end{equation}
with a positive exponent $\nu$. This expression fixes $a(0) = 1$ and the initial expansion rate~$H(0)=H_0$. The laboratory time is then given by
\begin{equation}
\labtime(\cosmictime) = \int_0^{\cosmictime} \frac{\diff t^\prime}{a^2(t^\prime)} = \frac{1}{H_0} \frac{\nu}{1-2\nu} \Bigl[ a(\cosmictime)^{(1-2\nu)/\nu}-1 \Bigr].
\end{equation}
If $\nu>1/2$, such as in the case of expansion in a matter-dominated universe (${\nu = 2/3}$) \cite{Kolb:1990vq}, the limit ${a=\infty}$ and ${\cosmictime=\infty}$ is reached within a finite laboratory time, given by
\begin{equation}
\label{eq:physicalcomp}
\labtime(\cosmictime = \infty) = \frac{\nu}{(2\nu-1) H_0} ,
\end{equation}
in line with the intuition that a larger expansion rate leads to a faster freeze-out.

Remarkably, \cref{eq:physicalcomp} associates the state of the simulating system at each instant of the laboratory time $\labtime$ with the asymptotic infinite-time state of a simulated system characterized by a certain value of the initial expansion rate $H_0$.
This is illustrated in \cref{fig:spmdcosmic}, where we have used ${H_0 = \num{0.0045} \times t_0^{-1}}$ and ${\nu=2/3}$, leading to the freeze-out time ${t(\tau = \infty) \approx 444 \times t_0}$.

The ability to associate the dynamics of a single simulation with arbitrary expansion schemes is a special case of the \gls{2d} geometry, where, owing to scale invariance, the equations of motion become independent of the scale factor $a$. The situation is different in three spatial dimensions, where, according to \cref{eq:coupling}, the coupling becomes time-dependent as $g \propto a^{-1}$. In the presence of a decreasing coupling, freeze-out may already occur during the simulation before reaching $\labtime(\cosmictime = \infty)$, in which case the expanding system will never thermalize.

\section{\label{sec:experiment}Experimental perspectives}

In this section, we summarize the experimental requirements for observing the salient features of reheating dynamics discussed in this paper.

The simulating system is a single-component \gls{bec} that is parametrically excited by modulating the interaction according to \cref{eq:modulation} around some positive offset value. This can be realized experimentally with the help of a Feshbach resonance, which allows one to tune the $s$-wave scattering length using an external magnetic field~\cite{Chin2010,Staliunas2002,PhysRevA.81.053627,PhysRevX.9.011052,Zhang2020}.
The interaction strength $g_{\mathrm{3D}}$ of a \gls{3d} Bose gas is related to the $s$-wave scattering length $a_s$ via ${g_{\mathrm{3D}} = 4 \pi \hbar^2 a_s / m}$. Our general approach is independent of dimensionality (but see our remarks towards the end of this section about the absence of a turbulent cascade in one dimension). For concreteness, we have focused here on \gls{2d} geometries, which may be realized through a tight confinement of the atomic gas in the vertical direction by a harmonic potential. In the quasi-\gls{2d} regime, the effective interaction strength $g_{\mathrm{2D}} = \tilde{g} \hbar^2 / m$ is characterized by the dimensionless parameter $\tilde{g} = \sqrt{8 \pi} a_s / a_{\mathrm{HO}}$, where ${a_{\mathrm{HO}} = \sqrt{\hbar / m \omega_{\mathrm{HO}}}}$ is the oscillator length of the confining harmonic potential with frequency $\omega_{\mathrm{HO}}$~\cite{Hadzibabic2011}.
Besides via Feshbach resonances, \gls{2d} Bose gases with tunable effective interactions can therefore alternatively be realized by changing the frequency of the harmonic trap~\cite{Ville2017}.

The numerical simulations presented in this paper have been conducted for the fixed value $N \tilde{g} = \num{2.5e3}$ of the product of the particle number~$N$ and the interaction strength~$\tilde{g}$. This value is readily achievable in present-day experiments with uniform \gls{2d} Bose gases~\cite{saint2019dynamical}. On the classical level, the parameters $N$ and $\tilde{g}$ enter the equations of motion only via the product $N \tilde{g}$ (cf.~\cref{app:dimensionlessgpe}), but this no longer holds on the quantum level. In fact, our classical-statistical simulations are sensitive to the individual values of $N$ and $\tilde{g}$ since they control the relative magnitude of mode occupancies with respect to quantum fluctuations mimicked in form of vacuum noise. As discussed in \cref{app:validity}, classical-statistical simulations are restricted to large occupancies and weak interactions, which is why our simulations of turbulent reheating dynamics have been conducted for a higher particle number than currently realizable in experiments. To estimate characteristic quantities for more realistic experimental setups, we rescale our numerical results to the reference parameters $N_{\mathrm{ref}} = \num{1e6}$ and $\tilde{g}_{\mathrm{ref}} = \num{2.5e-3}$. Pure \glspl{bec} with almost $\num{e6}$ atoms can be reached in state-of-the-art experimental setups, e.g., for $^{39}$K ~\cite{CabreraCordova2018,ParticleNumberCampbell2010,ParticleNumberLandini2012}, where a wide range of interactions is accessible through the broad Feshbach resonance at the magnetic flux $B = \SI{560.7}{G}$~\cite{d2007feshbach}. We emphasize that extrapolating weak coupling results to stronger couplings is \textit{a priori} not justified.
Yet, this procedure is often the only way to be consistent with experimental aspects and commonly used, e.g., in the context of \gls{qcd}~\cite{Kurkela2015}. In the field of ultracold atoms, a number of positive examples exist, where weak coupling expectations of non-equilibrium quantum dynamics have been found experimentally at strong couplings~\cite{Prufer:2018hto,Pruefer2019}.

For the above choice $N \tilde{g} = \num{2.5e3}$ and a \gls{2d} system of $^{39}$K atoms in a square box of length $L = \SI{50}{\micro \meter}$, one obtains the typical values $\lengthunit = \SI{1}{\micro \meter}$ and $\timeunit \approx \SI{0.6}{\milli \second}$ for the characteristic length and time scales $\lengthunit = \hbar / \sqrt{m n_0 g_0}$ and $\timeunit = \hbar / n_0 g_0$, respectively. Here, $g_0$ is the offset value around which the interaction is modulated and $n_0 = N / L^2$ is the atomic density.

As shown in the previous sections, the reheating dynamics studied in this work can be subdivided into two distinct stages, namely, the early stage of preheating, where occupancies of resonant momenta grow exponentially as a result of parametric instabilities, and the subsequent stage of turbulent thermalization, characterized by a self-similar transport of energy towards higher momenta. The time scales at which these phenomena can be observed depend on the choice of the modulation frequency $\omega$ and the relative modulation amplitude $r$ in \cref{eq:modulation}.
The choice $r = 1$, i.e., modulating the scattering length with an amplitude close to its offset value, is efficient for rapidly driving the system out of equilibrium. (Note that the perturbative expression for the growth rate \cref{eq:growthrateres} is no longer applicable in this case.)
The direct cascade is observed best if occupancies of the initial characteristic momentum, where energy is injected in the system, are as high as possible. To this end, the modulation frequency $\omega$ should be chosen close to the frequency of the lowest non-zero momentum mode supported by the system, which is typically deep in the phonon regime where the Bogoliubov dispersion is linear.
In our simulations of reheating dynamics presented in \cref{fig:turbulence,fig:prescaling}, we have chosen the modulation frequency such that the primary resonance occurs at the momentum $p_{\mathrm{res}} = 3 \times 2 \pi \hbar / L$, i.e., $\omega = 2 \epsilon_{\momentum_{\mathrm{res}}} / \hbar \approx 2 \pi \times \SI{199}{\hertz}$ for the example parameters mentioned above.
The onset of turbulence, marking the end of the preheating stage, occurs at around $t \approx 80 \times \timeunit$, corresponding to roughly $\num{10}$ oscillations of the scattering length.
The subsequent turbulent scaling regime extends up to times of about $\reheatingtime \approx \num{1500} \times \timeunit$, which can be deduced from \cref{fig:turbulence,fig:prescaling} with the help of \cref{eq:reheatingtimeestimate} after rescaling the occupancies to the reference particle number $N = \num{1e6}$ (cf.~\cref{app:validity}).
The estimated reheating time $\reheatingtime$ is well within the reach of modern experiments, where typical lifetimes of \glspl{bec} are on the order of a few seconds~\cite{ParticleNumberCampbell2010}.

In our analysis, we have considered the momentum distribution ${f(t, \vect{\momentum}) = \braket{\bosefield_{\vect{p}}^\dagger(t) \bosefield_{\vect{p}}(t)}}$ as the main observable. Experimentally, this quantity can be accessed in \gls{tof} measurements~\cite{Hadzibabic2011}.
In the case of driven turbulence [cf.~\cref{fig:driventurbulence}], by virtue of \cref{eq:finalcharacteristicmomentumestimate}, we find that the characteristic momentum at the reheating time $\reheatingtime$ is roughly given by $p_f \approx \num{1.5} \times p_{\xi}$, which is on the order of the momentum corresponding to the healing length $\xi$. In general, this value depends on the duration of the regime of driven turbulence and tends to be smaller in the case of free turbulence [cf.~\cref{fig:freeturbulence}] or in the prescaling scenario [cf.~\cref{fig:prescaling}], where less energy is injected into the system.
It is a fortunate circumstance that higher momenta, which are increasingly populated as the direct cascade progresses, are typically easier to resolve in \gls{tof} measurements since they require a shorter expansion time. As an alternative means of detection, \textit{in situ} images of the density profile may be used. We have checked that all relevant signatures of both the parametric resonance and the turbulent cascade can be extracted from the quantity $\braket{ \hat{n}_{\vect{p}}^\dagger(t) \hat{n}_{\vect{p}}(t) }$, where $\hat{n}_{\vect{p}}(t)$ is the Fourier transform of the operator ${\hat{n}(t, \vect{x}) = \bosefield^\dagger(t, \vect{x}) \bosefield(t, \vect{x})}$ describing the spatial density profile of the Bose gas.

Although in this paper we have put our main focus on uniform systems, nowadays available in many laboratories~\cite{Gaunt2013,Ville2017}, we expect the characteristic features of the dynamics to be prevalent even in the presence of a harmonic trap. In general, such an external trapping potential couples different momenta and thereby changes the nature of Bogoliubov quasi-particles. However, within a local density approximation, a harmonically trapped \gls{bec} in the Thomas--Fermi regime can be regarded as locally uniform around the center~\cite{Pitaevskii2016}. Thus, while inhomogeneities of the trap may distort long-range correlations in the system, the dynamics on small scales or at large momenta, relevant for the direct cascade, remains unaffected. We have confirmed this explicitly by means of \gls{gpe} simulations of a harmonically trapped system (not shown).

We conclude this section with some remarks on dimensionality. The observation of a turbulent cascade in a scalar \gls{bec} requires at least two spatial dimensions. This is owed to the fact that in a strictly \gls{1d} system, due to kinematic constraints, there can be no $2 \leftrightarrow 2$ scattering processes redistributing momenta. If, however, the system is elongated, but not as strongly confined in the transversal direction to be considered in the quasi-\gls{1d} regime, turbulent energy transport sets in once transversal modes are excited. We have confirmed this by numerically simulating a \gls{3d} system in an elongated periodic box.

Compared to the \gls{3d} case, the study of reheating dynamics in two dimensions is somewhat simpler for at least two reasons. First, as discussed in \cref{sec:special2d}, the effective interaction in \gls{2d} is independent of the cosmic scale factor and there is no need to continuously adjust the scattering length according to the specific expansion model chosen. Second, an absorption image of an atomic cloud, taken after a \gls{tof} expansion or \textit{in situ}, will always be a \gls{2d} projection on the plane transversal to the optical axis. In three dimensions, this means that the distribution is integrated along the optical axis, thereby mixing momenta of different magnitudes. We have mimicked this in a simulation of a \gls{3d} system in a periodic box, where we checked in particular that the scaling is robust if we integrate the momentum distribution along one spatial dimension before performing the radial average. Thus, our scheme can be readily applied to experimentally investigate reheating dynamics in three dimensions, which is of fundamental interest from a cosmological point of view.

\section{\label{sec:conclusion}Conclusion}

In this work, we have demonstrated by means of classical-statistical simulations how single-component ultracold Bose gases with tunable self-interactions can be employed for simulating the characteristic stages of the dynamics of post-inflationary reheating.
We have shown how expansion can be encoded in the time dependence of the interaction, and that in the special case of \gls{2d}, arbitrary expanding backgrounds can be associated with a single experiment in post-processing.
The resonant production of particles in the preheating stage of the early universe, driven by the oscillating inflaton, is mimicked by a modulation of the scattering length inducing parametric instabilities in the \gls{bec}.

Following the proliferation of secondary instabilities due to non-linear effects, the system enters a turbulent state, characterized by a self-similar transport of energy towards higher momenta. We have analyzed the universal scaling exponents describing this direct cascade, which in general undergoes a transition from a driven to a free regime after the inflaton has decayed.
This transition is characterized by a prescaling stage, where the momentum distribution has already attained its universal scaling form, but the scaling exponents are non-universal and time-dependent, yet fully describe the dynamics of the Bose gas. Our set-up opens the door to the experimental observation of both the universal direct cascade and the prescaling regime.

The approach to a non-thermal fixed point as well as prescaling represent a dramatic reduction of sensitivity to the initial state as well as a gradual loss of complexity in the system's dynamics, leading to the emergence of universal physics. Thus, from a more general point of view, our set-up provides a distinct platform for exploring the fundamental mechanisms of thermalization dynamics in quantum many-body systems far from equilibrium.

The final stage of the dynamics, when the system relaxes to thermal equilibrium, is dominated by quantum fluctuations and therefore not captured by the classical-statistical approximation. In contrast, it can be accessed by experimental studies in cold atomic systems, which are quantum-mechanical by nature and not restricted to the weak-coupling regime.

\appendix

\begin{acknowledgments}
	The authors thank Uwe R.~Fischer, Zoran Hadzibabic, Maurus Hans, Aleksas Mazeliauskas, Aleksandr N.~Mikheev, Julius Mildenberger, Christian-Marcel Schmied, Marius Sparn, Daniel Spitz, Helmut Strobel, Malo Tarpin, Gerard Valentí-Rojas, Celia Viermann, and Torsten V.~Zache for discussions and collaborations on related topics.
	This work is funded by Deutsche Forschungsgemeinschaft (DFG, German Research Foundation) under SFB 1225 ISOQUANT (project ID 27381115) as well as BE 2795/4-1, the European Research Council (ERC) under the European Union’s Horizon 2020 research and innovation programme (ERC StG StrEnQTh, Grant agreement No.\ 804305, and ERC AdG EntangleGen, Grant agreement No.\ 694561), Q@TN --- Quantum Science and Technology in Trento, and Provincia Autonoma di Trento.
	The authors acknowledge support by the state of Baden-Württemberg through bwHPC.
\end{acknowledgments}

A.C. and K.T.G. contributed equally to this work.\\

\section{\label{app:nonrelativisticlimit}Non-relativistic limit of the Klein--Gordon equation in expanding spacetime}

In this appendix, we derive the non-relativistic limit of the equations of motion of the inflaton field~\eqref{eq:eomrel} on the quantum level and discuss in detail the relevant approximations that enter the derivation.

Canonical quantization can be performed, as usually in quantum field theory on curved spacetime, by transforming to conformal time, ${\diff \conformaltime = \diff \cosmictime / a}$, and conformal field variables, ${\conformalinflaton = a^{(d - 1) / 2} \cosmicinflaton}$,~\cite{mukhanov2007introduction}. The equations of motion then take the same form as in Minkowski spacetime,
\begin{equation}
	\label{eq:eomconf}
	\frac{1}{c^2} \conformalinflaton^{\prime\prime} - \nabla^2 \conformalinflaton + \frac{m_{\mathrm{eff}}^2 c^2}{\hbar ^2} \conformalinflaton + \frac{\lambda a^{3 - d}}{6} \conformalinflaton^3 = 0,
\end{equation}
with a time-dependent coupling and effective mass
\begin{equation}
	m_{\rm eff}^2(a) = m^2a^2 - \frac{\hbar^2 (d - 1)}{2 c^{4}} \Bigl[ \Bigl(\frac{a'}{a}\Bigr)^2\frac{(d-3)}{2} + \frac{a''}{a} \Bigr].
\end{equation}
Here, $(\cdot)^\prime$ denotes the derivate with respect to conformal time $\conformaltime$. The field $\conformalinflaton$ and its conjugate momentum ${\conformalinflatonmomentum = \conformalinflaton^\prime / c}$ are then promoted to operators that satisfy the canonical commutation relations
\begin{equation}
	\label{eq:canonicalcomm}
	\begin{split}
		\commutator{\hat{\conformalinflaton}(\conformaltime, \vect{x})}{\hat{\conformalinflatonmomentum}(\conformaltime, \vect{y})} &= i \hbar \delta(\mathbf{x} - \mathbf{y}) ,\\
		\commutator{\hat{\conformalinflaton}(\conformaltime, \vect{x})}{\hat{\conformalinflaton}(\conformaltime, \vect{y})} &= \commutator{\hat{\conformalinflatonmomentum}(\conformaltime, \vect{x})}{\hat{\conformalinflatonmomentum}(\conformaltime, \vect{y})} = 0 .
	\end{split}
\end{equation}

We have used the transformation to conformal variables in order to quantize the system. We now switch back to the original variables $\cosmictime$ and $\hat{\cosmicinflaton} = a^{-(d - 1) / 2}\hat{\conformalinflaton}$, since the non-relativistic limit is most conveniently studied in terms of these.

To this end, we consider the complex field operator~$\cosmicbosefield(\cosmictime, \vect{x})$ defined in \cref{eq:slowlyvarying}, where we have factored out the rapid oscillations of the inflation field on the frequency scale determined by its rest mass. For taking the non-relativistic limit, we require the field to change much slower than these oscillations, i.e., we demand the mass term to be the dominant contribution on the right-hand side of the equations of motion~\eqref{eq:eomrel}. This requires the following conditions to hold.
First, the typical field values should be not too large, ${\lambda  \langle \hat \cosmicinflaton^2 \rangle \ll m^2c^2/ \hbar^2}$, which implies that the self-interactions are relatively weak. Similarly, the typical physical momenta should be non-relativistic, ${|\mathbf{p}| \ll mc}$. Finally, the expansion should not be too rapid, ${H = \dot a/a \ll mc^2/\hbar}$, since otherwise the field would be in the over-damped regime and not perform oscillations at all, as during inflation. Note that the first two assumptions are not necessarily satisfied by the inflaton field, especially at the early stages of preheating (see \cref{sec:preheating}).

The outlined assumptions imply the hierarchy
\begin{equation}
	\label{eq:nonrelhierarchy}
	\Big|\braket{  \cosmicbosefield \cdots }\Big| \gg \Big|\frac{\hbar}{mc^2} \braket{ \dot \cosmicbosefield \cdots }\Big| \gg \Big|\frac{\hbar^2}{m^2c^4} \braket{ \ddot \cosmicbosefield \cdots }\Big|,
\end{equation}
where $\braket{\cdots}$ denotes the expectation value. These inequalities are to be understood to hold for any correlation function involving the field or its derivatives.

Inserting the ansatz~\eqref{eq:slowlyvarying} into \cref{eq:eomrel} without any approximation yields
\begin{widetext}
\begin{equation}
	\label{eq:eomnonrelfull}
	\Biggl[ \frac{1}{c^2} \left( \ddot \cosmicbosefield - \frac{2imc^2}{\hbar} \dot \cosmicbosefield + dH \left( \dot \cosmicbosefield - \frac{imc^2}{\hbar} \cosmicbosefield \right) \right) -  \frac{\nabla^2}{a^2} \cosmicbosefield
	+ \frac{\lambda \hbar^2}{12 mc} \Bigl(  {\cosmicbosefield}^3 \etothepowerof{-2imc^2 \cosmictime / \hbar} + 3 \cosmicbosefield\cosmicbosefield^\dagger \cosmicbosefield \Bigr)    \Biggr]
	\etothepowerof{-imc^2 \cosmictime / \hbar}  + \mathrm{h.c.} = 0 .
\end{equation}
\end{widetext}

Employing the inequalities~\eqref{eq:nonrelhierarchy} allows us to neglect two of the first four terms in \cref{eq:eomnonrelfull}. Furthermore, we can use relations~\eqref{eq:nonrelhierarchy} to neglect the rapidly oscillating terms proportional to ${\cosmicbosefield^3 \etothepowerof{\pm 2i mc^2 \cosmictime / \hbar}}$ in the last bracket. The latter describe number-changing processes in the relativistic theory and neglecting them leads to an emergent $U(1)$ symmetry implying particle number conservation. As a result, \cref{eq:eomnonrelfull} reduces to \cref{eq:eomnonherm}, where we have normal-ordered the operators, dropping an irrelevant energy shift\footnote{Indeed, normal-ordering $g\cosmicbosefield \cosmicbosefield^\dagger \cosmicbosefield $ produces the commutator $g[\cosmicbosefield, \cosmicbosefield^\dagger] \cosmicbosefield = g a^{-d} \delta(0) \cosmicbosefield$, as follows from \cref{eq:canonicalcomm,eq:slowlyvarying}. This term can then be absorbed into an irrelevant dynamical phase via the transformation ${\cosmicbosefield \rightarrow \cosmicbosefield \exp{ \{ i \hbar \delta(0) \int g a^{-d} d\tau \} }}$.}.

By virtue of the relations~\eqref{eq:nonrelhierarchy}, it is straightforward to show that within the outlined approximations the canonical commutation relations~\eqref{eq:canonicalcomm} imply bosonic equal-time commutation relations for the field ${\labbosefield  = \cosmicbosefield a^{d/2}}$,
\begin{equation}
	\label{eq:bosecomm}
	\begin{split}
		\commutator{\labbosefield(\labtime, \vect{x})}{\labbosefield^\dagger(\labtime, \vect{y})} &= \delta(\vect{x} - \vect{y}) , \\
		\commutator{\labbosefield(\labtime, \vect{x})}{\labbosefield(\labtime, \vect{y})} &= \commutator{\labbosefield^\dagger(\labtime, \vect{x})}{\labbosefield^\dagger(\labtime, \vect{y})} = 0 .
	\end{split}
\end{equation}

\section{\label{app:analogpreheating}Formal differences of parametric oscillations in the relativistic and non-relativistic model}

To pinpoint the formal differences between the parametric oscillations described by \cref{eq:eomrelfluct} and those induced in the Bose gas by periodically modulating the interaction, we consider \cref{eq:eomnonrelfull}, which is an exact rewriting of the original relativistic equation of motion~\eqref{eq:eomrel}.
Linearizing the field operator around the homogeneous condensate, ${\cosmicbosefield(\cosmictime, \vect{x}) = \cosmiccondensate_0(\cosmictime) + \delta \cosmicbosefield(\cosmictime, \vect{x})}$, the interaction term takes the form
\begin{equation}
	\label{eq:interactiontermfull}
	\sim g \left( 3 \cosmiccondensate_0^2 \delta \cosmicbosefield \etothepowerof{-2imc^2 \cosmictime / \hbar} + 2 \left| \cosmiccondensate_0 \right|^2 \delta \cosmicbosefield + \cosmiccondensate_0^2 \delta \cosmicbosefield^\dagger \right) ,
\end{equation}
which corresponds to the term proportional to ${\lambda \inflaton_0^2 \delta \inflaton}$ in \cref{eq:eomrelfluct}.
When taking the non-relativistic limit to derive \cref{eq:eomnonrel}, we have neglected the rapidly oscillating $U(1)$-violating term proportional to $\etothepowerof{-2imc^2\cosmictime/\hbar}$, keeping only the $U(1)$-conserving, slowly-varying part. However, since the former reflects the oscillations of the inflaton responsible for parametric resonance, this effect is no longer present in the non-relativistic model.

To re-introduce parametric instabilities in the non-relativistic model, we add a periodic modulation of the coupling with frequency $\omega$, such that the linearized interaction term in \cref{eq:eomnonrel} takes the form
\begin{equation}
	\label{eq:interactiontermmodulated}
	\sim g \left(1 + \sin(\omega \cosmictime) \right) \left(2 \left| \cosmiccondensate_0 \right|^2 \delta \cosmicbosefield + \cosmiccondensate_0^2 \delta \cosmicbosefield^\dagger \right) .
\end{equation}
On a formal level, this is similar to \cref{eq:interactiontermfull} in the sense that we have an oscillating prefactor multiplying a term cubic in the field, although it does not have exactly the same form and therefore the mapping is not exact.

\section{\label{app:numerical}Numerical methods}

\subsection{\label{app:truncatedwigner}Classical-statistical simulations}

We simulate the dynamics of an ultracold Bose gas, governed by the Heisenberg equations of motion~\eqref{eq:eomnonrel}, by means of classical-statistical simulations~\cite{Khlebnikov:1996mc,Berges:2007ym}, known in the literature also under the name truncated Wigner simulations~\cite{Sinatra2002,Blakie2008}. This method takes into account quantum fluctuations in form of stochastic initial conditions, but relies on a deterministic time evolution governed by semi-classical equations of motion. Quantum expectation values are obtained as statistical averages over multiple realizations. The following summary of the truncated Wigner method is mainly based on Refs.~\cite{Sinatra2002,Blakie2008}.

For each realization, the initial field configuration is sampled from the Wigner distribution of the corresponding quantum state, which is commonly taken as the Bogoliubov state in equilibrium. Here, we consider a homogeneous scalar \gls{bec} of $N$ atoms at zero temperature in a box of volume $V$ with periodic boundary conditions. The initial wave function is sampled as
\begin{equation}
\label{eq:initialstatetwa}
\condensate_{\rm cl}(0, \vect{x}) =
\sqrt{n_0} \etothepowerof{i \phase_0}
+ \sum_{\vect{p} \ne 0}
\left[ \alpha_{\vect{p}} u_{\vect{p}}(\vect{x}) - \alpha_{\vect{p}}^{*} v_{\vect{p}}^{*}(\vect{x}) \right] .
\end{equation}
Here, the first term represents the condensate with particle density ${n_0 = N / V}$ and phase $\phase_0$. Due to the large occupancy of the condensate mode, the finite width of its Wigner function can be neglected, and thus the same value of the density can be used in each realization. To preserve the $U(1)$ symmetry, the phase is randomly sampled from the uniform distribution over the interval $[0, 2 \pi)$ for each realization.
The mode functions $u_{\vect{p}}(\vect{x}) = u_{\vect{p}} \etothepowerof{i \vect{p} \vect{x} / \hbar} / \sqrt{V}$ and $v_{\vect{p}}(\vect{x}) = v_{\vect{p}} \etothepowerof{i \vect{p} \vect{x} / \hbar} / \sqrt{V}$ are solutions of the Bogoliubov--de Gennes equations for a uniform system in a periodic box, with real coefficients ${u_{\vect{p}} = \sqrt{(\epsilon_{\vect{\momentum}, 0} + n_0 g_0) / 2 \epsilon_{\vect{\momentum}} + 1 / 2}}$ and $v_{\vect{p}}$ determined by the normalization $u_{\vect{p}}^2 - v_{\vect{p}}^2 = 1$ (cf.~\cref{sec:lininst} for details of the notation)~\cite{Pethick2008}.
In order to mimic quantum fluctuations, the quasi-particle amplitudes $\alpha_{\vect{\momentum}}$ are sampled as complex Gaussian random variables with zero mean, satisfying $\mean{\alpha_{\vect{p}}^* \alpha_{\vect{q}}} = \delta_{\vect{p}, \vect{q}} / 2$ at zero temperature. Here, $\mean{(\dots)}$ denotes the ensemble average over all realizations. This vacuum noise corresponds to an average occupancy of half a particle per mode, which is also known as the \enquote{quantum half}. Unless stated otherwise, the vacuum noise is cut off at the highest lattice momentum.

Here, we use a simplified approach, setting $u_{\vect{p}} = 1$ and $v_{\vect{p}} = 0$, such that the mode functions reduce to ordinary plane waves. Effectively, this corresponds to preparing the Bogoliubov ground state of an ideal gas, which is appropriate for our application since the precise nature of the quasi-particles seeding the parametric resonance is unimportant. As a side effect, we observe a transient growth of population at low momenta in our simulations at early times (cf.~\cref{fig:spmdpreheating}). The resulting $\momentum^{-1}$ behavior of the momentum distribution at low momenta matches the behavior of $u_{\vect{p}}^2$ and $v_{\vect{p}}^2$ for $|\vect{p}| \to 0$. Therefore, the observed growth can be interpreted as an artifact of the effective quench from an ideal Bose gas to a system with finite interaction at time $t = 0$.

In the truncated Wigner method, quantum expectation values are replaced by statistical averages over the Wigner distribution. The latter correspond to expectation values of symmetrically ordered quantum operators. Thus, the momentum distribution obtained from classical-statistical simulations corresponds to the one defined in \cref{eq:spmd} plus an extra contribution in form of the \enquote{quantum half} stemming from the commutation relations,
\begin{equation}
	\frac{1}{V} \overline{| \condensate_{\rm cl}(t, \vect{\momentum}) |^2} = f(t, \vect{\momentum}) + \frac{1}{2} .
\end{equation}
Here, $\condensate_{\rm cl}(t, \vect{\momentum}) = \int \diff^d x \, \condensate_{\rm cl}(t, \vect{x}) \etothepowerof{-i \vect{\momentum} \vect{x} / \hbar}$ denotes the Fourier transform of the classical field $\condensate_{\rm cl}(t, \vect{x})$.

The classical-statistical simulations presented in this work have typically been averaged over at least $64$ runs. The statistics is even further enhanced through radial averages due to the isotropy of the system. The statistical error bars are thus typically smaller than the line width in the plots.

\subsection{\label{app:dimensionlessgpe}Dimensionless Gross--Pitaevskii equation}

Each realization is propagated in time according to the \gls{gpe}~\eqref{eq:gpe}. To cast this equation into a dimensionless form suitable for numerical simulations, we express time and length in units of the characteristic scales ${t_0 = \hbar / n_0 g_0}$ and ${x_0 = \hbar / \sqrt{m n_0 g_0}}$, respectively. In terms of the dimensionless time ${\tilde{t} = t / t_0}$, position ${\tilde{\vect{x}} = \vect{x} / x_0}$, and field ${\tilde{\condensate}_{\rm cl} = x_0^{d/2} \condensate_{\rm cl}}$, the \gls{gpe} takes the form
\begin{equation}
\label{eq:gpedimensionless}
i \partial_{\tilde{t}} \tilde{\condensate}_{\rm cl} = \left( -\frac{1}{2} \tilde{\nabla}^2 + \tilde{g} \left( \tilde{t} \right) |\tilde{\condensate}_{\rm cl}|^2 \right) \tilde{\condensate}_{\rm cl} ,
\end{equation}
with the dimensionless coupling $\tilde{g} = g \times t_0 / \hbar x_0^d$.

In a quasi-\gls{2d} system (and in the absence of time-dependent modulations), the dimensionless coupling is, up to logarithmic corrections, given by ${\tilde{g} = \sqrt{8 \pi} a_s / a_{\mathrm{HO}}}$, where $a_s$ is the $s$-wave scattering length and $a_{\mathrm{HO}}$ is the oscillator length of the confining harmonic potential in vertical direction~\cite{Hadzibabic2011}. If the wave function is normalized to unity, the coupling changes as $\tilde{g} \to N \tilde{g}$, where $N$ is the particle number. This shows that the parameters $N$ and $\tilde{g}$ enter the classical equations of motion~\eqref{eq:gpedimensionless} only through the product $N \tilde{g}$ as the single relevant model parameter. Moreover, also the dimensionless box length $L / x_0 = \sqrt{N \tilde{g}}$ is fixed by this quantity as a consequence of scale invariance. By contrast, on the quantum level, the relative magnitudes of the parameters $N$ and $\tilde{g}$ play a role (cf.~\cref{app:validity}).

The \gls{gpe}~\eqref{eq:gpedimensionless} is discretized in space by means of a Fourier pseudospectral discretization and propagated in time using the well-known split-step method~\cite{Antoine2013}. The accuracy of the time evolution is further enhanced with the help of the optimized fourth-order time splitting scheme given in Table 2 of Ref.~\cite{Blanes2002}.

In our \gls{2d} simulations of turbulent reheating dynamics (cf.~\cref{sec:reheating}), we use a spatial ${N_g \times N_g}$ grid with at least ${N_g = 512}$ grid points per dimension, while ${N_g = 256}$ turns out to be sufficient for simulating early-time preheating dynamics (cf.~\cref{sec:preheating}).
For a system of size ${L \times L}$ with ${L / x_0 = 50}$, the corresponding grid spacing ${\Delta x = L / N_g}$ ensures that the healing length ${\xi = x_0 / \sqrt{2}}$, which is the smallest physical length scale in the system, is well-resolved. Numerical stability for late-time dynamics is achieved by choosing the numerical time step $\Delta \tilde{t}$ such that ${\Delta \tilde{t} \tilde{k}_{\rm max}^2 \lesssim 2 \pi}$, where ${\tilde{k}_{\rm max} = \pi / \Delta \tilde{x}}$ with ${\Delta \tilde{x} = \Delta x / x_0}$ is the maximum numerical wave number supported by the grid~\cite{Chin2007}.

\subsection{\label{app:validity}Validity of classical-statistical simulations}

Quantum dynamics is well described by classical-statistical simulations in the regime of large occupancies and weak couplings~\cite{Berges:2007ym,Berges2014}.
In particular, the truncated Wigner method is known to produce unphysical results if the number of virtual particles added to mimic quantum fluctuations becomes comparable to the number of real particles in the system~\cite{Sinatra2002,Blakie2008}.
At zero temperature, the failure of the classical-statistical approximation is indicated by the instability of the vacuum and a resulting unphysical dependence of observables on the \gls{uv} cutoff~\cite{Berges2014}. This decay of the \enquote{quantum half}, which inevitably occurs at sufficiently late times, naturally restricts the classical-statistical method to the weak coupling regime, where the instability is mitigated via a separation of scales. Moreover, if the coupling is too strong, physical observables risk attaining a dependence on the \gls{uv} cutoff already at relatively early times through the spurious quantum pressure generated by the decaying vacuum. This is because the coupling controls the relative magnitude of mode occupancies with respect to the vacuum noise, which, in turn, is regulated by the \gls{uv} cutoff~\cite{Berges2014}.

\begin{figure}
	\includegraphics[width=\columnwidth]{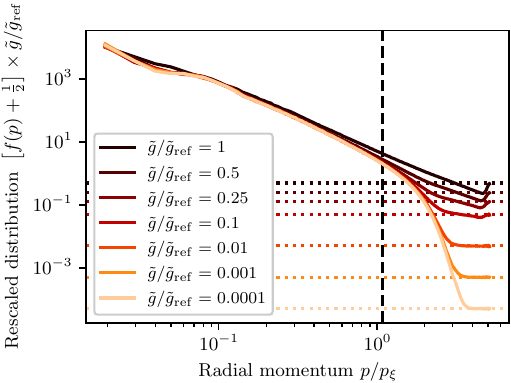}
	\caption{\label{fig:validity} Snapshot of the radially averaged, rescaled momentum distribution for the driven turbulent cascade [cf.~\cref{fig:driventurbulence}] at time $t = 640 \times t_0$ for different values of the coupling~$\tilde{g}$. The particle number is chosen according to ${N(\tilde{g}) = N_{\mathrm{ref}} \times (\tilde{g} / \tilde{g}_{\mathrm{ref}})^{-1}}$, such that the product ${N \tilde{g} = N_{\mathrm{ref}} \tilde{g}_{\mathrm{ref}}}$ remains constant. The vertical dashed line marks the characteristic momentum (cf.~\cref{sec:thermalization}), calculated for the distribution corresponding to the smallest value of $\tilde{g}$. The horizontal dotted lines mark the location of the \enquote{quantum half} after rescaling. Within the range of validity of classical-statistical simulations, the rescaled distributions are expected to collapse to a single curve. Deviations can be observed for $\tilde{g} / \tilde{g}_{\mathrm{ref}} \gtrsim \num{0.1}$, indicating a breakdown of the method for larger values of the coupling.}
\end{figure}

To ensure that our numerical simulations of reheating dynamics, in particular, in the turbulent scaling regime at late times, lie within the range of validity of the classical-statistical method, we follow a similar procedure as presented in Ref.~\cite{Berges2014}. Formally, the classical statistical limit is given by $N \to \infty$ and $g \to 0$, where $N$ is the particle number and $g$ is the coupling, such that $Ng = \mathrm{const.}$
To approach this limit in our \gls{2d} simulation, we fix the value $N \tilde{g} = (L / x_0)^2$ by the choice $L / x_0 = 50$ and repeat the simulation for different values of the coupling $\tilde{g}$, using the parametrization $N(\tilde{g}) = N_{\mathrm{ref}} \times (\tilde{g} / \tilde{g}_{\mathrm{ref}})^{-1}$ with reference parameters $N_{\mathrm{ref}} = \num{1e6}$ and $\tilde{g}_{\mathrm{ref}} = \num{2.5e-3}$.
\Cref{fig:validity} shows the radially averaged momentum distribution for the driven turbulence scenario [cf.~\cref{fig:driventurbulence}]  at time $t = 640 \times t_0$ for different values of the coupling $\tilde{g}$. All configurations have been averaged over at least $25$ single runs. To assess up to which value of the coupling the simulations are valid, we rescale the curves by $\tilde{g} / \tilde{g}_{\mathrm{ref}}$, in which case they are expected to lie on top of each other in the region $f(t, p) \times \tilde{g} / \tilde{g}_{\mathrm{ref}} \gtrsim 1$. (Note that regions where the occupancy drops below unity are outside of the range of validity \textit{per se}.) The vertical dashed line marks the position of the characteristic momentum (cf.~\cref{sec:thermalization}), calculated for the distribution with the lowest value of $\tilde{g}$, around which the \gls{uv} scaling properties of the distribution are most sensitive. It can be seen that deviations occur for $\tilde{g} / \tilde{g}_{\mathrm{ref}} \gtrsim \num{0.1}$. For larger values of the coupling, the shape of the power law changes and the distribution hits the \gls{uv} cutoff, such that cutoff-independence of the results is no longer guaranteed. Furthermore, the distribution at large momenta drops below the \enquote{quantum half}, as indicated by the horizontal dotted lines, which formally corresponds to the unphysical situation of negative occupancies. We have also checked, using a similar approach as in Ref.~\cite{Berges2014}, that the contribution of the vacuum noise to the total energy budget of the system becomes significant for $\tilde{g} / \tilde{g}_{\mathrm{ref}} \gtrsim \num{0.1}$, which coincides with the values of the coupling for which deviations in the momentum distribution are observed.

While the early-time preheating dynamics presented in \cref{sec:preheating} is only insignificantly affected by these deficiencies for the reference parameters $N_{\mathrm{ref}} = \num{1e6}$ and $\tilde{g}_{\mathrm{ref}} = \num{2.5e-3}$, this is no longer true for the late-time turbulent reheating dynamics, as illustrated in \cref{fig:validity}. The results presented in \cref{sec:reheating} have therefore been computed for $\tilde{g} / \tilde{g}_{\mathrm{ref}} = \num{0.01}$, which safely lies within the range of validity of classical-statistical simulations.

\section{\label{app:mathieu}Parametric resonance and Mathieu's equation}

To gain analytical insight in the mechanism behind parametric resonance, we rewrite the equation for the density fluctuations~\eqref{eq:parametricoscillatordensity} in the form
\begin{equation}
\label{eq:mathieu}
\frac{\partial^2 \density_{1 \, \vect{\momentum}}(s)}{\partial s^2} + \left[ A_{\vect{\momentum}} - 2 q_{\vect{\momentum}} \cos(2s) \right] \density_{1 \, \vect{\momentum}}(s) = 0
\end{equation}
with the dimensionless time ${s = \omega t / 2 + \pi / 4}$, and parameters ${A_{\vect{\momentum}} = \epsilon_{\vect{\momentum}}^2 / (\hbar \omega / 2)^2}$ and ${q_{\vect{\momentum}} = r \epsilon_{\vect{\momentum}, 0} n_0 g_0 / (\hbar \omega / 2)^2}$, where $\epsilon_{\vect{\momentum}} = \sqrt{\epsilon_{\vect{\momentum}, 0} (\epsilon_{\vect{\momentum}, 0} + 2 n_0 g_0)}$ denotes the Bogoliubov dispersion relation. \Cref{eq:mathieu} is the standard form of Mathieu's equation~\cite{McLachlan1951}. Importantly, this equation admits solutions which can be expressed as the product of an oscillatory function and an exponentially growing prefactor ${\sim \etothepowerof{\zeta_{\vect{\momentum}} t}}$, describing parametric resonance~\cite{Greene:1998pb,Berges:2002cz}. The condition for exact resonance is given by ${A_{\vect{\momentum}} = 1}$, or equivalently $\epsilon_{\vect{\momentum}} = \hbar \omega / 2$.

The growth rate $\zeta_{\vect{\momentum}}$ of unstable modes can be estimated for small modulation amplitudes $q_{\vect{\momentum}} \ll 1$ using perturbation theory~\cite{Nayfeh1973}. To leading order, it is given by
\begin{equation}
\label{eq:growthrate}
\zeta_{\vect{\momentum}} = \frac{\omega}{4} \sqrt{q_{\vect{\momentum}}^2 - (A_{\vect{\momentum}} - 1)^2} ,
\end{equation}
which reduces to \cref{eq:growthrateres} for the resonant momentum $\vect{\momentum}_{\mathrm{res}}$.

Moreover, as it can be seen in \cref{eq:growthrate}, there is an entire range of modes around $\vect{\momentum}_{\mathrm{res}}$ that experience a positive growth rate, and which thus undergo parametric resonance.
The width of this instability band is delimited by the modes satisfying $A_{\vect{\momentum}} = 1 \pm q_{\vect{\momentum}}$, which increases with the amplitude $r$ of the modulation.

The growth rate predicted from \cref{eq:growthrate} is compared to our numerical simulations in \cref{fig:growthrates}.

\section{\label{app:secondaries}Secondary instabilities}

In this appendix, we discuss the secondary amplification of fluctuations due to the non-linear corrections in \cref{eq:momint}. For modes that are stable on the linear level, the integral on the right-hand side of \cref{eq:momint} acts as a source term. It has a complicated momentum structure, but its time evolution is straightforward since it is dominated by the contribution from the exponentially growing unstable modes $\vect{q}$, for which ${|\vect{p}-\vect{q}| \approx |\vect{q}| \approx p_{\mathrm{res}}}$. This condition can be satisfied if $p \lesssim 2 p_\mathrm{res}$, as can be seen from the triangle inequality,
$$p = |\vect{p} - \vect{q} + \vect{q}| \leq  |\vect{p}-\vect{q}|+ |\vect{q}| \approx 2 p_\mathrm{res} .$$
For $p \gg 2 p_{\mathrm{res}}$, on the other hand, the growth of the integral is exponentially suppressed. The narrower the primary resonance, the more rapid the integral drops at $p \approx 2 p_\mathrm{res}$.

As in the previous appendix, we assume here that the resonance is not too strong, such that $\zeta_{\vect{p}_{\rm res}} \ll \omega $. The time evolution of resonant modes can then be written as a product of an exponential growth and an oscillating function~\cite{McLachlan1951,Nayfeh1973},
$$n_{1\, \vect{p}} \propto \etothepowerof{\zeta_{\vect{p}_{\rm res}} t} \etothepowerof{\pm i  \omega_{\rm res}  t },\: \: \: \: \: \theta_{1\, \vect{p}} \propto \etothepowerof{\zeta_{\vect{p}_{\rm res}} t} \etothepowerof{\pm i  \omega_{\rm res}  t },$$
with the growth rate $\zeta_{\vect{p}_{\rm res}}$ given by \cref{eq:growthrate} and $\omega_{\rm res}$ being the frequency of the resonant modes. As a result, \cref{eq:momint} takes the form of a forced harmonic oscillator,
\begin{equation}
\label{eq:secgrowth}
\ddot n_{1\, \vect{p}} +\omega_{\vect{p}}^2 n_{1\, \vect{p}}  = \etothepowerof{2 \zeta_{\vect{p}_{\rm res}} t} \left( \mathcal{S}_{\vect{p}} \etothepowerof{2 i  \omega_{\mathrm{res}} t} + \mathcal{Q}_{\vect{p}}  \etothepowerof{-2 i  \omega_{\mathrm{res}} t} + \mathcal{P}_{\vect{p}} \right) .
\end{equation}
Here, $\mathcal{Q}_{\vect{p}}$ and $\mathcal{S}_{\vect{p}}$ contain the momentum-dependent contributions of the integrals in \cref{eq:momint} corresponding to positive and negative frequencies, respectively, and $\mathcal{P}_{\vect{p}}$ contains the non-oscillating part. The contribution to the amplitude of $n_{1\, \vect{p}}$ from the first term on the right-hand side of \cref{eq:secgrowth} proportional to $\mathcal{S}_{\vect{p}}$ is given by
\begin{equation}
\label{S_p}
|n_{1\, \vect{p}}| = \frac{\etothepowerof{2 \zeta_{\vect{p}_{\rm res}} t} |\mathcal{S}_{\vect{p}}|}{\sqrt{   (8\zeta_{\vect{p}_{\rm res}} \omega_{\rm res})^2 +(\omega_{\vect{p}}^2+(2\zeta_{\vect{p}_{\rm res}})^2-2\omega_{\mathrm{res}}^{2})^2}} ,
\end{equation}
and analogous expressions can be obtained for $\mathcal{Q}_{\vect{p}}$ and $\mathcal{P}_{\vect{p}}$, respectively. In other words, the modes in the above-mentioned momentum range are amplified as $\exp({2\zeta_{\vect{p}_{\rm res}} t})$ and a resonant amplification occurs if $\omega_{\vect{p}} \approx 2\omega_{\rm res}$. A smoother spectrum is expected in the case of strong driving, i.e., large $\zeta_{\vect{p}_{\rm res}}$.

The pattern described above is indeed observed in \cref{fig:spmdpreheating}, where the secondary and higher-order peaks at multiple frequencies are clearly visible.

\section{\label{app:instabilitiesquantum}Analysis of instabilities in terms of quantum equations of motion for correlation functions}

In this appendix, we study the instabilities arising in the presence of a periodically modulated coupling using quantum equations of motion for the two-point correlation functions. We validate the results of \cref{sec:preheating}, where quantum fluctuations have been treated within the classical-statistical approximation.
\bigskip

The classical action for a (non-relativistic) bosonic field~$\Psi$ with a quartic self-interaction is given by~\cite{Gasenzer:2005ze}
\begin{equation}
\label{classical:action}
S[\Psi] = \int_{x, C} \Bigl[ \frac{i\hbar }{2} (\Psi^{*} \dot \Psi - \dot \Psi^{*}  \Psi ) - \frac{\hbar^2}{2m} \nabla \Psi^{*}  \nabla \Psi-\frac{g}{2}(\Psi^{*} \Psi )^2 \Bigr].
\end{equation}
The integration is performed over the Schwinger--Keldysh closed-time contour~\cite{Keldysh:1964ud}. In the following discussion, we treat $\varphi_1 = \sqrt{2} \real \Psi$ and $ \varphi_2 = \sqrt{2} \imag \Psi$ as the two independent field components.

In quantum theory, the central objects are field correlation functions. The one-point correlation function is defined as $\phi_i(x) = \langle \hat \varphi_i(x) \rangle$. Throughout this appendix, we use the notation $x = (x_0, \vect{x}) \equiv (t, \vect{x})$. For homogeneous systems, the one-point function depends only on time and we can identify it with the wave function of the condensate. The time-ordered connected two-point function is defined as~\cite{Berges:2015kfa, Gasenzer:2005ze}
\begin{align}
\nonumber
G_{ij}(x,y) &= \langle \mathcal{T} \hat \varphi_i(x) \hat \varphi_j(y) \rangle - \phi_i(x) \phi_j(y)  \\
&= F_{ij}(x,y) - \frac{i}{2} \mathrm{sgn}_C(x_0-y_0) \rho_{ij}(x,y),
\end{align}
where $\mathcal{T}$ is the time-ordering operator along the closed-time contour and $\mathrm{sgn}_C(x_{0}-y_{0})$ takes the values $\pm 1$ depending on whether $x_0$ is after or before $y_0$ on the Schwinger--Keldysh contour. In the last step, the propagator has been decomposed into the spectral function~$\rho$ and the statistical propagator~$F$, defined as
\begin{eqnarray}
\label{defineF}
F_{ij}(x,y) &=& \frac{1}{2} \langle \{ \hat{\varphi}_i(x),  \hat{\varphi}_j(y) \}\rangle  - \phi_i(x)\phi_j(y), \\ 
\label{definerho}
\rho_{ij}(x,y)&=&  i \langle [ \hat{\varphi}_i(x),  \hat{\varphi}_j(y) ] \rangle.
\end{eqnarray}
Here, $[\cdot, \cdot]$ and $\{\cdot, \cdot\}$ denote the commutator and the anti-commutator, respectively.
The spectral function describes the spectrum of the excitations and, in particular, encodes the equal-time commutation relations, while the statistical propagator characterizes its occupations.

The equations of motion for the one- and two-point functions can be obtained by a variational principle from the \gls{2pi} effective action. For the two-point functions, they have the form~\cite{Berges:2015kfa}
\begin{equation}\label{linFrho}
\begin{split}
\mathcal{D}_{ik}(x) F_{kj}(x,y)= \int_{t_0}^{x_{0}}dz\Sigma_{ik}^{\rho}(x,z)F_{kj}(z,y)\\
- \int_{t_0}^{y_{0}}dz\Sigma_{ik}^{F}(x,z)\rho_{kj}(z,y),\\
\mathcal{D}_{ik}(x) \rho_{kj}(x,y)= \int_{y_{t_0}}^{x_{0}}dz \Sigma_{ik}^{\rho}(x,z)\rho_{kj}(z,y),
\end{split}
\end{equation}
where $\mathcal{D}_{ik}(x) = [-i\hbar \sigma_{2, ik} \partial_{x_0} - M_{ik}(x)]$,
\begin{eqnarray}
\label{M_ab}
\nonumber
M_{ij}(x) &=& \delta_{ij} \Bigl[ -\frac{\hbar^2 \Delta_{\mathbf{x}}}{2m} + \frac{g}{2} \Bigl( \phi_k(x) \phi_k(x) + F_{kk}(x, x)\Bigr) \Bigr] \\
&+&  g \Bigl( \phi_i(x) \phi_j(x) + F_{ij}(x, x) \Bigr),
\end{eqnarray}
and $\Sigma^F$ and $\Sigma^{\rho}$ are the components of proper self-energy ${\Sigma_{ab}(x,y) =  \Sigma_{ab}^{F}(x,y) -(i/2) \, \mathrm{sgn}_C(x_{0}-y_{0}) \Sigma_{ab}^{\rho}(x,y)}$,
which sums all non-local two-point one-particle-irreducible diagrams constructed from the interaction vertices in the action (\ref{classical:action}) and with lines associated to full time-ordered propagators $G$. Summation over repeated indices is implied in all expressions in this appendix. The presence of a non-vanishing one-point function generates effective cubic interactions in addition to the quartic one, which follows from the shift $\varphi_i \rightarrow \phi_i+\varphi_i$. The resulting interaction part of the action, in terms of the fields $\varphi_i$, reads
\begin{equation}
\label{SintGPE}
S_{\mathrm{int}}[\varphi, \phi] = -\frac{g}{8} \int_{x,C}  \Bigl[  \varphi_i \varphi_i \varphi_j \varphi_j+ 4 \phi_i \varphi_i\varphi_j \varphi_j \Bigr].
\end{equation}

Equations (\ref{linFrho}) are supplemented with analogous equations of motion for the one-point function $\phi$. For the discussion of instabilities, the back-reaction of produced excitations on $\phi$ is negligible, and, as a result, its evolution is approximately governed by the classical \gls{gpe}.

\subsection{Primary instabilities} 

In the linear regime, the self-energies in \cref{linFrho} as well the $F$-dependent corrections to the mass in (\ref{M_ab}) can be neglected. We consider the evolution of the statistical propagator $F$, which can be written as
\begin{align}
\label{Fkj}
\nonumber
\Bigl[-i\hbar \sigma_{2, ik} \partial_{x_0} +  \delta_{ik} \Bigl(  \frac{\hbar^2 \Delta_{\mathbf{x}}}{2m} - &\frac{g}{2}  \phi_l(x) \phi_l(x) \Bigr) \\ - g  \phi_i(x) \phi_k(x) \Bigr]&F_{kj}(x,y) =0.
\end{align}
It is convenient to introduce the \enquote{normal} and the \enquote{anomalous} correlation functions, defined as~\cite{Gasenzer:2005ze}
\begin{align}
\nonumber
\widetilde n(x,y) &= \frac{1}{2} \langle \{ \hat \Psi(x), \hat \Psi^{\dagger}(y) \} \rangle - \langle \hat \Psi(x) \rangle \langle \hat \Psi^{\dagger}(y) \rangle \\
\label{def_n}
 &= \frac{1}{2} F_{ij}(x, y) (\delta_{ij} + \sigma_{2, ij}) = \widetilde n^*(y, x),\\
\nonumber
\widetilde m(x,y) &= \frac{1}{2} \langle \{ \hat \Psi(x), \hat \Psi(y) \} \rangle -  \langle \hat \Psi(x) \rangle \langle \hat \Psi(y) \rangle \\
\label{def_m}
&=  \frac{1}{2} F_{ij}(x, y) (\sigma_{3, ij} + i \sigma_{1, ij}) = \widetilde m(y, x),
\end{align}
which are complex-valued functions. Assuming homogeneity, these functions depend only on the relative spatial coordinate, i.e., $\widetilde n = \widetilde n(t, t', \mathbf{x-y})$. The evolution equations (\ref{Fkj}) for the components of the statistical propagator, Fourier transformed with respect to this relative coordinate, are given by
\begin{equation}
\nonumber
\Bigl[ i\hbar\partial_{t} - \frac{p^2}{2m} - 2g(t) |\Psi(t)|^2 \Bigr] \widetilde n_{\vect{p}}(t, t') \\- g(t)\Psi^2(t) \widetilde m^{*}_{\vect{p}}(t, t') =0,\\
\end{equation}
\begin{equation}
\nonumber
\Bigl[ i\hbar\partial_{t} - \frac{p^2}{2m} - 2g(t) |\Psi(t)|^2 \Bigr] \widetilde m_{\vect{p}}(t, t') \\- g(t)\Psi^{2}(t) \widetilde n_{\vect{p}}^*(t, t') =0.
\end{equation}
These equations have the same form as the ones for the linearized (classical-statistical) fluctuation field $\delta \Psi_{\vect{p}}(t)$. In particular, the linear combinations $\psi^*(t) \widetilde n_{\vect{p}}(t, t')+ \psi(t) \widetilde m_{\vect{p}}^*(t, t')$ and $(i/2) (\widetilde m_{\vect{p}}^*(t, t')/\psi(t) -  \widetilde n_{\vect{p}}(t, t')/\psi^*(t) )$ satisfy the wave equations (\ref{eq:parametricoscillator}). Therefore, ignoring the time dependence of the coupling, one has
\begin{align}
\nonumber
\widetilde n_\mathbf{p}(t,t') \propto e^{i(\pm \omega_\mathbf{p} - g_0 n_0) (t-t')}, \\
\nonumber
\widetilde m_\mathbf{p}(t,t') \propto e^{i(\pm \omega_\mathbf{p} - g_0 n_0) (t+t')},
\end{align}
where $\omega_\mathbf{p}$ corresponds to the Bogoliubov dispersion relation. The dependence on $t'$ in the above expression is determined from the symmetry properties of the functions $\widetilde n_{\vect{p}}(t,t')$ and $\widetilde m_{\vect{p}}(t,t')$, given in (\ref{def_n}). 

\begin{figure}[!t]
	\centering
	\includegraphics[width=\columnwidth]{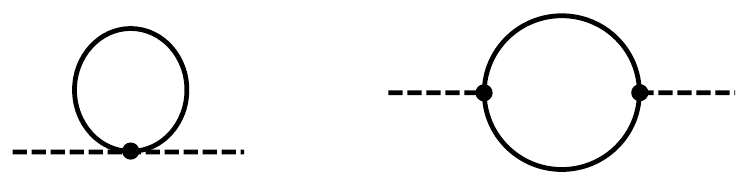}
	\caption{Diagrammatic illustration of the local (left) and non-local (right) one-loop contributions to the self-energy. The second diagram accounts for secondary instabilities.}
	\label{feynman}
\end{figure}

The unstable modes exhibit an exponential amplification, $\widetilde n_{\vect{p}}(t,t') \propto \widetilde m_{\vect{p}}(t,t') \propto \exp({\zeta_{\vect{p}} (t+t')})$, as explained in section (\ref{sec:lininst}). Here we again employed the symmetry properties of $\widetilde n$ and $\widetilde m$. The same holds for the components of the spectral function. However, at $t=t'$ the spectral function is determined solely by the commutation relations, as follows from its definition (\ref{definerho}),
\begin{equation}
\label{commut}
\rho_{ij}(x,y)|_{x_0=y_0} =  - i \, \sigma_{2, ij} \, \delta(\mathbf{x-y}).
\end{equation}
In the next subsection we include the first non-linear corrections to describe the secondary amplification of fluctuations.

\subsection{Secondary instabilities}

As the fluctuations become sufficiently strong, the linear approximation for the dynamics becomes inapplicable. The non-local one-loop correction, which is diagrammatically illustrated in the right part of \cref{feynman}, accounts for secondary instabilities~\cite{Berges:2002cz, Zache:2017dnz}. The corresponding self-energy, which can be calculated by means of standard Feynman diagram techniques, reads
\begin{widetext}
\begin{align}
\label{olse}
\nonumber
\Sigma_{ij}(x,y; \phi, G) = -\frac{ g(x_0) g(y_0) }{2\hbar} \Bigl[ &\phi_i(x) \phi_j(y)G_{ac}(x, y)G_{ac}(x, y)+ 2\phi_b(x) \phi_d(y) G_{ij}(x, y) G_{bd}(x, y)  \\ 
+ 2\phi_b(x) \phi_d(y)G_{id}(x, y)G_{bj}(x, y)+ 2&\phi_b(x) \phi_j(y)G_{ic}(x, y) G_{bc}(x, y) + 2\phi_i(x) \phi_d(y)G_{aj}(x, y)G_{ad}(x, y) \Bigr].
\end{align}
The other one-loop correction is the tadpole diagram, shown in the left part of \cref{feynman}, which is explicitly included in the definition (\ref{M_ab}) of the effective mass . As can be seen from that definition, this correction becomes relevant only later, when the components of $F_{ij}$ become of the order of the condensate density $|\Psi|^2$ (see also~\cite{Berges:2002cz}).

Including the one-loop self-energy in \cref{linFrho} leads to integrals over the past evolution, i.e., memory integrals, on the right-hand side. Due to the exponential growth of fluctuations, these integrals are, however, dominated by the latest times~\cite{Berges:2015kfa, Zache:2017dnz}, which allows to simplify them. Note that $\Sigma$ contains terms proportional to
\begin{eqnarray}
\nonumber
G_{\alpha \beta}(x,y)G_{\gamma \delta}(x,y)   = \underbrace{ F_{\alpha \beta}(x,y)F_{\gamma \delta}(x,y) - \frac{1}{4} \rho_{\alpha \beta}(x,y)\rho_{\gamma \delta}(x,y) }_\text{$\Sigma_F$} 
 -\frac{i}{2} \mathrm{sgn}_C(x_0-y_0)
    \underbrace{ F_{\alpha \beta}(x,y)\rho_{\gamma \delta}(x,y) + \rho_{\alpha \beta}(x,y) F_{\gamma \delta}(x,y)}_\text{$\Sigma_{\rho}$}.
\end{eqnarray}
The $\int \Sigma^{\rho} F$ term in the limit of $z_0 \rightarrow x_0$ represents a ``tadpole''-like contribution, which becomes important only at later times~\cite{Berges:2002cz}. For the $\int \Sigma^{F} \rho$ term one obtains 
\begin{equation}
\nonumber
- \int_{t_0}^{y_{0}}dz\Sigma^{F}_{ac}(x,z)\rho_{cb}(z,y) \approx - \int_{y_0-c}^{y_{0}}dz\Sigma^{F}_{ac}(x,z)\rho_{cb}(z,y) \approx  i\hbar c \Sigma^F_{ac}(x,y) \sigma_{2, cb},
\end{equation}
where $c^{-1} = \mathcal{O}(g_0n_0)$ is some parameter of the order of the oscillation frequency of the condensate. In $\Sigma^F$, there will be terms $\propto F^2$ and $\propto {\rho}^2$. We first consider the $\propto F^2$ part in $\Sigma^F$, which is equivalent to $\Sigma|_{G\rightarrow F}$. Inserting \cref{olse} into the equations of motion (\ref{linFrho}) and using the transformation (\ref{def_n}) and (\ref{def_m}), one arrives at the following equations for the momentum modes of the normal and the anomalous two-point functions,
\begin{eqnarray}
\nonumber
\Bigl[ i\hbar \partial_{t} - \frac{\mathbf{p}^2}{2m} - 2g(t) |\Psi(t)|^2 \Bigr] \widetilde n_{\vect{p}} &-& g(t)\Psi^2(t) \widetilde m_{\vect{p}}^{*}= i c g(t) g(t') \int_{\mathbf{q}} \Bigl[ 4 \Psi(t) \Psi^*(t')  \Bigl( \widetilde n_{\vect{q}} \widetilde n_{\vect{p-q}}^*+ \widetilde m_{\vect{q}} \widetilde m_{\vect{p-q}}^* \Bigr) \\
&+&4 \Psi^*(t) \Psi^*(t') \widetilde n_{\vect{q}} \widetilde m_{\vect{p-q}}+ 4 \Psi(t) \Psi(t')  \widetilde n_{\vect{q}} \widetilde m_{\vect{p-q}}^*
+ 2  \Psi^*(t) \Psi(t')  \widetilde  n_{\vect{q}} \widetilde n_{\vect{p-q}} \Bigr],\\
\nonumber
\Bigl[  i\hbar\partial_{t} - \frac{\mathbf{p}^2}{2m} - 2g(t) |\Psi(t)|^2 \Bigr] \widetilde m_{\vect{p}} &-& g(t)\Psi^{2}(t) \widetilde n_{\vect{p}}^* = -i c g(t) g(t') \int_{\mathbf{q}} \Bigl[ 4 \Psi(t) \Psi(t')  \Bigl( n_{\vect{q}} n_{\vect{p-q}}^*+ m_{\vect{q}} m_{\vect{p-q}}^* \Bigr)\\
&+&4 \Psi^*(t) \Psi(t') \widetilde m_{\vect{q}} \widetilde n_{\vect{p-q}}+ 4 \Psi(t) \Psi^*(t')  \widetilde m_{\vect{q}} \widetilde n_{\vect{p-q}}^*
+ 2  \Psi^*(t) \Psi^*(t')   \widetilde m_{\vect{q}} \widetilde m_{\vect{p-q}} \Bigr],
\end{eqnarray}
\end{widetext}
where we have omitted the arguments $t$ and $t'$ from the two-point functions.

The $\propto \rho^2$ part in $\Sigma^F$, which is equivalent to $\Sigma|_{G\rightarrow i \rho/2}$, leads to an analogous expression on the right-hand side. However, the components $\rho$ are of order one in the $t' \rightarrow t$ limit (in momentum space) due to \cref{commut}. As a result, that contribution is negligible compared to the exponentially growing one from the $\propto F^2$ part. This also justifies the use of classical-statistical simulations for studying the evolution of equal-time observables.

The momentum integrals on the right-hand side are analogous to the ones in \cref{eq:momint} and therefore lead to a secondary exponential amplification, i.e., $\widetilde n_{\vect{p}}(t,t') \propto \exp({2\zeta_{\vect{p}} (t+t')})$, in the momentum range $p \lesssim 2 p_{\mathrm{res}}$. One can check that the terms on the right-hand side oscillate\footnote{Here, we neglect the time dependence of $g$, which is suppressed for small values of $r$.} as functions of $t$ either as $\exp(-ig_0 n_0t)$ or as $\exp(-i(g_0 n_0 \pm 2\omega_{\mathrm{res}}) t)$. Hence, one expects enhanced amplification near $\omega_{\mathbf{p}} = 0$ and $\omega_{\mathbf{p}} =2 \omega_{\mathrm{res}}$. The resonant frequencies correspond to the conservation of energy (in addition to the conservation of momentum) in the vertices of the Feynman diagrams.

\bigskip
To conclude, in this appendix we have analyzed the primary and secondary instabilities within the non-equilibrium \gls{2pi} framework and explained the validity of the classical-statistical approximation, justifying its use in \cref{sec:preheating} to describe the regime when typical occupation numbers of excitations become sufficiently large.

\section{\label{app:extractscalingexponents}Maximum likelihood technique for extracting scaling exponents}

In this appendix, we describe the maximum likelihood technique used in \cref{sec:reheating} to extract the scaling exponents $\alpha$ and $\beta$ describing self-similar time evolution according to \cref{eq:selfsimilar} from our numerical data.
\bigskip

To extract the exponents, we use numerical fits, similar to the procedure in Ref.~\cite{Berges:2013fga}. More specifically, at different times $t$, the distribution functions at $t$ and $t/2$ are compared. Assuming the behavior of \cref{eq:selfsimilar}, we test different values of the exponents. For each pair of exponents, we define the (normalized) likelihood
\begin{align}
W(\alpha, \beta) \propto \mathrm{exp}\Bigl[ -\chi(\alpha, \beta)/(2 \bar \chi) \Bigr],
\end{align}
where
\begin{align}
 \chi(\alpha, \beta) = \frac{\int  [   (f(t', p) - 2^{\alpha}f(t'/2,2^{\beta} p) ) p^l   ]^2 \diff p}{\int  [f(t', p) p^l]^2 \diff p}.
\end{align}
Here $\bar \chi$ is the minimal observed value of $\chi(\alpha, \beta)$, which corresponds to the best fit. We use the third moment of the radial distribution function because for $d=2$ and $\omega_{p} = p^2/2m$, it is proportional to the energy density per momentum interval. The integrals are performed over large momenta for which $f(p) \gtrsim 1$. We find the best-fitting values of the exponents and identify the fitting errors with the Gaussian widths of the marginal likelihoods $\int W d\beta$ and $\int W d\alpha$.

The variable $t$ always denotes the time elapsed from the beginning of the simulation. This includes the stage of parametric resonance, which is unrelated to turbulence. While irrelevant for late times, a more reasonable choice of the initial time, closer to the beginning of turbulent dynamics, improves the scaling analysis at early times, by allowing to avoid the extraction of large exponents decreasing with time. For the considered set of parameters in \cref{sec:reheating}, we set $t = 63 \times t_0$ in the fitting routine described above.

\section{\label{app:methodofmoments}Analysis of prescaling using the method of moments}

To gain a deeper understanding of the phenomenon of prescaling, we provide here a complementary scaling analysis using the method of moments~\cite{Mazeliauskas:2018yef}.
As explained in \cref{app:extractscalingexponents}, the maximum likelihood technique used to extract the scaling exponents in \cref{fig:turbulence,fig:prescaling} locally compares the distributions at two reference times $t_1$ and $t_2 > t_1$. By iterating over all times $t_1$, time-dependent scaling exponents are obtained that best collapse the pairs of distributions $f(t_1, \vect{p})$ and $f(t_2, \vect{p})$ on top of each other. By contrast, the method introduced in Ref.~\cite{Mazeliauskas:2018yef} relies on the moments as global properties of the distribution and allows one to extract instantaneous scaling exponents $\alpha(t)$ and $\beta(t)$ that do not depend on a reference time. In what follows, we briefly outline the method of moments and apply it to the prescaling scenario discussed in \cref{sec:prescaling}.

The $n$-th moment of the distribution $f(t, \vect{p})$ is defined as
\begin{equation}
\label{eq:moments}
M_{(n)}(t) = V \int \frac{\diff^d p}{(2 \pi \hbar)^d} \left( \frac{p}{p_0} \right)^n f(t, \vect{p}) ,
\end{equation}
where $V$ is the volume, $p = |\vect{p}|$, and $p_0$ is an arbitrary momentum scale to make the moment dimensionless. Note that in an isotropic system, as considered here, the distribution in fact depends only on the magnitude of the momentum, $f(t, \vect{p}) = f(t, p)$. For each moment $M_{(n)}$, the integrand is peaked around a certain characteristic momentum whose scaling properties are probed. In particular, the zeroth moment is the total particle number and the second moment is proportional to the total kinetic energy for a system with quadratic dispersion.

\begin{figure}
	\centering
	\subfloat{\includegraphics[width=\columnwidth]{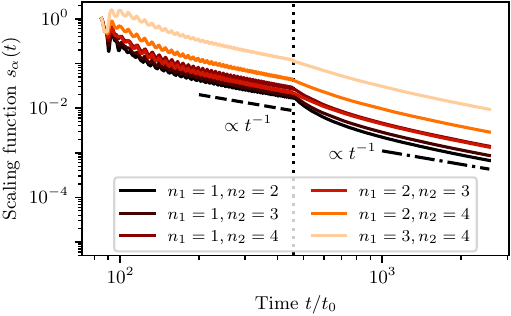}}\\
	\subfloat{\includegraphics[width=\columnwidth]{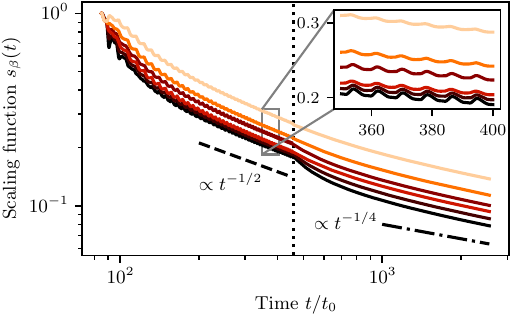}}
	\caption{\label{fig:prescalingmoments}Scaling functions $s_{\alpha}(t)$ (upper panel) and $s_{\beta}(t)$ (lower panel) with respect to the reference time $t_{\mathrm{ref}} = 85 \times \timeunit$ extracted from the moments of orders ${1 \le n_1 < n_2 \le 4}$. The vertical dotted line marks the time $t \approx \num{458.6} \times t_0$ when the modulation is switched off instantaneously. Before this point, the oscillatory behavior of the moments is directly reflected in the evolution of the scaling functions. As driven turbulence develops, their time averages approach power laws with exponents close to the predictions from kinetic theory, $\alpha_{\mathrm{driven}} = -1$ and $\beta_{\mathrm{driven}} = -1/2$ (dashed lines). After the modulation is switched off, the scaling functions exhibit a kink and the oscillations vanish. In the subsequent evolution, the scaling functions extracted from different moments evolve asynchronously, until they adopt a power-law behavior again for times $t \gtrsim 1000 \times \timeunit$ with exponents close to the predictions from kinetic theory in the regime of free turbulence, $\alpha_{\mathrm{free}} = -1$ and $\beta_{\mathrm{free}} = -1/4$ (dashed-dotted lines). The system evolves self-similarly where all curves have the same slope, as analyzed in \cref{fig:prescalingmomentsexponents}.}
\end{figure}

The most straightforward way of introducing time-dependent scaling exponents $\alpha(t)$ and $\beta(t)$ is to make the replacements $\alpha \to \alpha(t)$ and $\beta \to \beta(t)$ in the scaling ansatz \cref{eq:selfsimilar}, such that the latter reads $f(t, \vect{p}) = s^{\alpha(t)} f_{\mathrm{S}} \left( s^{\beta(t)} \vect{p} \right)$ with $s = t / t_{\mathrm{ref}}$. However, the exponents defined in this way depend on the reference time $t_{\mathrm{ref}}$. To lift this dependency, it is advantageous to define $\alpha(t)$ and $\beta(t)$ instead in terms of the scaling functions
\begin{equation}
	s_{\alpha}(t) = \exp \left[ \int_{t_{\mathrm{ref}}}^{t} \alpha(t^\prime) \frac{\diff t^\prime}{t^\prime} \right]
\end{equation}
and $s_{\beta}(t)$ defined in an analogous way. The scaling ansatz in \cref{eq:selfsimilar} then generalizes to
\begin{equation}
\label{eq:selfsimilartimedependent}
f(t, \vect{p}) = s_{\alpha}(t) f_{\mathrm{S}} \left[ s_{\beta}(t) \vect{p} \right] .
\end{equation}
For constant exponents $\alpha$ and $\beta$, the power-law scaling of \cref{eq:selfsimilar} is recovered.

Inserting the above scaling ansatz into \cref{eq:moments}, it is straightforward to derive that the moments scale with time as
\begin{equation}
\label{eq:momentsscaling}
M_{(n)}(t) = s_{\alpha}(t) s_{\beta}^{d + n}(t) M_{(n)}(t_{\mathrm{ref}}) .
\end{equation}
Given a pair of moments $M_{(n_1)}(t)$ and $M_{(n_2)}(t)$ with ${n_1 \neq n_2}$, it is thus possible to express the scaling functions $s_{\alpha}(t)$ and $s_{\beta}(t)$ in terms of these moments,
\begin{subequations}
\begin{align}
s_{\alpha}(t) &= \left[ \frac{M_{(n_1)}^{d + n_2}(t) / M_{(n_1)}^{d + n_2}(t_{\mathrm{ref}})}{M_{(n_2)}^{d + n_1}(t) / M_{(n_2)}^{d + n_1}(t_{\mathrm{ref}})} \right]^{1 / (n_2 - n_1)} , \\
s_{\beta}(t) &= \left[ \frac{M_{(n_1)}(t) / M_{(n_1)}(t_{\mathrm{ref}})}{M_{(n_2)}(t) / M_{(n_2)}(t_{\mathrm{ref}})} \right]^{1 / (n_2 - n_1)} ,
\end{align}
\end{subequations}
which allows one to access the scaling exponents as
\begin{subequations}
\label{eq:exponentsmethodofmoments}
\begin{align}
\alpha(t) &= \frac{1}{n_2 - n_1} \frac{\diff}{\diff \ln t} \ln \frac{M^{d + n_2}_{(n_1)}(t)}{M^{d + n_1}_{(n_2)}(t)} , \\
\beta(t) &= \frac{1}{n_2 - n_1} \frac{\diff}{\diff \ln t} \ln \frac{M_{(n_1)}(t)}{M_{(n_2)}(t)} .
\end{align}
\end{subequations}
Note that the dependency on the reference time $t_{\mathrm{ref}}$ drops out as a consequence of the derivatives.

\begin{figure}
	\centering
	\subfloat{\includegraphics[width=\columnwidth]{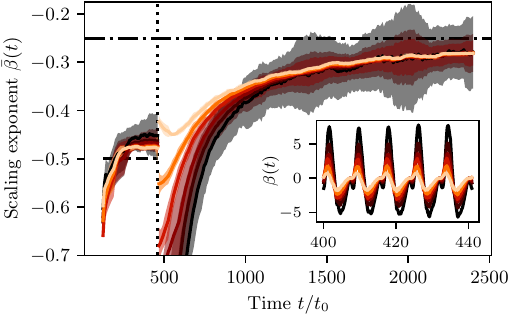}}\\
	\subfloat{\includegraphics[width=\columnwidth]{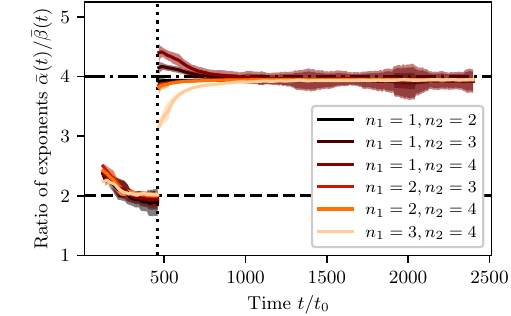}}
	\caption{\label{fig:prescalingmomentsexponents}Time-dependent scaling exponents extracted from the moments of orders $1 \le n_1 < n_2 \le 4$. The inset in the upper panel shows the instantaneous exponent $\beta(t)$, which strongly oscillates due to the modulation. The time-averaged quantities $\bar{\beta}(t)$ and $\bar{\alpha}(t) / \bar{\beta}(t)$ are displayed in the main plots in the upper and lower panel, respectively. The data have been smoothed using simple moving means and the shaded regions show the corresponding moving standard deviations. The vertical dotted line represents the time when the modulation is switched off. Before this point, in the regime of driven turbulence, the exponents are approximately constant and close to the analytical predictions from kinetic theory, $\beta_{\mathrm{driven}} = -1/2$ and $\alpha_{\mathrm{driven}} / \beta_{\mathrm{driven}} = 2$ (dashed lines). After the modulation is stopped, the exponents jump discontinuously and the exponents extracted from different combinations of moments exhibit discrepancies. For $t \gtrsim 1000 \times t_0$, they converge and continue evolving as a single curve, certifying the existence of a prescaling regime of self-similar time evolution. The exponent $\beta(t)$ gradually approaches the universal value $\beta_{\mathrm{free}} = -1/4$ (dashed-dotted line in the upper panel) predicted from kinetic theory in the regime of free turbulence, while the ratio of the exponents quickly adjusts to the prediction $\alpha_{\mathrm{free}} / \beta_{\mathrm{free}} = 4$ (dashed-dotted line in the lower panel), reflecting energy conservation.}
\end{figure}

Since each moment condenses the information about the distribution into a single number, it is necessary to examine many different moments, each sensitive to a different characteristic momentum, to certify that a given distribution scales as a whole. \Cref{fig:prescalingmoments,fig:prescalingmomentsexponents} show the scaling functions and time-dependent scaling exponents, respectively, extracted using the method of moments outlined above for all possible combinations of moments ${1 \le n_1 < n_2 \le 4}$. We exclude the zeroth moment from our analysis since the associated particle number conservation is incompatible with the energy-conserving direct cascade. All simulation parameters are the same as in \cref{fig:prescaling}, with the exception that the modulation is switched off at a later time, $t = 56 \times 2 \pi / \omega \approx \num{458.6} \times t_0$, allowing one to better distinguish the regimes of driven and free turbulence. The classical-statistical simulation presented in \cref{fig:prescalingmoments,fig:prescalingmomentsexponents} has been conducted on a spatial grid of $768 \times 768$ grid points to increase the \gls{uv} resolution and the data have been averaged over $100$ runs. To reduce numerical errors of the calculated moments due to instabilities near the \gls{uv} cutoff, we restrict the integral in \cref{eq:moments} to absolute momenta $|\vect{p}| \le 3 \times p_{\xi}$ ($|\vect{p}| \le \pi \hbar / \Delta x$, where $\Delta x$ is the lattice spacing) before (after) switching off the modulation. As explained in \cref{app:extractscalingexponents}, we shift the origin of time as $t \to t - 63 \times t_0$ to improve the scaling analysis at early times.

\Cref{fig:prescalingmoments} shows the scaling functions $s_{\alpha}(t)$ and $s_{\beta}(t)$ with respect to the reference time $t_{\mathrm{ref}} = \num{85} \times \timeunit$. As turbulence develops, the scaling functions approach power laws $s_{\alpha}(t) \propto (t / t_{\mathrm{ref}})^{\alpha}$ and $s_{\beta}(t) \propto (t / t_{\mathrm{ref}})^{\beta}$. In the driven regime, before switching off the modulation, the corresponding exponents are close to the analytical predictions from kinetic theory, $\alpha_{\mathrm{driven}} = -1$ and $\beta_{\mathrm{driven}} = -1/2$ (see \cref{sec:kinetictheory}). At the time when the modulation is switched off instantaneously, the scaling functions exhibit a kink and continue evolving asynchronously for some time. This indicates that the shape of the distribution is not preserved exactly, but slightly readjusts during the transition from driven to free turbulence. The changes in the shape of the distribution are more pronounced at lower momenta, which is reflected by the fact that those curves in \cref{fig:prescalingmoments} extracted from lower moments react stronger at the transition. Despite these local adjustments, it is remarkable that the global scaling form of the distribution is approximately preserved during the transition from driven to free turbulence, as can be seen in \cref{fig:prescaling}. For $t \gtrsim 1000 \times t_0$, the scaling functions again approach a power law with exponents close to the analytical predictions for free turbulence, $\alpha_{\mathrm{free}} = -1$ and $\beta_{\mathrm{free}} = -1/4$. In the region where the scaling functions extracted from different combinations of moments all evolve in parallel, the distribution scales self-similarly as a whole. To certify that the distribution also exhibits prescaling, i.e., a self-similar evolution with time-dependent scaling exponents, we analyze the change of the scaling functions as quantified by the exponents $\alpha(t)$ and $\beta(t)$ in what follows.

In the driven regime, the scaling functions oscillate in time, as indicated in the inset in the lower panel of \cref{fig:prescalingmoments}. These oscillations originate from the modulated interaction, which causes the distribution and therefore the moments to oscillate in time (cf.~\cref{fig:growthrates}). As a result, the instantaneous scaling exponents $\alpha(t)$ and $\beta(t)$, which are derivatives of the scaling functions according to \cref{eq:exponentsmethodofmoments}, exhibit the strong oscillatory behavior shown for $\beta(t)$ in the inset in the upper panel of \cref{fig:prescalingmomentsexponents}.
To connect with our previous results obtained using the maximum likelihood technique (cf~\cref{fig:prescaling}), it is convenient to consider instead the time-averaged exponents $\bar{\alpha}(t)$ and $\bar{\beta}(t)$, defined as
\begin{equation}
\label{eq:exponentstimeaveraged}
\bar{\beta}(t) = \frac{1}{\ln (1 + T / t)} \int_{t}^{t + T} \beta(t^\prime) \frac{\diff t^\prime}{t^\prime} ,
\end{equation}
and analogously for $\bar{\alpha}(t)$, where $T = 2 \pi / \omega$ is the modulation period. After switching off the modulation, the oscillations of the moments vanish and the time-averaging of the exponents is not required any more.

\Cref{fig:prescalingmomentsexponents} shows the quantities $\bar{\beta}(t)$ (upper panel) and $\bar{\alpha}(t) / \bar{\beta}(t)$ (lower panel) as a function of time. The derivatives in \cref{eq:exponentsmethodofmoments} are sensitive to small fluctuations in the data due to statistical uncertainties, resulting in a large spread of the data points. This is especially true where moments of lower orders are involved since they probe the distribution at smaller momenta with a lower density of states. To better visualize the trend of the exponents, the data have been smoothed by calculating simple moving means (separately in the regimes of driven and free turbulence) using a window size of $8$ ($16$)  periods involving $512$ ($1024$) data points for $n_1 > 1$ ($n_1 = 1$). The shaded regions represent the corresponding moving standard deviations.

Before the modulation is switched off, the system is in the state of driven turbulence. Indeed, for $t \gtrsim 200 \times t_0$, both $\bar{\beta}(t)$ and $\bar{\alpha}(t) / \bar{\beta}(t)$ are approximately constant and  for all considered moments close to the analytical predictions $\beta_{\mathrm{driven}} = -1/2$ and $\alpha_{\mathrm{driven}} / \beta_{\mathrm{driven}} = 2$ (cf.~\cref{sec:kinetictheory}).
After suddenly switching off the modulation, the exponents jump discontinuously, reflecting the kink in the scaling functions (cf.~\cref{fig:prescalingmoments}). Furthermore, there are discrepancies in the values of the exponents extracted from different combinations of moments after the modulation is stopped. As discussed above, this indicates a slight readjustment of the shape of the distribution, which is consistent with the behavior in \cref{fig:prescaling}, where the exponents exhibit large error bars in this regime.
For times $t \gtrsim 1000 \times t_0$, the exponents extracted from different moments converge to a single curve, certifying a self-similar evolution of the distribution as a whole. During this prescaling stage, the exponent $\bar{\beta}(t)$ evolves gradually towards the analytical prediction for free turbulence, $\beta_{\mathrm{free}} = -1/4$.
By contrast, the ratio $\bar{\alpha}(t) / \bar{\beta}(t)$ adjusts rather quickly to the value $\alpha_{\mathrm{free}} / \beta_{\mathrm{free}} = 4$, expressing energy conservation.

In conclusion, our analysis certifies the existence of a prescaling regime during the transition from driven to free turbulence, where the momentum distribution scales self-similarly with time-dependent scaling exponents approaching gradually their universal values.

\bibliography{references}

\begin{thebibliography}{85}%
\makeatletter
\providecommand \@ifxundefined [1]{%
 \@ifx{#1\undefined}
}%
\providecommand \@ifnum [1]{%
 \ifnum #1\expandafter \@firstoftwo
 \else \expandafter \@secondoftwo
 \fi
}%
\providecommand \@ifx [1]{%
 \ifx #1\expandafter \@firstoftwo
 \else \expandafter \@secondoftwo
 \fi
}%
\providecommand \natexlab [1]{#1}%
\providecommand \enquote  [1]{``#1''}%
\providecommand \bibnamefont  [1]{#1}%
\providecommand \bibfnamefont [1]{#1}%
\providecommand \citenamefont [1]{#1}%
\providecommand \href@noop [0]{\@secondoftwo}%
\providecommand \href [0]{\begingroup \@sanitize@url \@href}%
\providecommand \@href[1]{\@@startlink{#1}\@@href}%
\providecommand \@@href[1]{\endgroup#1\@@endlink}%
\providecommand \@sanitize@url [0]{\catcode `\\12\catcode `\$12\catcode
  `\&12\catcode `\#12\catcode `\^12\catcode `\_12\catcode `\%12\relax}%
\providecommand \@@startlink[1]{}%
\providecommand \@@endlink[0]{}%
\providecommand \url  [0]{\begingroup\@sanitize@url \@url }%
\providecommand \@url [1]{\endgroup\@href {#1}{\urlprefix }}%
\providecommand \urlprefix  [0]{URL }%
\providecommand \Eprint [0]{\href }%
\providecommand \doibase [0]{https://doi.org/}%
\providecommand \selectlanguage [0]{\@gobble}%
\providecommand \bibinfo  [0]{\@secondoftwo}%
\providecommand \bibfield  [0]{\@secondoftwo}%
\providecommand \translation [1]{[#1]}%
\providecommand \BibitemOpen [0]{}%
\providecommand \bibitemStop [0]{}%
\providecommand \bibitemNoStop [0]{.\EOS\space}%
\providecommand \EOS [0]{\spacefactor3000\relax}%
\providecommand \BibitemShut  [1]{\csname bibitem#1\endcsname}%
\let\auto@bib@innerbib\@empty
\bibitem [{\citenamefont {Guth}(1981)}]{Guth:1980zm}%
  \BibitemOpen
  \bibfield  {author} {\bibinfo {author} {\bibfnamefont {A.~H.}\ \bibnamefont
  {Guth}},\ }\bibfield  {title} {\bibinfo {title} {Inflationary universe: A
  possible solution to the horizon and flatness problems},\ }\href
  {https://doi.org/10.1103/PhysRevD.23.347} {\bibfield  {journal} {\bibinfo
  {journal} {Phys. Rev. D}\ }\textbf {\bibinfo {volume} {23}},\ \bibinfo
  {pages} {347} (\bibinfo {year} {1981})}\BibitemShut {NoStop}%
\bibitem [{\citenamefont {Starobinsky}(1980)}]{Starobinsky:1980te}%
  \BibitemOpen
  \bibfield  {author} {\bibinfo {author} {\bibfnamefont {A.}~\bibnamefont
  {Starobinsky}},\ }\bibfield  {title} {\bibinfo {title} {A new type of
  isotropic cosmological models without singularity},\ }\href
  {https://doi.org/https://doi.org/10.1016/0370-2693(80)90670-X} {\bibfield
  {journal} {\bibinfo  {journal} {Phys. Lett. B}\ }\textbf {\bibinfo {volume}
  {91}},\ \bibinfo {pages} {99 } (\bibinfo {year} {1980})}\BibitemShut
  {NoStop}%
\bibitem [{\citenamefont {Starobinsky}(1982)}]{Starobinsky:1982ee}%
  \BibitemOpen
  \bibfield  {author} {\bibinfo {author} {\bibfnamefont {A.~A.}\ \bibnamefont
  {Starobinsky}},\ }\bibfield  {title} {\bibinfo {title} {Dynamics of phase
  transition in the new inflationary universe scenario and generation of
  perturbations},\ }\href {https://doi.org/10.1016/0370-2693(82)90541-X}
  {\bibfield  {journal} {\bibinfo  {journal} {Phys. Lett.}\ }\textbf {\bibinfo
  {volume} {117B}},\ \bibinfo {pages} {175} (\bibinfo {year}
  {1982})}\BibitemShut {NoStop}%
\bibitem [{\citenamefont {Kofman}\ \emph {et~al.}(1994)\citenamefont {Kofman},
  \citenamefont {Linde},\ and\ \citenamefont {Starobinsky}}]{Kofman:1994rk}%
  \BibitemOpen
  \bibfield  {author} {\bibinfo {author} {\bibfnamefont {L.}~\bibnamefont
  {Kofman}}, \bibinfo {author} {\bibfnamefont {A.~D.}\ \bibnamefont {Linde}},\
  and\ \bibinfo {author} {\bibfnamefont {A.~A.}\ \bibnamefont {Starobinsky}},\
  }\bibfield  {title} {\bibinfo {title} {Reheating after inflation},\ }\href
  {https://doi.org/10.1103/PhysRevLett.73.3195} {\bibfield  {journal} {\bibinfo
   {journal} {Phys. Rev. Lett.}\ }\textbf {\bibinfo {volume} {73}},\ \bibinfo
  {pages} {3195} (\bibinfo {year} {1994})},\ \Eprint
  {https://arxiv.org/abs/hep-th/9405187} {arXiv:hep-th/9405187 [hep-th]}
  \BibitemShut {NoStop}%
\bibitem [{\citenamefont {Kofman}\ \emph {et~al.}(1997)\citenamefont {Kofman},
  \citenamefont {Linde},\ and\ \citenamefont {Starobinsky}}]{Kofman:1997yn}%
  \BibitemOpen
  \bibfield  {author} {\bibinfo {author} {\bibfnamefont {L.}~\bibnamefont
  {Kofman}}, \bibinfo {author} {\bibfnamefont {A.~D.}\ \bibnamefont {Linde}},\
  and\ \bibinfo {author} {\bibfnamefont {A.~A.}\ \bibnamefont {Starobinsky}},\
  }\bibfield  {title} {\bibinfo {title} {Towards the theory of reheating after
  inflation},\ }\href {https://doi.org/10.1103/PhysRevD.56.3258} {\bibfield
  {journal} {\bibinfo  {journal} {Phys. Rev.}\ }\textbf {\bibinfo {volume}
  {D56}},\ \bibinfo {pages} {3258} (\bibinfo {year} {1997})},\ \Eprint
  {https://arxiv.org/abs/hep-ph/9704452} {arXiv:hep-ph/9704452 [hep-ph]}
  \BibitemShut {NoStop}%
\bibitem [{\citenamefont {Amin}\ \emph {et~al.}(2014)\citenamefont {Amin},
  \citenamefont {Hertzberg}, \citenamefont {Kaiser},\ and\ \citenamefont
  {Karouby}}]{Amin:2014eta}%
  \BibitemOpen
  \bibfield  {author} {\bibinfo {author} {\bibfnamefont {M.~A.}\ \bibnamefont
  {Amin}}, \bibinfo {author} {\bibfnamefont {M.~P.}\ \bibnamefont {Hertzberg}},
  \bibinfo {author} {\bibfnamefont {D.~I.}\ \bibnamefont {Kaiser}},\ and\
  \bibinfo {author} {\bibfnamefont {J.}~\bibnamefont {Karouby}},\ }\bibfield
  {title} {\bibinfo {title} {Nonperturbative dynamics of reheating after
  inflation: A review},\ }\href {https://doi.org/10.1142/S0218271815300037}
  {\bibfield  {journal} {\bibinfo  {journal} {Int. J. Mod. Phys.}\ }\textbf
  {\bibinfo {volume} {D24}},\ \bibinfo {pages} {1530003} (\bibinfo {year}
  {2014})},\ \Eprint {https://arxiv.org/abs/1410.3808} {arXiv:1410.3808
  [hep-ph]} \BibitemShut {NoStop}%
\bibitem [{\citenamefont {Micha}\ and\ \citenamefont
  {Tkachev}(2004)}]{Micha:2004bv}%
  \BibitemOpen
  \bibfield  {author} {\bibinfo {author} {\bibfnamefont {R.}~\bibnamefont
  {Micha}}\ and\ \bibinfo {author} {\bibfnamefont {I.~I.}\ \bibnamefont
  {Tkachev}},\ }\bibfield  {title} {\bibinfo {title} {Turbulent
  thermalization},\ }\href {https://doi.org/10.1103/PhysRevD.70.043538}
  {\bibfield  {journal} {\bibinfo  {journal} {Phys. Rev.}\ }\textbf {\bibinfo
  {volume} {D70}},\ \bibinfo {pages} {043538} (\bibinfo {year} {2004})},\
  \Eprint {https://arxiv.org/abs/hep-ph/0403101} {arXiv:hep-ph/0403101
  [hep-ph]} \BibitemShut {NoStop}%
\bibitem [{\citenamefont {Berges}\ \emph {et~al.}(2015)\citenamefont {Berges},
  \citenamefont {Boguslavski}, \citenamefont {Schlichting},\ and\ \citenamefont
  {Venugopalan}}]{Berges:2014bba}%
  \BibitemOpen
  \bibfield  {author} {\bibinfo {author} {\bibfnamefont {J.}~\bibnamefont
  {Berges}}, \bibinfo {author} {\bibfnamefont {K.}~\bibnamefont {Boguslavski}},
  \bibinfo {author} {\bibfnamefont {S.}~\bibnamefont {Schlichting}},\ and\
  \bibinfo {author} {\bibfnamefont {R.}~\bibnamefont {Venugopalan}},\
  }\bibfield  {title} {\bibinfo {title} {Universality far from equilibrium:
  From superfluid {Bose} gases to heavy-ion collisions},\ }\href
  {https://doi.org/10.1103/PhysRevLett.114.061601} {\bibfield  {journal}
  {\bibinfo  {journal} {Phys. Rev. Lett.}\ }\textbf {\bibinfo {volume} {114}},\
  \bibinfo {pages} {061601} (\bibinfo {year} {2015})},\ \Eprint
  {https://arxiv.org/abs/1408.1670} {arXiv:1408.1670 [hep-ph]} \BibitemShut
  {NoStop}%
\bibitem [{\citenamefont {Nowak}\ and\ \citenamefont
  {Gasenzer}(2014)}]{Nowak:2012gd}%
  \BibitemOpen
  \bibfield  {author} {\bibinfo {author} {\bibfnamefont {B.}~\bibnamefont
  {Nowak}}\ and\ \bibinfo {author} {\bibfnamefont {T.}~\bibnamefont
  {Gasenzer}},\ }\bibfield  {title} {\bibinfo {title} {Universal dynamics on
  the way to thermalization},\ }\href
  {https://doi.org/10.1088/1367-2630/16/9/093052} {\bibfield  {journal}
  {\bibinfo  {journal} {New J. Phys.}\ }\textbf {\bibinfo {volume} {16}},\
  \bibinfo {pages} {093052} (\bibinfo {year} {2014})},\ \Eprint
  {https://arxiv.org/abs/1206.3181} {arXiv:1206.3181 [cond-mat.quant-gas]}
  \BibitemShut {NoStop}%
\bibitem [{\citenamefont {Pi\~neiro Orioli}\ \emph {et~al.}(2015)\citenamefont
  {Pi\~neiro Orioli}, \citenamefont {Boguslavski},\ and\ \citenamefont
  {Berges}}]{Orioli:2015dxa}%
  \BibitemOpen
  \bibfield  {author} {\bibinfo {author} {\bibfnamefont {A.}~\bibnamefont
  {Pi\~neiro Orioli}}, \bibinfo {author} {\bibfnamefont {K.}~\bibnamefont
  {Boguslavski}},\ and\ \bibinfo {author} {\bibfnamefont {J.}~\bibnamefont
  {Berges}},\ }\bibfield  {title} {\bibinfo {title} {Universal self-similar
  dynamics of relativistic and nonrelativistic field theories near nonthermal
  fixed points},\ }\href {https://doi.org/10.1103/PhysRevD.92.025041}
  {\bibfield  {journal} {\bibinfo  {journal} {Phys. Rev. D}\ }\textbf {\bibinfo
  {volume} {92}},\ \bibinfo {pages} {025041} (\bibinfo {year} {2015})},\
  \Eprint {https://arxiv.org/abs/1503.02498} {arXiv:1503.02498 [hep-ph]}
  \BibitemShut {NoStop}%
\bibitem [{\citenamefont {Prüfer}\ \emph {et~al.}(2018)\citenamefont
  {Prüfer}, \citenamefont {Kunkel}, \citenamefont {Strobel}, \citenamefont
  {Lannig}, \citenamefont {Linnemann}, \citenamefont {Schmied}, \citenamefont
  {Berges}, \citenamefont {Gasenzer},\ and\ \citenamefont
  {Oberthaler}}]{Prufer:2018hto}%
  \BibitemOpen
  \bibfield  {author} {\bibinfo {author} {\bibfnamefont {M.}~\bibnamefont
  {Prüfer}}, \bibinfo {author} {\bibfnamefont {P.}~\bibnamefont {Kunkel}},
  \bibinfo {author} {\bibfnamefont {H.}~\bibnamefont {Strobel}}, \bibinfo
  {author} {\bibfnamefont {S.}~\bibnamefont {Lannig}}, \bibinfo {author}
  {\bibfnamefont {D.}~\bibnamefont {Linnemann}}, \bibinfo {author}
  {\bibfnamefont {C.-M.}\ \bibnamefont {Schmied}}, \bibinfo {author}
  {\bibfnamefont {J.}~\bibnamefont {Berges}}, \bibinfo {author} {\bibfnamefont
  {T.}~\bibnamefont {Gasenzer}},\ and\ \bibinfo {author} {\bibfnamefont
  {M.~K.}\ \bibnamefont {Oberthaler}},\ }\bibfield  {title} {\bibinfo {title}
  {Observation of universal dynamics in a spinor {Bose} gas far from
  equilibrium},\ }\href {https://doi.org/10.1038/s41586-018-0659-0} {\bibfield
  {journal} {\bibinfo  {journal} {\nat}\ }\textbf {\bibinfo {volume} {563}},\
  \bibinfo {pages} {217} (\bibinfo {year} {2018})},\ \Eprint
  {https://arxiv.org/abs/1805.11881} {arXiv:1805.11881 [cond-mat.quant-gas]}
  \BibitemShut {NoStop}%
\bibitem [{\citenamefont {Erne}\ \emph {et~al.}(2018)\citenamefont {Erne},
  \citenamefont {Bücker}, \citenamefont {Gasenzer}, \citenamefont {Berges},\
  and\ \citenamefont {Schmiedmayer}}]{Erne:2018gmz}%
  \BibitemOpen
  \bibfield  {author} {\bibinfo {author} {\bibfnamefont {S.}~\bibnamefont
  {Erne}}, \bibinfo {author} {\bibfnamefont {R.}~\bibnamefont {Bücker}},
  \bibinfo {author} {\bibfnamefont {T.}~\bibnamefont {Gasenzer}}, \bibinfo
  {author} {\bibfnamefont {J.}~\bibnamefont {Berges}},\ and\ \bibinfo {author}
  {\bibfnamefont {J.}~\bibnamefont {Schmiedmayer}},\ }\bibfield  {title}
  {\bibinfo {title} {Universal dynamics in an isolated one-dimensional {Bose}
  gas far from equilibrium},\ }\href
  {https://doi.org/10.1038/s41586-018-0667-0} {\bibfield  {journal} {\bibinfo
  {journal} {\nat}\ }\textbf {\bibinfo {volume} {563}},\ \bibinfo {pages} {225}
  (\bibinfo {year} {2018})},\ \Eprint {https://arxiv.org/abs/1805.12310}
  {arXiv:1805.12310 [cond-mat.quant-gas]} \BibitemShut {NoStop}%
\bibitem [{\citenamefont {Fedichev}\ and\ \citenamefont
  {Fischer}(2004)}]{Fedichev:2003bv}%
  \BibitemOpen
  \bibfield  {author} {\bibinfo {author} {\bibfnamefont {P.~O.}\ \bibnamefont
  {Fedichev}}\ and\ \bibinfo {author} {\bibfnamefont {U.~R.}\ \bibnamefont
  {Fischer}},\ }\bibfield  {title} {\bibinfo {title} {\enquote{Cosmological}
  quasiparticle production in harmonically trapped superfluid gases},\ }\href
  {https://doi.org/10.1103/PhysRevA.69.033602} {\bibfield  {journal} {\bibinfo
  {journal} {Phys. Rev.}\ }\textbf {\bibinfo {volume} {A69}},\ \bibinfo {pages}
  {033602} (\bibinfo {year} {2004})},\ \Eprint
  {https://arxiv.org/abs/cond-mat/0303063} {arXiv:cond-mat/0303063 [cond-mat]}
  \BibitemShut {NoStop}%
\bibitem [{\citenamefont {Fischer}\ and\ \citenamefont
  {Schutzhold}(2004)}]{Fischer:2004bf}%
  \BibitemOpen
  \bibfield  {author} {\bibinfo {author} {\bibfnamefont {U.~R.}\ \bibnamefont
  {Fischer}}\ and\ \bibinfo {author} {\bibfnamefont {R.}~\bibnamefont
  {Schutzhold}},\ }\bibfield  {title} {\bibinfo {title} {Quantum simulation of
  cosmic inflation in two-component {Bose--Einstein} condensates},\ }\href
  {https://doi.org/10.1103/PhysRevA.70.063615} {\bibfield  {journal} {\bibinfo
  {journal} {Phys. Rev.}\ }\textbf {\bibinfo {volume} {A70}},\ \bibinfo {pages}
  {063615} (\bibinfo {year} {2004})},\ \Eprint
  {https://arxiv.org/abs/cond-mat/0406470} {arXiv:cond-mat/0406470 [cond-mat]}
  \BibitemShut {NoStop}%
\bibitem [{\citenamefont {Uhlmann}\ \emph {et~al.}(2005)\citenamefont
  {Uhlmann}, \citenamefont {Xu},\ and\ \citenamefont
  {Schutzhold}}]{Uhlmann:2005hf}%
  \BibitemOpen
  \bibfield  {author} {\bibinfo {author} {\bibfnamefont {M.}~\bibnamefont
  {Uhlmann}}, \bibinfo {author} {\bibfnamefont {Y.}~\bibnamefont {Xu}},\ and\
  \bibinfo {author} {\bibfnamefont {R.}~\bibnamefont {Schutzhold}},\ }\bibfield
   {title} {\bibinfo {title} {Aspects of cosmic inflation in expanding
  {Bose--Einstein} condensates},\ }\href
  {https://doi.org/10.1088/1367-2630/7/1/248} {\bibfield  {journal} {\bibinfo
  {journal} {New J. Phys.}\ }\textbf {\bibinfo {volume} {7}},\ \bibinfo {pages}
  {248} (\bibinfo {year} {2005})},\ \Eprint
  {https://arxiv.org/abs/quant-ph/0509063} {arXiv:quant-ph/0509063}
  \BibitemShut {NoStop}%
\bibitem [{\citenamefont {Neuenhahn}\ and\ \citenamefont
  {Marquardt}(2015)}]{Neuenhahn:2012dz}%
  \BibitemOpen
  \bibfield  {author} {\bibinfo {author} {\bibfnamefont {C.}~\bibnamefont
  {Neuenhahn}}\ and\ \bibinfo {author} {\bibfnamefont {F.}~\bibnamefont
  {Marquardt}},\ }\bibfield  {title} {\bibinfo {title} {Quantum simulation of
  expanding space–time with tunnel-coupled condensates},\ }\href
  {https://doi.org/10.1088/1367-2630/17/12/125007} {\bibfield  {journal}
  {\bibinfo  {journal} {New J. Phys.}\ }\textbf {\bibinfo {volume} {17}},\
  \bibinfo {pages} {125007} (\bibinfo {year} {2015})},\ \Eprint
  {https://arxiv.org/abs/1208.2255} {arXiv:1208.2255 [cond-mat.quant-gas]}
  \BibitemShut {NoStop}%
\bibitem [{\citenamefont {{Neuenhahn}}\ \emph {et~al.}(2012)\citenamefont
  {{Neuenhahn}}, \citenamefont {{Polkovnikov}},\ and\ \citenamefont
  {{Marquardt}}}]{neuenhahn2012localized}%
  \BibitemOpen
  \bibfield  {author} {\bibinfo {author} {\bibfnamefont {C.}~\bibnamefont
  {{Neuenhahn}}}, \bibinfo {author} {\bibfnamefont {A.}~\bibnamefont
  {{Polkovnikov}}},\ and\ \bibinfo {author} {\bibfnamefont {F.}~\bibnamefont
  {{Marquardt}}},\ }\bibfield  {title} {\bibinfo {title} {Localized phase
  structures growing out of quantum fluctuations in a quench of tunnel-coupled
  atomic condensates},\ }\href {https://doi.org/10.1103/PhysRevLett.109.085304}
  {\bibfield  {journal} {\bibinfo  {journal} {\prl}\ }\textbf {\bibinfo
  {volume} {109}},\ \bibinfo {eid} {085304} (\bibinfo {year} {2012})},\ \Eprint
  {https://arxiv.org/abs/1112.5982} {arXiv:1112.5982 [cond-mat.quant-gas]}
  \BibitemShut {NoStop}%
\bibitem [{\citenamefont {Posazhennikova}\ \emph {et~al.}(2016)\citenamefont
  {Posazhennikova}, \citenamefont {Trujillo-Martinez},\ and\ \citenamefont
  {Kroha}}]{Posazhennikova:2016nut}%
  \BibitemOpen
  \bibfield  {author} {\bibinfo {author} {\bibfnamefont {A.}~\bibnamefont
  {Posazhennikova}}, \bibinfo {author} {\bibfnamefont {M.}~\bibnamefont
  {Trujillo-Martinez}},\ and\ \bibinfo {author} {\bibfnamefont
  {J.}~\bibnamefont {Kroha}},\ }\bibfield  {title} {\bibinfo {title}
  {Inflationary quasiparticle creation and thermalization dynamics in coupled
  {Bose--Einstein} condensates},\ }\href
  {https://doi.org/10.1103/PhysRevLett.116.225304} {\bibfield  {journal}
  {\bibinfo  {journal} {Phys. Rev. Lett.}\ }\textbf {\bibinfo {volume} {116}},\
  \bibinfo {pages} {225304} (\bibinfo {year} {2016})},\ \Eprint
  {https://arxiv.org/abs/1603.04898} {arXiv:1603.04898 [cond-mat.quant-gas]}
  \BibitemShut {NoStop}%
\bibitem [{\citenamefont {Zache}\ \emph {et~al.}(2017)\citenamefont {Zache},
  \citenamefont {Kasper},\ and\ \citenamefont {Berges}}]{Zache:2017dnz}%
  \BibitemOpen
  \bibfield  {author} {\bibinfo {author} {\bibfnamefont {T.~V.}\ \bibnamefont
  {Zache}}, \bibinfo {author} {\bibfnamefont {V.}~\bibnamefont {Kasper}},\ and\
  \bibinfo {author} {\bibfnamefont {J.}~\bibnamefont {Berges}},\ }\bibfield
  {title} {\bibinfo {title} {Inflationary preheating dynamics with two-species
  condensates},\ }\href {https://doi.org/10.1103/PhysRevA.95.063629} {\bibfield
   {journal} {\bibinfo  {journal} {Phys. Rev.}\ }\textbf {\bibinfo {volume}
  {A95}},\ \bibinfo {pages} {063629} (\bibinfo {year} {2017})},\ \Eprint
  {https://arxiv.org/abs/1704.02271} {arXiv:1704.02271 [cond-mat.quant-gas]}
  \BibitemShut {NoStop}%
\bibitem [{\citenamefont {Robertson}\ \emph {et~al.}(2018)\citenamefont
  {Robertson}, \citenamefont {Michel},\ and\ \citenamefont
  {Parentani}}]{Robertson:2018gwi}%
  \BibitemOpen
  \bibfield  {author} {\bibinfo {author} {\bibfnamefont {S.}~\bibnamefont
  {Robertson}}, \bibinfo {author} {\bibfnamefont {F.}~\bibnamefont {Michel}},\
  and\ \bibinfo {author} {\bibfnamefont {R.}~\bibnamefont {Parentani}},\
  }\bibfield  {title} {\bibinfo {title} {Nonlinearities induced by parametric
  resonance in effectively {1D} atomic {Bose} condensates},\ }\href
  {https://doi.org/10.1103/PhysRevD.98.056003} {\bibfield  {journal} {\bibinfo
  {journal} {Phys. Rev.}\ }\textbf {\bibinfo {volume} {D98}},\ \bibinfo {pages}
  {056003} (\bibinfo {year} {2018})},\ \Eprint
  {https://arxiv.org/abs/1802.00739} {arXiv:1802.00739 [cond-mat.quant-gas]}
  \BibitemShut {NoStop}%
\bibitem [{\citenamefont {Wittemer}\ \emph {et~al.}(2019)\citenamefont
  {Wittemer}, \citenamefont {Hakelberg}, \citenamefont {Kiefer}, \citenamefont
  {Schr\"oder}, \citenamefont {Fey}, \citenamefont {Sch\"utzhold},
  \citenamefont {Warring},\ and\ \citenamefont {Schaetz}}]{Wittemer:2019agm}%
  \BibitemOpen
  \bibfield  {author} {\bibinfo {author} {\bibfnamefont {M.}~\bibnamefont
  {Wittemer}}, \bibinfo {author} {\bibfnamefont {F.}~\bibnamefont {Hakelberg}},
  \bibinfo {author} {\bibfnamefont {P.}~\bibnamefont {Kiefer}}, \bibinfo
  {author} {\bibfnamefont {J.-P.}\ \bibnamefont {Schr\"oder}}, \bibinfo
  {author} {\bibfnamefont {C.}~\bibnamefont {Fey}}, \bibinfo {author}
  {\bibfnamefont {R.}~\bibnamefont {Sch\"utzhold}}, \bibinfo {author}
  {\bibfnamefont {U.}~\bibnamefont {Warring}},\ and\ \bibinfo {author}
  {\bibfnamefont {T.}~\bibnamefont {Schaetz}},\ }\bibfield  {title} {\bibinfo
  {title} {Phonon pair creation by inflating quantum fluctuations in an ion
  trap},\ }\href {https://doi.org/10.1103/PhysRevLett.123.180502} {\bibfield
  {journal} {\bibinfo  {journal} {Phys. Rev. Lett.}\ }\textbf {\bibinfo
  {volume} {123}},\ \bibinfo {pages} {180502} (\bibinfo {year} {2019})},\
  \Eprint {https://arxiv.org/abs/1903.05523} {arXiv:1903.05523 [quant-ph]}
  \BibitemShut {NoStop}%
\bibitem [{\citenamefont {Chin}\ \emph {et~al.}(2010)\citenamefont {Chin},
  \citenamefont {Grimm}, \citenamefont {Julienne},\ and\ \citenamefont
  {Tiesinga}}]{Chin2010}%
  \BibitemOpen
  \bibfield  {author} {\bibinfo {author} {\bibfnamefont {C.}~\bibnamefont
  {Chin}}, \bibinfo {author} {\bibfnamefont {R.}~\bibnamefont {Grimm}},
  \bibinfo {author} {\bibfnamefont {P.}~\bibnamefont {Julienne}},\ and\
  \bibinfo {author} {\bibfnamefont {E.}~\bibnamefont {Tiesinga}},\ }\bibfield
  {title} {\bibinfo {title} {{Feshbach} resonances in ultracold gases},\ }\href
  {https://doi.org/10.1103/RevModPhys.82.1225} {\bibfield  {journal} {\bibinfo
  {journal} {Rev. Mod. Phys.}\ }\textbf {\bibinfo {volume} {82}},\ \bibinfo
  {pages} {1225} (\bibinfo {year} {2010})},\ \Eprint
  {https://arxiv.org/abs/0812.1496} {arXiv:0812.1496 [cond-mat.other]}
  \BibitemShut {NoStop}%
\bibitem [{\citenamefont {Eckel}\ \emph {et~al.}(2018)\citenamefont {Eckel},
  \citenamefont {Kumar}, \citenamefont {Jacobson}, \citenamefont {Spielman},\
  and\ \citenamefont {Campbell}}]{Eckel:2017uqx}%
  \BibitemOpen
  \bibfield  {author} {\bibinfo {author} {\bibfnamefont {S.}~\bibnamefont
  {Eckel}}, \bibinfo {author} {\bibfnamefont {A.}~\bibnamefont {Kumar}},
  \bibinfo {author} {\bibfnamefont {T.}~\bibnamefont {Jacobson}}, \bibinfo
  {author} {\bibfnamefont {I.~B.}\ \bibnamefont {Spielman}},\ and\ \bibinfo
  {author} {\bibfnamefont {G.~K.}\ \bibnamefont {Campbell}},\ }\bibfield
  {title} {\bibinfo {title} {A rapidly expanding {Bose-Einstein} condensate: an
  expanding universe in the lab},\ }\href
  {https://doi.org/10.1103/PhysRevX.8.021021} {\bibfield  {journal} {\bibinfo
  {journal} {Phys. Rev.}\ }\textbf {\bibinfo {volume} {X8}},\ \bibinfo {pages}
  {021021} (\bibinfo {year} {2018})},\ \Eprint
  {https://arxiv.org/abs/1710.05800} {arXiv:1710.05800 [cond-mat.quant-gas]}
  \BibitemShut {NoStop}%
\bibitem [{\citenamefont {Gritsev}\ \emph {et~al.}(2010)\citenamefont
  {Gritsev}, \citenamefont {Barmettler},\ and\ \citenamefont
  {Demler}}]{Gritsev:2009gf}%
  \BibitemOpen
  \bibfield  {author} {\bibinfo {author} {\bibfnamefont {V.}~\bibnamefont
  {Gritsev}}, \bibinfo {author} {\bibfnamefont {P.}~\bibnamefont
  {Barmettler}},\ and\ \bibinfo {author} {\bibfnamefont {E.}~\bibnamefont
  {Demler}},\ }\bibfield  {title} {\bibinfo {title} {Scaling approach to
  quantum non-equilibrium dynamics of many-body systems},\ }\href
  {https://doi.org/10.1088/1367-2630/12/11/113005} {\bibfield  {journal}
  {\bibinfo  {journal} {New J. Phys.}\ }\textbf {\bibinfo {volume} {12}},\
  \bibinfo {pages} {113005} (\bibinfo {year} {2010})},\ \Eprint
  {https://arxiv.org/abs/0912.2744} {arXiv:0912.2744 [cond-mat.quant-gas]}
  \BibitemShut {NoStop}%
\bibitem [{\citenamefont {Ch\"a}\ and\ \citenamefont
  {Fischer}(2017)}]{Cha:2016esj}%
  \BibitemOpen
  \bibfield  {author} {\bibinfo {author} {\bibfnamefont {S.-Y.}\ \bibnamefont
  {Ch\"a}}\ and\ \bibinfo {author} {\bibfnamefont {U.~R.}\ \bibnamefont
  {Fischer}},\ }\bibfield  {title} {\bibinfo {title} {Probing the scale
  invariance of the inflationary power spectrum in expanding
  quasi-two-dimensional dipolar condensates},\ }\href
  {https://doi.org/10.1103/PhysRevLett.118.130404} {\bibfield  {journal}
  {\bibinfo  {journal} {Phys. Rev. Lett.}\ }\textbf {\bibinfo {volume} {118}},\
  \bibinfo {pages} {130404} (\bibinfo {year} {2017})},\ \Eprint
  {https://arxiv.org/abs/1609.06155} {arXiv:1609.06155 [cond-mat.quant-gas]}
  \BibitemShut {NoStop}%
\bibitem [{\citenamefont {{Saint-Jalm}}\ \emph {et~al.}(2019)\citenamefont
  {{Saint-Jalm}}, \citenamefont {{Castilho}}, \citenamefont {{Le Cerf}},
  \citenamefont {{Bakkali-Hassani}}, \citenamefont {{Ville}}, \citenamefont
  {{Nascimbene}}, \citenamefont {{Beugnon}},\ and\ \citenamefont
  {{Dalibard}}}]{saint2019dynamical}%
  \BibitemOpen
  \bibfield  {author} {\bibinfo {author} {\bibfnamefont {R.}~\bibnamefont
  {{Saint-Jalm}}}, \bibinfo {author} {\bibfnamefont {P.~C.~M.}\ \bibnamefont
  {{Castilho}}}, \bibinfo {author} {\bibfnamefont {{\'E}.}~\bibnamefont {{Le
  Cerf}}}, \bibinfo {author} {\bibfnamefont {B.}~\bibnamefont
  {{Bakkali-Hassani}}}, \bibinfo {author} {\bibfnamefont {J.~L.}\ \bibnamefont
  {{Ville}}}, \bibinfo {author} {\bibfnamefont {S.}~\bibnamefont
  {{Nascimbene}}}, \bibinfo {author} {\bibfnamefont {J.}~\bibnamefont
  {{Beugnon}}},\ and\ \bibinfo {author} {\bibfnamefont {J.}~\bibnamefont
  {{Dalibard}}},\ }\bibfield  {title} {\bibinfo {title} {Dynamical symmetry and
  breathers in a two-dimensional {Bose} gas},\ }\href
  {https://doi.org/10.1103/PhysRevX.9.021035} {\bibfield  {journal} {\bibinfo
  {journal} {Phys. Rev. X}\ }\textbf {\bibinfo {volume} {9}},\ \bibinfo {eid}
  {021035} (\bibinfo {year} {2019})},\ \Eprint
  {https://arxiv.org/abs/1903.04528} {arXiv:1903.04528 [cond-mat.quant-gas]}
  \BibitemShut {NoStop}%
\bibitem [{\citenamefont {Khlebnikov}\ and\ \citenamefont
  {Tkachev}(1996)}]{Khlebnikov:1996mc}%
  \BibitemOpen
  \bibfield  {author} {\bibinfo {author} {\bibfnamefont {S.~{\relax Yu}.}\
  \bibnamefont {Khlebnikov}}\ and\ \bibinfo {author} {\bibfnamefont {I.~I.}\
  \bibnamefont {Tkachev}},\ }\bibfield  {title} {\bibinfo {title} {Classical
  decay of inflaton},\ }\href {https://doi.org/10.1103/PhysRevLett.77.219}
  {\bibfield  {journal} {\bibinfo  {journal} {Phys. Rev. Lett.}\ }\textbf
  {\bibinfo {volume} {77}},\ \bibinfo {pages} {219} (\bibinfo {year} {1996})},\
  \Eprint {https://arxiv.org/abs/hep-ph/9603378} {arXiv:hep-ph/9603378
  [hep-ph]} \BibitemShut {NoStop}%
\bibitem [{\citenamefont {Berges}\ and\ \citenamefont
  {Gasenzer}(2007)}]{Berges:2007ym}%
  \BibitemOpen
  \bibfield  {author} {\bibinfo {author} {\bibfnamefont {J.}~\bibnamefont
  {Berges}}\ and\ \bibinfo {author} {\bibfnamefont {T.}~\bibnamefont
  {Gasenzer}},\ }\bibfield  {title} {\bibinfo {title} {Quantum versus classical
  statistical dynamics of an ultracold {Bose} gas},\ }\href
  {https://doi.org/10.1103/PhysRevA.76.033604} {\bibfield  {journal} {\bibinfo
  {journal} {Phys. Rev.}\ }\textbf {\bibinfo {volume} {A76}},\ \bibinfo {pages}
  {033604} (\bibinfo {year} {2007})},\ \Eprint
  {https://arxiv.org/abs/cond-mat/0703163} {arXiv:cond-mat/0703163
  [cond-mat.other]} \BibitemShut {NoStop}%
\bibitem [{\citenamefont {{Sinatra}}\ \emph {et~al.}(2002)\citenamefont
  {{Sinatra}}, \citenamefont {{Lobo}},\ and\ \citenamefont
  {{Castin}}}]{Sinatra2002}%
  \BibitemOpen
  \bibfield  {author} {\bibinfo {author} {\bibfnamefont {A.}~\bibnamefont
  {{Sinatra}}}, \bibinfo {author} {\bibfnamefont {C.}~\bibnamefont {{Lobo}}},\
  and\ \bibinfo {author} {\bibfnamefont {Y.}~\bibnamefont {{Castin}}},\
  }\bibfield  {title} {\bibinfo {title} {The truncated {Wigner} method for
  {Bose}-condensed gases: limits of validity and applications},\ }\href
  {https://doi.org/10.1088/0953-4075/35/17/301} {\bibfield  {journal} {\bibinfo
   {journal} {J. Phys. B}\ }\textbf {\bibinfo {volume} {35}},\ \bibinfo {pages}
  {3599} (\bibinfo {year} {2002})},\ \Eprint
  {https://arxiv.org/abs/cond-mat/0201217} {arXiv:cond-mat/0201217
  [cond-mat.stat-mech]} \BibitemShut {NoStop}%
\bibitem [{\citenamefont {Blakie}\ \emph {et~al.}(2008)\citenamefont {Blakie},
  \citenamefont {Bradley}, \citenamefont {Davis}, \citenamefont {Ballagh},\
  and\ \citenamefont {Gardiner}}]{Blakie2008}%
  \BibitemOpen
  \bibfield  {author} {\bibinfo {author} {\bibfnamefont {P.}~\bibnamefont
  {Blakie}}, \bibinfo {author} {\bibfnamefont {A.}~\bibnamefont {Bradley}},
  \bibinfo {author} {\bibfnamefont {M.}~\bibnamefont {Davis}}, \bibinfo
  {author} {\bibfnamefont {R.}~\bibnamefont {Ballagh}},\ and\ \bibinfo {author}
  {\bibfnamefont {C.}~\bibnamefont {Gardiner}},\ }\bibfield  {title} {\bibinfo
  {title} {Dynamics and statistical mechanics of ultra-cold {Bose} gases using
  c-field techniques},\ }\href {https://doi.org/10.1080/00018730802564254}
  {\bibfield  {journal} {\bibinfo  {journal} {Adv. Phys.}\ }\textbf {\bibinfo
  {volume} {57}},\ \bibinfo {pages} {363} (\bibinfo {year} {2008})},\ \Eprint
  {https://arxiv.org/abs/0809.1487} {arXiv:0809.1487 [cond-mat.stat-mech]}
  \BibitemShut {NoStop}%
\bibitem [{\citenamefont {Barcel{\'o}}\ \emph {et~al.}(2011)\citenamefont
  {Barcel{\'o}}, \citenamefont {Liberati},\ and\ \citenamefont
  {Visser}}]{Barcelo2011}%
  \BibitemOpen
  \bibfield  {author} {\bibinfo {author} {\bibfnamefont {C.}~\bibnamefont
  {Barcel{\'o}}}, \bibinfo {author} {\bibfnamefont {S.}~\bibnamefont
  {Liberati}},\ and\ \bibinfo {author} {\bibfnamefont {M.}~\bibnamefont
  {Visser}},\ }\bibfield  {title} {\bibinfo {title} {Analogue gravity},\ }\href
  {https://doi.org/10.12942/lrr-2011-3} {\bibfield  {journal} {\bibinfo
  {journal} {Living Rev. Relativ.}\ }\textbf {\bibinfo {volume} {14}},\
  \bibinfo {pages} {3} (\bibinfo {year} {2011})},\ \Eprint
  {https://arxiv.org/abs/gr-qc/0505065} {arXiv:gr-qc/0505065 [gr-qc]}
  \BibitemShut {NoStop}%
\bibitem [{\citenamefont {Jain}\ \emph {et~al.}(2007)\citenamefont {Jain},
  \citenamefont {Weinfurtner}, \citenamefont {Visser},\ and\ \citenamefont
  {Gardiner}}]{Jain:2007gg}%
  \BibitemOpen
  \bibfield  {author} {\bibinfo {author} {\bibfnamefont {P.}~\bibnamefont
  {Jain}}, \bibinfo {author} {\bibfnamefont {S.}~\bibnamefont {Weinfurtner}},
  \bibinfo {author} {\bibfnamefont {M.}~\bibnamefont {Visser}},\ and\ \bibinfo
  {author} {\bibfnamefont {C.~W.}\ \bibnamefont {Gardiner}},\ }\bibfield
  {title} {\bibinfo {title} {Analogue model of a {FRW} universe in
  {Bose--Einstein} condensates: Application of the classical field method},\
  }\href {https://doi.org/10.1103/PhysRevA.76.033616} {\bibfield  {journal}
  {\bibinfo  {journal} {Phys. Rev.}\ }\textbf {\bibinfo {volume} {A76}},\
  \bibinfo {pages} {033616} (\bibinfo {year} {2007})},\ \Eprint
  {https://arxiv.org/abs/0705.2077} {arXiv:0705.2077 [cond-mat.other]}
  \BibitemShut {NoStop}%
\bibitem [{\citenamefont {{Staliunas}}\ \emph {et~al.}(2002)\citenamefont
  {{Staliunas}}, \citenamefont {{Longhi}},\ and\ \citenamefont {{de
  Valc{\'a}rcel}}}]{Staliunas2002}%
  \BibitemOpen
  \bibfield  {author} {\bibinfo {author} {\bibfnamefont {K.}~\bibnamefont
  {{Staliunas}}}, \bibinfo {author} {\bibfnamefont {S.}~\bibnamefont
  {{Longhi}}},\ and\ \bibinfo {author} {\bibfnamefont {G.~J.}\ \bibnamefont
  {{de Valc{\'a}rcel}}},\ }\bibfield  {title} {\bibinfo {title} {{Faraday}
  patterns in {Bose--Einstein} condensates},\ }\href
  {https://doi.org/10.1103/PhysRevLett.89.210406} {\bibfield  {journal}
  {\bibinfo  {journal} {\prl}\ }\textbf {\bibinfo {volume} {89}},\ \bibinfo
  {eid} {210406} (\bibinfo {year} {2002})},\ \Eprint
  {https://arxiv.org/abs/cond-mat/0204517} {arXiv:cond-mat/0204517
  [cond-mat.stat-mech]} \BibitemShut {NoStop}%
\bibitem [{\citenamefont {Engels}\ \emph {et~al.}(2007)\citenamefont {Engels},
  \citenamefont {Atherton},\ and\ \citenamefont {Hoefer}}]{Engels:2007zz}%
  \BibitemOpen
  \bibfield  {author} {\bibinfo {author} {\bibfnamefont {P.}~\bibnamefont
  {Engels}}, \bibinfo {author} {\bibfnamefont {C.}~\bibnamefont {Atherton}},\
  and\ \bibinfo {author} {\bibfnamefont {M.~A.}\ \bibnamefont {Hoefer}},\
  }\bibfield  {title} {\bibinfo {title} {Observation of {Faraday} waves in a
  {Bose--Einstein} condensate},\ }\href
  {https://doi.org/10.1103/PhysRevLett.98.095301} {\bibfield  {journal}
  {\bibinfo  {journal} {Phys. Rev. Lett.}\ }\textbf {\bibinfo {volume} {98}},\
  \bibinfo {pages} {095301} (\bibinfo {year} {2007})},\ \Eprint
  {https://arxiv.org/abs/cond-mat/0701028} {arXiv:cond-mat/0701028
  [cond-mat.other]} \BibitemShut {NoStop}%
\bibitem [{\citenamefont {{Pollack}}\ \emph {et~al.}(2010)\citenamefont
  {{Pollack}}, \citenamefont {{Dries}}, \citenamefont {{Hulet}}, \citenamefont
  {{Magalh{\~a}es}}, \citenamefont {{Henn}}, \citenamefont {{Ramos}},
  \citenamefont {{Caracanhas}},\ and\ \citenamefont
  {{Bagnato}}}]{PhysRevA.81.053627}%
  \BibitemOpen
  \bibfield  {author} {\bibinfo {author} {\bibfnamefont {S.~E.}\ \bibnamefont
  {{Pollack}}}, \bibinfo {author} {\bibfnamefont {D.}~\bibnamefont {{Dries}}},
  \bibinfo {author} {\bibfnamefont {R.~G.}\ \bibnamefont {{Hulet}}}, \bibinfo
  {author} {\bibfnamefont {K.~M.~F.}\ \bibnamefont {{Magalh{\~a}es}}}, \bibinfo
  {author} {\bibfnamefont {E.~A.~L.}\ \bibnamefont {{Henn}}}, \bibinfo {author}
  {\bibfnamefont {E.~R.~F.}\ \bibnamefont {{Ramos}}}, \bibinfo {author}
  {\bibfnamefont {M.~A.}\ \bibnamefont {{Caracanhas}}},\ and\ \bibinfo {author}
  {\bibfnamefont {V.~S.}\ \bibnamefont {{Bagnato}}},\ }\bibfield  {title}
  {\bibinfo {title} {Collective excitation of a {Bose--Einstein} condensate by
  modulation of the atomic scattering length},\ }\href
  {https://doi.org/10.1103/PhysRevA.81.053627} {\bibfield  {journal} {\bibinfo
  {journal} {\pra}\ }\textbf {\bibinfo {volume} {81}},\ \bibinfo {eid} {053627}
  (\bibinfo {year} {2010})},\ \Eprint {https://arxiv.org/abs/1004.2887}
  {arXiv:1004.2887 [cond-mat.quant-gas]} \BibitemShut {NoStop}%
\bibitem [{\citenamefont {{Vidanovi{\'c}}}\ \emph {et~al.}(2011)\citenamefont
  {{Vidanovi{\'c}}}, \citenamefont {{Bala{\v{z}}}}, \citenamefont
  {{Al-Jibbouri}},\ and\ \citenamefont {{Pelster}}}]{PhysRevA.84.013618}%
  \BibitemOpen
  \bibfield  {author} {\bibinfo {author} {\bibfnamefont {I.}~\bibnamefont
  {{Vidanovi{\'c}}}}, \bibinfo {author} {\bibfnamefont {A.}~\bibnamefont
  {{Bala{\v{z}}}}}, \bibinfo {author} {\bibfnamefont {H.}~\bibnamefont
  {{Al-Jibbouri}}},\ and\ \bibinfo {author} {\bibfnamefont {A.}~\bibnamefont
  {{Pelster}}},\ }\bibfield  {title} {\bibinfo {title} {Nonlinear
  {Bose--Einstein}-condensate dynamics induced by a harmonic modulation of the
  s-wave scattering length},\ }\href
  {https://doi.org/10.1103/PhysRevA.84.013618} {\bibfield  {journal} {\bibinfo
  {journal} {\pra}\ }\textbf {\bibinfo {volume} {84}},\ \bibinfo {eid} {013618}
  (\bibinfo {year} {2011})},\ \Eprint {https://arxiv.org/abs/1106.4686}
  {arXiv:1106.4686 [cond-mat.quant-gas]} \BibitemShut {NoStop}%
\bibitem [{\citenamefont {Jaskula}\ \emph {et~al.}(2012)\citenamefont
  {Jaskula}, \citenamefont {Partridge}, \citenamefont {Bonneau}, \citenamefont
  {Lopes}, \citenamefont {Ruaudel}, \citenamefont {Boiron},\ and\ \citenamefont
  {Westbrook}}]{Jaskula:2012ab}%
  \BibitemOpen
  \bibfield  {author} {\bibinfo {author} {\bibfnamefont {J.~C.}\ \bibnamefont
  {Jaskula}}, \bibinfo {author} {\bibfnamefont {G.~B.}\ \bibnamefont
  {Partridge}}, \bibinfo {author} {\bibfnamefont {M.}~\bibnamefont {Bonneau}},
  \bibinfo {author} {\bibfnamefont {R.}~\bibnamefont {Lopes}}, \bibinfo
  {author} {\bibfnamefont {J.}~\bibnamefont {Ruaudel}}, \bibinfo {author}
  {\bibfnamefont {D.}~\bibnamefont {Boiron}},\ and\ \bibinfo {author}
  {\bibfnamefont {C.~I.}\ \bibnamefont {Westbrook}},\ }\bibfield  {title}
  {\bibinfo {title} {An acoustic analog to the dynamical {Casimir} effect in a
  {Bose-Einstein} condensate},\ }\href
  {https://doi.org/10.1103/PhysRevLett.109.220401} {\bibfield  {journal}
  {\bibinfo  {journal} {Phys. Rev. Lett.}\ }\textbf {\bibinfo {volume} {109}},\
  \bibinfo {pages} {220401} (\bibinfo {year} {2012})},\ \Eprint
  {https://arxiv.org/abs/1207.1338} {arXiv:1207.1338 [cond-mat.quant-gas]}
  \BibitemShut {NoStop}%
\bibitem [{\citenamefont {Nguyen}\ \emph {et~al.}(2019)\citenamefont {Nguyen},
  \citenamefont {Tsatsos}, \citenamefont {Luo}, \citenamefont {Lode},
  \citenamefont {Telles}, \citenamefont {Bagnato},\ and\ \citenamefont
  {Hulet}}]{PhysRevX.9.011052}%
  \BibitemOpen
  \bibfield  {author} {\bibinfo {author} {\bibfnamefont {J.~H.~V.}\
  \bibnamefont {Nguyen}}, \bibinfo {author} {\bibfnamefont {M.~C.}\
  \bibnamefont {Tsatsos}}, \bibinfo {author} {\bibfnamefont {D.}~\bibnamefont
  {Luo}}, \bibinfo {author} {\bibfnamefont {A.~U.~J.}\ \bibnamefont {Lode}},
  \bibinfo {author} {\bibfnamefont {G.~D.}\ \bibnamefont {Telles}}, \bibinfo
  {author} {\bibfnamefont {V.~S.}\ \bibnamefont {Bagnato}},\ and\ \bibinfo
  {author} {\bibfnamefont {R.~G.}\ \bibnamefont {Hulet}},\ }\bibfield  {title}
  {\bibinfo {title} {Parametric excitation of a {Bose--Einstein} condensate:
  From {Faraday} waves to granulation},\ }\href
  {https://doi.org/10.1103/PhysRevX.9.011052} {\bibfield  {journal} {\bibinfo
  {journal} {Phys. Rev. X}\ }\textbf {\bibinfo {volume} {9}},\ \bibinfo {pages}
  {011052} (\bibinfo {year} {2019})},\ \Eprint
  {https://arxiv.org/abs/1707.04055} {arXiv:1707.04055 [cond-mat.quant-gas]}
  \BibitemShut {NoStop}%
\bibitem [{\citenamefont {{Zhang}}\ \emph {et~al.}(2020)\citenamefont
  {{Zhang}}, \citenamefont {{Yao}}, \citenamefont {{Feng}}, \citenamefont
  {{Hu}},\ and\ \citenamefont {{Chin}}}]{Zhang2020}%
  \BibitemOpen
  \bibfield  {author} {\bibinfo {author} {\bibfnamefont {Z.}~\bibnamefont
  {{Zhang}}}, \bibinfo {author} {\bibfnamefont {K.-X.}\ \bibnamefont {{Yao}}},
  \bibinfo {author} {\bibfnamefont {L.}~\bibnamefont {{Feng}}}, \bibinfo
  {author} {\bibfnamefont {J.}~\bibnamefont {{Hu}}},\ and\ \bibinfo {author}
  {\bibfnamefont {C.}~\bibnamefont {{Chin}}},\ }\bibfield  {title} {\bibinfo
  {title} {Pattern formation in a driven {Bose--Einstein} condensate},\ }\href
  {https://doi.org/10.1038/s41567-020-0839-3} {\bibfield  {journal} {\bibinfo
  {journal} {Nat. Phys.}\ }\textbf {\bibinfo {volume} {16}},\ \bibinfo {pages}
  {652} (\bibinfo {year} {2020})},\ \Eprint {https://arxiv.org/abs/1909.05536}
  {arXiv:1909.05536 [cond-mat.quant-gas]} \BibitemShut {NoStop}%
\bibitem [{\citenamefont {Berges}\ and\ \citenamefont
  {Serreau}(2003)}]{Berges:2002cz}%
  \BibitemOpen
  \bibfield  {author} {\bibinfo {author} {\bibfnamefont {J.}~\bibnamefont
  {Berges}}\ and\ \bibinfo {author} {\bibfnamefont {J.}~\bibnamefont
  {Serreau}},\ }\bibfield  {title} {\bibinfo {title} {Parametric resonance in
  quantum field theory},\ }\href
  {https://doi.org/10.1103/PhysRevLett.91.111601} {\bibfield  {journal}
  {\bibinfo  {journal} {Phys. Rev. Lett.}\ }\textbf {\bibinfo {volume} {91}},\
  \bibinfo {pages} {111601} (\bibinfo {year} {2003})},\ \Eprint
  {https://arxiv.org/abs/hep-ph/0208070} {arXiv:hep-ph/0208070 [hep-ph]}
  \BibitemShut {NoStop}%
\bibitem [{\citenamefont {Berges}\ and\ \citenamefont
  {Hoffmeister}(2009)}]{Berges:2008sr}%
  \BibitemOpen
  \bibfield  {author} {\bibinfo {author} {\bibfnamefont {J.}~\bibnamefont
  {Berges}}\ and\ \bibinfo {author} {\bibfnamefont {G.}~\bibnamefont
  {Hoffmeister}},\ }\bibfield  {title} {\bibinfo {title} {Nonthermal fixed
  points and the functional renormalization group},\ }\href
  {https://doi.org/10.1016/j.nuclphysb.2008.12.017} {\bibfield  {journal}
  {\bibinfo  {journal} {Nucl. Phys.}\ }\textbf {\bibinfo {volume} {B813}},\
  \bibinfo {pages} {383} (\bibinfo {year} {2009})},\ \Eprint
  {https://arxiv.org/abs/0809.5208} {arXiv:0809.5208 [hep-th]} \BibitemShut
  {NoStop}%
\bibitem [{\citenamefont {Schmied}\ \emph
  {et~al.}(2019{\natexlab{a}})\citenamefont {Schmied}, \citenamefont
  {Mikheev},\ and\ \citenamefont {Gasenzer}}]{Schmied:2018mte}%
  \BibitemOpen
  \bibfield  {author} {\bibinfo {author} {\bibfnamefont {C.-M.}\ \bibnamefont
  {Schmied}}, \bibinfo {author} {\bibfnamefont {A.~N.}\ \bibnamefont
  {Mikheev}},\ and\ \bibinfo {author} {\bibfnamefont {T.}~\bibnamefont
  {Gasenzer}},\ }\bibfield  {title} {\bibinfo {title} {Non-thermal fixed
  points: Universal dynamics far from equilibrium},\ }\href
  {https://doi.org/10.1142/S0217751X19410069} {\bibfield  {journal} {\bibinfo
  {journal} {Int. J. Mod. Phys. A}\ }\textbf {\bibinfo {volume} {34}},\
  \bibinfo {pages} {1941006} (\bibinfo {year} {2019}{\natexlab{a}})},\ \Eprint
  {https://arxiv.org/abs/1810.08143} {arXiv:1810.08143 [cond-mat.quant-gas]}
  \BibitemShut {NoStop}%
\bibitem [{\citenamefont {Schmied}\ \emph
  {et~al.}(2019{\natexlab{b}})\citenamefont {Schmied}, \citenamefont
  {Pr\"ufer}, \citenamefont {Oberthaler},\ and\ \citenamefont
  {Gasenzer}}]{Schmied:2018osf}%
  \BibitemOpen
  \bibfield  {author} {\bibinfo {author} {\bibfnamefont {C.-M.}\ \bibnamefont
  {Schmied}}, \bibinfo {author} {\bibfnamefont {M.}~\bibnamefont {Pr\"ufer}},
  \bibinfo {author} {\bibfnamefont {M.~K.}\ \bibnamefont {Oberthaler}},\ and\
  \bibinfo {author} {\bibfnamefont {T.}~\bibnamefont {Gasenzer}},\ }\bibfield
  {title} {\bibinfo {title} {Bidirectional universal dynamics in a spinor
  {Bose} gas close to a nonthermal fixed point},\ }\href
  {https://doi.org/10.1103/PhysRevA.99.033611} {\bibfield  {journal} {\bibinfo
  {journal} {Phys. Rev. A}\ }\textbf {\bibinfo {volume} {99}},\ \bibinfo
  {pages} {033611} (\bibinfo {year} {2019}{\natexlab{b}})},\ \Eprint
  {https://arxiv.org/abs/1812.08571} {arXiv:1812.08571 [cond-mat.quant-gas]}
  \BibitemShut {NoStop}%
\bibitem [{\citenamefont {Glidden}\ \emph {et~al.}(2021)\citenamefont
  {Glidden}, \citenamefont {Eigen}, \citenamefont {Dogra}, \citenamefont
  {Hilker}, \citenamefont {Smith},\ and\ \citenamefont
  {Hadzibabic}}]{Glidden:2020qmu}%
  \BibitemOpen
  \bibfield  {author} {\bibinfo {author} {\bibfnamefont {J.~A.~P.}\
  \bibnamefont {Glidden}}, \bibinfo {author} {\bibfnamefont {C.}~\bibnamefont
  {Eigen}}, \bibinfo {author} {\bibfnamefont {L.~H.}\ \bibnamefont {Dogra}},
  \bibinfo {author} {\bibfnamefont {T.~A.}\ \bibnamefont {Hilker}}, \bibinfo
  {author} {\bibfnamefont {R.~P.}\ \bibnamefont {Smith}},\ and\ \bibinfo
  {author} {\bibfnamefont {Z.}~\bibnamefont {Hadzibabic}},\ }\bibfield  {title}
  {\bibinfo {title} {Bidirectional dynamic scaling in an isolated {Bose} gas
  far from equilibrium},\ }\href {https://doi.org/10.1038/s41567-020-01114-x}
  {\bibfield  {journal} {\bibinfo  {journal} {Nat. Phys.}\ }\textbf {\bibinfo
  {volume} {17}},\ \bibinfo {pages} {457} (\bibinfo {year} {2021})},\ \Eprint
  {https://arxiv.org/abs/2006.01118} {arXiv:2006.01118 [cond-mat.quant-gas]}
  \BibitemShut {NoStop}%
\bibitem [{\citenamefont {Mazeliauskas}\ and\ \citenamefont
  {Berges}(2019)}]{Mazeliauskas:2018yef}%
  \BibitemOpen
  \bibfield  {author} {\bibinfo {author} {\bibfnamefont {A.}~\bibnamefont
  {Mazeliauskas}}\ and\ \bibinfo {author} {\bibfnamefont {J.}~\bibnamefont
  {Berges}},\ }\bibfield  {title} {\bibinfo {title} {Prescaling and
  far-from-equilibrium hydrodynamics in the quark-gluon plasma},\ }\href
  {https://doi.org/10.1103/PhysRevLett.122.122301} {\bibfield  {journal}
  {\bibinfo  {journal} {Phys. Rev. Lett.}\ }\textbf {\bibinfo {volume} {122}},\
  \bibinfo {pages} {122301} (\bibinfo {year} {2019})},\ \Eprint
  {https://arxiv.org/abs/1810.10554} {arXiv:1810.10554 [hep-ph]} \BibitemShut
  {NoStop}%
\bibitem [{\citenamefont {Schmied}\ \emph
  {et~al.}(2019{\natexlab{c}})\citenamefont {Schmied}, \citenamefont
  {Mikheev},\ and\ \citenamefont {Gasenzer}}]{Schmied2019}%
  \BibitemOpen
  \bibfield  {author} {\bibinfo {author} {\bibfnamefont {C.-M.}\ \bibnamefont
  {Schmied}}, \bibinfo {author} {\bibfnamefont {A.~N.}\ \bibnamefont
  {Mikheev}},\ and\ \bibinfo {author} {\bibfnamefont {T.}~\bibnamefont
  {Gasenzer}},\ }\bibfield  {title} {\bibinfo {title} {Prescaling in a
  far-from-equilibrium {Bose} gas},\ }\href
  {https://doi.org/10.1103/PhysRevLett.122.170404} {\bibfield  {journal}
  {\bibinfo  {journal} {Phys. Rev. Lett.}\ }\textbf {\bibinfo {volume} {122}},\
  \bibinfo {pages} {170404} (\bibinfo {year} {2019}{\natexlab{c}})},\ \Eprint
  {https://arxiv.org/abs/1807.07514} {arXiv:1807.07514 [cond-mat.quant-gas]}
  \BibitemShut {NoStop}%
\bibitem [{\citenamefont {Berges}\ \emph
  {et~al.}(2014{\natexlab{a}})\citenamefont {Berges}, \citenamefont
  {Boguslavski}, \citenamefont {Schlichting},\ and\ \citenamefont
  {Venugopalan}}]{Berges2014}%
  \BibitemOpen
  \bibfield  {author} {\bibinfo {author} {\bibfnamefont {J.}~\bibnamefont
  {Berges}}, \bibinfo {author} {\bibfnamefont {K.}~\bibnamefont {Boguslavski}},
  \bibinfo {author} {\bibfnamefont {S.}~\bibnamefont {Schlichting}},\ and\
  \bibinfo {author} {\bibfnamefont {R.}~\bibnamefont {Venugopalan}},\
  }\bibfield  {title} {\bibinfo {title} {Basin of attraction for turbulent
  thermalization and the range of validity of classical-statistical
  simulations},\ }\href {https://doi.org/10.1007/JHEP05(2014)054} {\bibfield
  {journal} {\bibinfo  {journal} {J. High Energy Phys.}\ }\textbf {\bibinfo
  {volume} {2014}}\bibfield  {number} {\bibinfo  {number} { (5)},\ \bibinfo
  {pages} {54}},\ }\Eprint {https://arxiv.org/abs/1312.5216} {arXiv:1312.5216
  [hep-ph]} \BibitemShut {NoStop}%
\bibitem [{\citenamefont {Kolb}\ and\ \citenamefont
  {Turner}(1990)}]{Kolb:1990vq}%
  \BibitemOpen
  \bibfield  {author} {\bibinfo {author} {\bibfnamefont {E.~W.}\ \bibnamefont
  {Kolb}}\ and\ \bibinfo {author} {\bibfnamefont {M.~S.}\ \bibnamefont
  {Turner}},\ }\href@noop {} {\emph {\bibinfo {title} {The Early Universe}}},\
  \bibinfo {series} {Frontiers in Physics}, Vol.~\bibinfo {volume} {69}\
  (\bibinfo  {publisher} {Addison-Wesley},\ \bibinfo {address} {Redwood City,
  CA},\ \bibinfo {year} {1990})\BibitemShut {NoStop}%
\bibitem [{\citenamefont {Pitaevskii}\ and\ \citenamefont
  {Stringari}(2016)}]{Pitaevskii2016}%
  \BibitemOpen
  \bibfield  {author} {\bibinfo {author} {\bibfnamefont {L.~P.}\ \bibnamefont
  {Pitaevskii}}\ and\ \bibinfo {author} {\bibfnamefont {S.}~\bibnamefont
  {Stringari}},\ }\href@noop {} {\emph {\bibinfo {title} {{Bose}--{Einstein}
  Condensation and Superfluidity}}},\ International Series of Monographs on
  Physics\ (\bibinfo  {publisher} {Oxford University Press},\ \bibinfo
  {address} {Oxford},\ \bibinfo {year} {2016})\BibitemShut {NoStop}%
\bibitem [{\citenamefont {Su\'arez}\ and\ \citenamefont
  {Chavanis}(2015)}]{Suarez:2015fga}%
  \BibitemOpen
  \bibfield  {author} {\bibinfo {author} {\bibfnamefont {A.}~\bibnamefont
  {Su\'arez}}\ and\ \bibinfo {author} {\bibfnamefont {P.-H.}\ \bibnamefont
  {Chavanis}},\ }\bibfield  {title} {\bibinfo {title} {Hydrodynamic
  representation of the {Klein--Gordon--Einstein} equations in the weak field
  limit: General formalism and perturbations analysis},\ }\href
  {https://doi.org/10.1103/PhysRevD.92.023510} {\bibfield  {journal} {\bibinfo
  {journal} {Phys. Rev. D}\ }\textbf {\bibinfo {volume} {92}},\ \bibinfo
  {pages} {023510} (\bibinfo {year} {2015})},\ \Eprint
  {https://arxiv.org/abs/1503.07437} {arXiv:1503.07437 [gr-qc]} \BibitemShut
  {NoStop}%
\bibitem [{\citenamefont {Mukhanov}\ and\ \citenamefont
  {Winitzki}(2007)}]{mukhanov2007introduction}%
  \BibitemOpen
  \bibfield  {author} {\bibinfo {author} {\bibfnamefont {V.}~\bibnamefont
  {Mukhanov}}\ and\ \bibinfo {author} {\bibfnamefont {S.}~\bibnamefont
  {Winitzki}},\ }\href {https://doi.org/10.1017/CBO9780511809149} {\emph
  {\bibinfo {title} {Introduction to Quantum Effects in Gravity}}}\ (\bibinfo
  {publisher} {Cambridge University Press},\ \bibinfo {address} {Cambridge},\
  \bibinfo {year} {2007})\BibitemShut {NoStop}%
\bibitem [{\citenamefont {{Olshanii}}\ \emph {et~al.}(2010)\citenamefont
  {{Olshanii}}, \citenamefont {{Perrin}},\ and\ \citenamefont
  {{Lorent}}}]{Olshanii2010}%
  \BibitemOpen
  \bibfield  {author} {\bibinfo {author} {\bibfnamefont {M.}~\bibnamefont
  {{Olshanii}}}, \bibinfo {author} {\bibfnamefont {H.}~\bibnamefont
  {{Perrin}}},\ and\ \bibinfo {author} {\bibfnamefont {V.}~\bibnamefont
  {{Lorent}}},\ }\bibfield  {title} {\bibinfo {title} {Example of a quantum
  anomaly in the physics of ultracold gases},\ }\href
  {https://doi.org/10.1103/PhysRevLett.105.095302} {\bibfield  {journal}
  {\bibinfo  {journal} {\prl}\ }\textbf {\bibinfo {volume} {105}},\ \bibinfo
  {eid} {095302} (\bibinfo {year} {2010})},\ \Eprint
  {https://arxiv.org/abs/1006.1072} {arXiv:1006.1072 [cond-mat.quant-gas]}
  \BibitemShut {NoStop}%
\bibitem [{\citenamefont {Holten}\ \emph {et~al.}(2018)\citenamefont {Holten},
  \citenamefont {Bayha}, \citenamefont {Klein}, \citenamefont {Murthy},
  \citenamefont {Preiss},\ and\ \citenamefont {Jochim}}]{Holten2018}%
  \BibitemOpen
  \bibfield  {author} {\bibinfo {author} {\bibfnamefont {M.}~\bibnamefont
  {Holten}}, \bibinfo {author} {\bibfnamefont {L.}~\bibnamefont {Bayha}},
  \bibinfo {author} {\bibfnamefont {A.~C.}\ \bibnamefont {Klein}}, \bibinfo
  {author} {\bibfnamefont {P.~A.}\ \bibnamefont {Murthy}}, \bibinfo {author}
  {\bibfnamefont {P.~M.}\ \bibnamefont {Preiss}},\ and\ \bibinfo {author}
  {\bibfnamefont {S.}~\bibnamefont {Jochim}},\ }\bibfield  {title} {\bibinfo
  {title} {Anomalous breaking of scale invariance in a two-dimensional {Fermi}
  gas},\ }\href {https://doi.org/10.1103/PhysRevLett.121.120401} {\bibfield
  {journal} {\bibinfo  {journal} {Phys. Rev. Lett.}\ }\textbf {\bibinfo
  {volume} {121}},\ \bibinfo {pages} {120401} (\bibinfo {year} {2018})},\
  \Eprint {https://arxiv.org/abs/1803.08879} {arXiv:1803.08879
  [cond-mat.quant-gas]} \BibitemShut {NoStop}%
\bibitem [{\citenamefont {{Hung}}\ \emph {et~al.}(2011)\citenamefont {{Hung}},
  \citenamefont {{Zhang}}, \citenamefont {{Gemelke}},\ and\ \citenamefont
  {{Chin}}}]{Hung2011}%
  \BibitemOpen
  \bibfield  {author} {\bibinfo {author} {\bibfnamefont {C.-L.}\ \bibnamefont
  {{Hung}}}, \bibinfo {author} {\bibfnamefont {X.}~\bibnamefont {{Zhang}}},
  \bibinfo {author} {\bibfnamefont {N.}~\bibnamefont {{Gemelke}}},\ and\
  \bibinfo {author} {\bibfnamefont {C.}~\bibnamefont {{Chin}}},\ }\bibfield
  {title} {\bibinfo {title} {Observation of scale invariance and universality
  in two-dimensional {Bose} gases},\ }\href
  {https://doi.org/10.1038/nature09722} {\bibfield  {journal} {\bibinfo
  {journal} {\nat}\ }\textbf {\bibinfo {volume} {470}},\ \bibinfo {pages} {236}
  (\bibinfo {year} {2011})},\ \Eprint {https://arxiv.org/abs/1009.0016}
  {arXiv:1009.0016 [cond-mat.quant-gas]} \BibitemShut {NoStop}%
\bibitem [{\citenamefont {Greene}\ \emph {et~al.}(1999)\citenamefont {Greene},
  \citenamefont {Kofman},\ and\ \citenamefont {Starobinsky}}]{Greene:1998pb}%
  \BibitemOpen
  \bibfield  {author} {\bibinfo {author} {\bibfnamefont {P.~B.}\ \bibnamefont
  {Greene}}, \bibinfo {author} {\bibfnamefont {L.}~\bibnamefont {Kofman}},\
  and\ \bibinfo {author} {\bibfnamefont {A.~A.}\ \bibnamefont {Starobinsky}},\
  }\bibfield  {title} {\bibinfo {title} {{Sine--Gordon} parametric resonance},\
  }\href {https://doi.org/10.1016/S0550-3213(99)00018-8} {\bibfield  {journal}
  {\bibinfo  {journal} {Nucl. Phys.}\ }\textbf {\bibinfo {volume} {B543}},\
  \bibinfo {pages} {423} (\bibinfo {year} {1999})},\ \Eprint
  {https://arxiv.org/abs/hep-ph/9808477} {arXiv:hep-ph/9808477 [hep-ph]}
  \BibitemShut {NoStop}%
\bibitem [{Sup()}]{SuppMatArXiv}%
  \BibitemOpen
  \href@noop {} {}\bibinfo {note} {See section \enquote{ancillary files} on the
  article's arXiv page for a video illustrating the preheating dynamics of a
  single realization.}\BibitemShut {Stop}%
\bibitem [{\citenamefont {Hadzibabic}\ and\ \citenamefont
  {Dalibard}(2011)}]{Hadzibabic2011}%
  \BibitemOpen
  \bibfield  {author} {\bibinfo {author} {\bibfnamefont {Z.}~\bibnamefont
  {Hadzibabic}}\ and\ \bibinfo {author} {\bibfnamefont {J.}~\bibnamefont
  {Dalibard}},\ }\bibfield  {title} {\bibinfo {title} {Two-dimensional {Bose}
  fluids: An atomic physics perspective},\ }\href
  {https://doi.org/10.1393/ncr/i2011-10066-3} {\bibfield  {journal} {\bibinfo
  {journal} {Riv. del Nuovo Cimento}\ }\textbf {\bibinfo {volume} {34}},\
  \bibinfo {pages} {389} (\bibinfo {year} {2011})},\ \Eprint
  {https://arxiv.org/abs/0912.1490} {arXiv:0912.1490 [cond-mat.quant-gas]}
  \BibitemShut {NoStop}%
\bibitem [{\citenamefont {Bogoliubov}(1947)}]{Bogolyubov:1947zz}%
  \BibitemOpen
  \bibfield  {author} {\bibinfo {author} {\bibfnamefont {N.}~\bibnamefont
  {Bogoliubov}},\ }\bibfield  {title} {\bibinfo {title} {On the theory of
  superfluidity},\ }\href@noop {} {\bibfield  {journal} {\bibinfo  {journal}
  {Izv. Akad. Nauk Ser. Fiz.}\ }\textbf {\bibinfo {volume} {11}},\ \bibinfo
  {pages} {77} (\bibinfo {year} {1947})},\ \bibinfo {note} {[J. Phys. (USSR)
  \textbf{11}, 23 (1947)]}\BibitemShut {NoStop}%
\bibitem [{\citenamefont {McLachlan}(1951)}]{McLachlan1951}%
  \BibitemOpen
  \bibfield  {author} {\bibinfo {author} {\bibfnamefont {N.~W.}\ \bibnamefont
  {McLachlan}},\ }\href@noop {} {\emph {\bibinfo {title} {Theory and
  Application of {Mathieu} Functions}}}\ (\bibinfo  {publisher} {Clarendon
  Press},\ \bibinfo {address} {Oxford},\ \bibinfo {year} {1951})\BibitemShut
  {NoStop}%
\bibitem [{\citenamefont {Nayfeh}(1973)}]{Nayfeh1973}%
  \BibitemOpen
  \bibfield  {author} {\bibinfo {author} {\bibfnamefont {A.~H.}\ \bibnamefont
  {Nayfeh}},\ }\href@noop {} {\emph {\bibinfo {title} {Perturbation Methods}}}\
  (\bibinfo  {publisher} {Wiley},\ \bibinfo {address} {New York},\ \bibinfo
  {year} {1973})\BibitemShut {NoStop}%
\bibitem [{\citenamefont {Zakharov}\ \emph {et~al.}(1992)\citenamefont
  {Zakharov}, \citenamefont {L'vov},\ and\ \citenamefont
  {Falkovich}}]{Zakharov1992}%
  \BibitemOpen
  \bibfield  {author} {\bibinfo {author} {\bibfnamefont {V.~E.}\ \bibnamefont
  {Zakharov}}, \bibinfo {author} {\bibfnamefont {V.~S.}\ \bibnamefont
  {L'vov}},\ and\ \bibinfo {author} {\bibfnamefont {G.}~\bibnamefont
  {Falkovich}},\ }\href {https://doi.org/10.1007/978-3-642-50052-7} {\emph
  {\bibinfo {title} {{Kolmogorov} Spectra of Turbulence I: Wave Turbulence}}},\
  Springer Series in Nonlinear Dynamics\ (\bibinfo  {publisher} {Springer},\
  \bibinfo {address} {Berlin, Heidelberg},\ \bibinfo {year} {1992})\BibitemShut
  {NoStop}%
\bibitem [{\citenamefont {Berges}\ \emph {et~al.}(2017)\citenamefont {Berges},
  \citenamefont {Boguslavski}, \citenamefont {Chatrchyan},\ and\ \citenamefont
  {Jaeckel}}]{Berges:2017ldx}%
  \BibitemOpen
  \bibfield  {author} {\bibinfo {author} {\bibfnamefont {J.}~\bibnamefont
  {Berges}}, \bibinfo {author} {\bibfnamefont {K.}~\bibnamefont {Boguslavski}},
  \bibinfo {author} {\bibfnamefont {A.}~\bibnamefont {Chatrchyan}},\ and\
  \bibinfo {author} {\bibfnamefont {J.}~\bibnamefont {Jaeckel}},\ }\bibfield
  {title} {\bibinfo {title} {Attractive versus repulsive interactions in the
  {Bose--Einstein} condensation dynamics of relativistic field theories},\
  }\href {https://doi.org/10.1103/PhysRevD.96.076020} {\bibfield  {journal}
  {\bibinfo  {journal} {Phys. Rev.}\ }\textbf {\bibinfo {volume} {D96}},\
  \bibinfo {pages} {076020} (\bibinfo {year} {2017})},\ \Eprint
  {https://arxiv.org/abs/1707.07696} {arXiv:1707.07696 [hep-ph]} \BibitemShut
  {NoStop}%
\bibitem [{\citenamefont {Berges}\ \emph {et~al.}(2008)\citenamefont {Berges},
  \citenamefont {Rothkopf},\ and\ \citenamefont {Schmidt}}]{Berges:2008wm}%
  \BibitemOpen
  \bibfield  {author} {\bibinfo {author} {\bibfnamefont {J.}~\bibnamefont
  {Berges}}, \bibinfo {author} {\bibfnamefont {A.}~\bibnamefont {Rothkopf}},\
  and\ \bibinfo {author} {\bibfnamefont {J.}~\bibnamefont {Schmidt}},\
  }\bibfield  {title} {\bibinfo {title} {Non-thermal fixed points: Effective
  weak-coupling for strongly correlated systems far from equilibrium},\ }\href
  {https://doi.org/10.1103/PhysRevLett.101.041603} {\bibfield  {journal}
  {\bibinfo  {journal} {Phys. Rev. Lett.}\ }\textbf {\bibinfo {volume} {101}},\
  \bibinfo {pages} {041603} (\bibinfo {year} {2008})},\ \Eprint
  {https://arxiv.org/abs/0803.0131} {arXiv:0803.0131 [hep-ph]} \BibitemShut
  {NoStop}%
\bibitem [{\citenamefont {Scheppach}\ \emph {et~al.}(2010)\citenamefont
  {Scheppach}, \citenamefont {Berges},\ and\ \citenamefont
  {Gasenzer}}]{Scheppach:2009wu}%
  \BibitemOpen
  \bibfield  {author} {\bibinfo {author} {\bibfnamefont {C.}~\bibnamefont
  {Scheppach}}, \bibinfo {author} {\bibfnamefont {J.}~\bibnamefont {Berges}},\
  and\ \bibinfo {author} {\bibfnamefont {T.}~\bibnamefont {Gasenzer}},\
  }\bibfield  {title} {\bibinfo {title} {Matter wave turbulence: Beyond kinetic
  scaling},\ }\href {https://doi.org/10.1103/PhysRevA.81.033611} {\bibfield
  {journal} {\bibinfo  {journal} {Phys. Rev.}\ }\textbf {\bibinfo {volume}
  {A81}},\ \bibinfo {pages} {033611} (\bibinfo {year} {2010})},\ \Eprint
  {https://arxiv.org/abs/0912.4183} {arXiv:0912.4183 [cond-mat.quant-gas]}
  \BibitemShut {NoStop}%
\bibitem [{\citenamefont {Dyachenko}\ \emph {et~al.}(1992)\citenamefont
  {Dyachenko}, \citenamefont {Newell}, \citenamefont {Pushkarev},\ and\
  \citenamefont {Zakharov}}]{dyachenko1992optical}%
  \BibitemOpen
  \bibfield  {author} {\bibinfo {author} {\bibfnamefont {S.}~\bibnamefont
  {Dyachenko}}, \bibinfo {author} {\bibfnamefont {A.}~\bibnamefont {Newell}},
  \bibinfo {author} {\bibfnamefont {A.}~\bibnamefont {Pushkarev}},\ and\
  \bibinfo {author} {\bibfnamefont {V.}~\bibnamefont {Zakharov}},\ }\bibfield
  {title} {\bibinfo {title} {Optical turbulence: Weak turbulence, condensates
  and collapsing filaments in the nonlinear {Schrödinger} equation},\ }\href
  {https://doi.org/10.1016/0167-2789(92)90090-A} {\bibfield  {journal}
  {\bibinfo  {journal} {Physica D}\ }\textbf {\bibinfo {volume} {57}},\
  \bibinfo {pages} {96 } (\bibinfo {year} {1992})}\BibitemShut {NoStop}%
\bibitem [{\citenamefont {Nowak}\ \emph {et~al.}(2012)\citenamefont {Nowak},
  \citenamefont {Schole}, \citenamefont {Sexty},\ and\ \citenamefont
  {Gasenzer}}]{Nowak:2011sk}%
  \BibitemOpen
  \bibfield  {author} {\bibinfo {author} {\bibfnamefont {B.}~\bibnamefont
  {Nowak}}, \bibinfo {author} {\bibfnamefont {J.}~\bibnamefont {Schole}},
  \bibinfo {author} {\bibfnamefont {D.}~\bibnamefont {Sexty}},\ and\ \bibinfo
  {author} {\bibfnamefont {T.}~\bibnamefont {Gasenzer}},\ }\bibfield  {title}
  {\bibinfo {title} {Nonthermal fixed points, vortex statistics, and superfluid
  turbulence in an ultracold {Bose} gas},\ }\href
  {https://doi.org/10.1103/PhysRevA.85.043627} {\bibfield  {journal} {\bibinfo
  {journal} {Phys. Rev.}\ }\textbf {\bibinfo {volume} {A85}},\ \bibinfo {pages}
  {043627} (\bibinfo {year} {2012})},\ \Eprint
  {https://arxiv.org/abs/1111.6127} {arXiv:1111.6127 [cond-mat.quant-gas]}
  \BibitemShut {NoStop}%
\bibitem [{\citenamefont {{Navon}}\ \emph {et~al.}(2016)\citenamefont
  {{Navon}}, \citenamefont {{Gaunt}}, \citenamefont {{Smith}},\ and\
  \citenamefont {{Hadzibabic}}}]{navon2016emergence}%
  \BibitemOpen
  \bibfield  {author} {\bibinfo {author} {\bibfnamefont {N.}~\bibnamefont
  {{Navon}}}, \bibinfo {author} {\bibfnamefont {A.~L.}\ \bibnamefont
  {{Gaunt}}}, \bibinfo {author} {\bibfnamefont {R.~P.}\ \bibnamefont
  {{Smith}}},\ and\ \bibinfo {author} {\bibfnamefont {Z.}~\bibnamefont
  {{Hadzibabic}}},\ }\bibfield  {title} {\bibinfo {title} {Emergence of a
  turbulent cascade in a quantum gas},\ }\href
  {https://doi.org/10.1038/nature20114} {\bibfield  {journal} {\bibinfo
  {journal} {\nat}\ }\textbf {\bibinfo {volume} {539}},\ \bibinfo {pages} {72}
  (\bibinfo {year} {2016})},\ \Eprint {https://arxiv.org/abs/1609.01271}
  {arXiv:1609.01271 [cond-mat.quant-gas]} \BibitemShut {NoStop}%
\bibitem [{\citenamefont {{Proukakis}}\ \emph {et~al.}(1998)\citenamefont
  {{Proukakis}}, \citenamefont {{Burnett}},\ and\ \citenamefont
  {{Stoof}}}]{Proukakis:1998zz}%
  \BibitemOpen
  \bibfield  {author} {\bibinfo {author} {\bibfnamefont {N.~P.}\ \bibnamefont
  {{Proukakis}}}, \bibinfo {author} {\bibfnamefont {K.}~\bibnamefont
  {{Burnett}}},\ and\ \bibinfo {author} {\bibfnamefont {H.~T.~C.}\ \bibnamefont
  {{Stoof}}},\ }\bibfield  {title} {\bibinfo {title} {Microscopic treatment of
  binary interactions in the nonequilibrium dynamics of partially
  {Bose}-condensed trapped gases},\ }\href
  {https://doi.org/10.1103/PhysRevA.57.1230} {\bibfield  {journal} {\bibinfo
  {journal} {\pra}\ }\textbf {\bibinfo {volume} {57}},\ \bibinfo {pages} {1230}
  (\bibinfo {year} {1998})},\ \Eprint {https://arxiv.org/abs/cond-mat/9703199}
  {arXiv:cond-mat/9703199 [cond-mat.stat-mech]} \BibitemShut {NoStop}%
\bibitem [{\citenamefont {Felder}\ and\ \citenamefont
  {Tkachev}(2008)}]{Felder:2000hq}%
  \BibitemOpen
  \bibfield  {author} {\bibinfo {author} {\bibfnamefont {G.~N.}\ \bibnamefont
  {Felder}}\ and\ \bibinfo {author} {\bibfnamefont {I.}~\bibnamefont
  {Tkachev}},\ }\bibfield  {title} {\bibinfo {title} {{LATTICEEASY}: A program
  for lattice simulations of scalar fields in an expanding universe},\ }\href
  {https://doi.org/10.1016/j.cpc.2008.02.009} {\bibfield  {journal} {\bibinfo
  {journal} {Comput. Phys. Commun.}\ }\textbf {\bibinfo {volume} {178}},\
  \bibinfo {pages} {929} (\bibinfo {year} {2008})},\ \Eprint
  {https://arxiv.org/abs/hep-ph/0011159} {arXiv:hep-ph/0011159 [hep-ph]}
  \BibitemShut {NoStop}%
\bibitem [{\citenamefont {{Ville}}\ \emph {et~al.}(2017)\citenamefont
  {{Ville}}, \citenamefont {{Bienaim{\'e}}}, \citenamefont {{Saint-Jalm}},
  \citenamefont {{Corman}}, \citenamefont {{Aidelsburger}}, \citenamefont
  {{Chomaz}}, \citenamefont {{Kleinlein}}, \citenamefont {{Perconte}},
  \citenamefont {{Nascimb{\`e}ne}}, \citenamefont {{Dalibard}},\ and\
  \citenamefont {{Beugnon}}}]{Ville2017}%
  \BibitemOpen
  \bibfield  {author} {\bibinfo {author} {\bibfnamefont {J.~L.}\ \bibnamefont
  {{Ville}}}, \bibinfo {author} {\bibfnamefont {T.}~\bibnamefont
  {{Bienaim{\'e}}}}, \bibinfo {author} {\bibfnamefont {R.}~\bibnamefont
  {{Saint-Jalm}}}, \bibinfo {author} {\bibfnamefont {L.}~\bibnamefont
  {{Corman}}}, \bibinfo {author} {\bibfnamefont {M.}~\bibnamefont
  {{Aidelsburger}}}, \bibinfo {author} {\bibfnamefont {L.}~\bibnamefont
  {{Chomaz}}}, \bibinfo {author} {\bibfnamefont {K.}~\bibnamefont
  {{Kleinlein}}}, \bibinfo {author} {\bibfnamefont {D.}~\bibnamefont
  {{Perconte}}}, \bibinfo {author} {\bibfnamefont {S.}~\bibnamefont
  {{Nascimb{\`e}ne}}}, \bibinfo {author} {\bibfnamefont {J.}~\bibnamefont
  {{Dalibard}}},\ and\ \bibinfo {author} {\bibfnamefont {J.}~\bibnamefont
  {{Beugnon}}},\ }\bibfield  {title} {\bibinfo {title} {Loading and compression
  of a single two-dimensional {Bose} gas in an optical accordion},\ }\href
  {https://doi.org/10.1103/PhysRevA.95.013632} {\bibfield  {journal} {\bibinfo
  {journal} {\pra}\ }\textbf {\bibinfo {volume} {95}},\ \bibinfo {eid} {013632}
  (\bibinfo {year} {2017})},\ \Eprint {https://arxiv.org/abs/1611.07681}
  {arXiv:1611.07681 [cond-mat.quant-gas]} \BibitemShut {NoStop}%
\bibitem [{\citenamefont {Cabrera~C{\'{o}}rdova}(2018)}]{CabreraCordova2018}%
  \BibitemOpen
  \bibfield  {author} {\bibinfo {author} {\bibfnamefont {C.~R.}\ \bibnamefont
  {Cabrera~C{\'{o}}rdova}},\ }\emph {\bibinfo {title} {Quantum liquid droplets
  in a mixture of {Bose--Einstein} condensates}},\ \href
  {http://hdl.handle.net/10803/663331} {Ph.D. thesis},\ \bibinfo  {school}
  {ICFO – Institute of Photonic Sciences, Universidad Politécnica de
  Cataluña}, \bibinfo {address} {Barcelona} (\bibinfo {year}
  {2018})\BibitemShut {NoStop}%
\bibitem [{\citenamefont {Campbell}\ \emph {et~al.}(2010)\citenamefont
  {Campbell}, \citenamefont {Smith}, \citenamefont {Tammuz}, \citenamefont
  {Beattie}, \citenamefont {Moulder},\ and\ \citenamefont
  {Hadzibabic}}]{ParticleNumberCampbell2010}%
  \BibitemOpen
  \bibfield  {author} {\bibinfo {author} {\bibfnamefont {R.~L.~D.}\
  \bibnamefont {Campbell}}, \bibinfo {author} {\bibfnamefont {R.~P.}\
  \bibnamefont {Smith}}, \bibinfo {author} {\bibfnamefont {N.}~\bibnamefont
  {Tammuz}}, \bibinfo {author} {\bibfnamefont {S.}~\bibnamefont {Beattie}},
  \bibinfo {author} {\bibfnamefont {S.}~\bibnamefont {Moulder}},\ and\ \bibinfo
  {author} {\bibfnamefont {Z.}~\bibnamefont {Hadzibabic}},\ }\bibfield  {title}
  {\bibinfo {title} {Efficient production of large ${}^{39}\mathrm{K}$
  {Bose--Einstein} condensates},\ }\href
  {https://doi.org/10.1103/PhysRevA.82.063611} {\bibfield  {journal} {\bibinfo
  {journal} {Phys. Rev. A}\ }\textbf {\bibinfo {volume} {82}},\ \bibinfo
  {pages} {063611} (\bibinfo {year} {2010})},\ \Eprint
  {https://arxiv.org/abs/1010.3739} {arXiv:1010.3739 [cond-mat.quant-gas]}
  \BibitemShut {NoStop}%
\bibitem [{\citenamefont {Landini}\ \emph {et~al.}(2012)\citenamefont
  {Landini}, \citenamefont {Roy}, \citenamefont {Roati}, \citenamefont
  {Simoni}, \citenamefont {Inguscio}, \citenamefont {Modugno},\ and\
  \citenamefont {Fattori}}]{ParticleNumberLandini2012}%
  \BibitemOpen
  \bibfield  {author} {\bibinfo {author} {\bibfnamefont {M.}~\bibnamefont
  {Landini}}, \bibinfo {author} {\bibfnamefont {S.}~\bibnamefont {Roy}},
  \bibinfo {author} {\bibfnamefont {G.}~\bibnamefont {Roati}}, \bibinfo
  {author} {\bibfnamefont {A.}~\bibnamefont {Simoni}}, \bibinfo {author}
  {\bibfnamefont {M.}~\bibnamefont {Inguscio}}, \bibinfo {author}
  {\bibfnamefont {G.}~\bibnamefont {Modugno}},\ and\ \bibinfo {author}
  {\bibfnamefont {M.}~\bibnamefont {Fattori}},\ }\bibfield  {title} {\bibinfo
  {title} {Direct evaporative cooling of ${}^{39}${K} atoms to {Bose--Einstein}
  condensation},\ }\href {https://doi.org/10.1103/PhysRevA.86.033421}
  {\bibfield  {journal} {\bibinfo  {journal} {Phys. Rev. A}\ }\textbf {\bibinfo
  {volume} {86}},\ \bibinfo {pages} {033421} (\bibinfo {year} {2012})},\
  \Eprint {https://arxiv.org/abs/1207.4713} {arXiv:1207.4713
  [cond-mat.quant-gas]} \BibitemShut {NoStop}%
\bibitem [{\citenamefont {{D'Errico}}\ \emph {et~al.}(2007)\citenamefont
  {{D'Errico}}, \citenamefont {{Zaccanti}}, \citenamefont {{Fattori}},
  \citenamefont {{Roati}}, \citenamefont {{Inguscio}}, \citenamefont
  {{Modugno}},\ and\ \citenamefont {{Simoni}}}]{d2007feshbach}%
  \BibitemOpen
  \bibfield  {author} {\bibinfo {author} {\bibfnamefont {C.}~\bibnamefont
  {{D'Errico}}}, \bibinfo {author} {\bibfnamefont {M.}~\bibnamefont
  {{Zaccanti}}}, \bibinfo {author} {\bibfnamefont {M.}~\bibnamefont
  {{Fattori}}}, \bibinfo {author} {\bibfnamefont {G.}~\bibnamefont {{Roati}}},
  \bibinfo {author} {\bibfnamefont {M.}~\bibnamefont {{Inguscio}}}, \bibinfo
  {author} {\bibfnamefont {G.}~\bibnamefont {{Modugno}}},\ and\ \bibinfo
  {author} {\bibfnamefont {A.}~\bibnamefont {{Simoni}}},\ }\bibfield  {title}
  {\bibinfo {title} {{Feshbach} resonances in ultracold $^{39}${K}},\ }\href
  {https://doi.org/10.1088/1367-2630/9/7/223} {\bibfield  {journal} {\bibinfo
  {journal} {New J. Phys.}\ }\textbf {\bibinfo {volume} {9}},\ \bibinfo {pages}
  {223} (\bibinfo {year} {2007})},\ \Eprint {https://arxiv.org/abs/0705.3036}
  {arXiv:0705.3036 [physics.atom-ph]} \BibitemShut {NoStop}%
\bibitem [{\citenamefont {Kurkela}\ and\ \citenamefont
  {Zhu}(2015)}]{Kurkela2015}%
  \BibitemOpen
  \bibfield  {author} {\bibinfo {author} {\bibfnamefont {A.}~\bibnamefont
  {Kurkela}}\ and\ \bibinfo {author} {\bibfnamefont {Y.}~\bibnamefont {Zhu}},\
  }\bibfield  {title} {\bibinfo {title} {Isotropization and hydrodynamization
  in weakly coupled heavy-ion collisions},\ }\href
  {https://doi.org/10.1103/PhysRevLett.115.182301} {\bibfield  {journal}
  {\bibinfo  {journal} {Phys. Rev. Lett.}\ }\textbf {\bibinfo {volume} {115}},\
  \bibinfo {pages} {182301} (\bibinfo {year} {2015})},\ \Eprint
  {https://arxiv.org/abs/1506.06647} {arXiv:1506.06647 [hep-ph]} \BibitemShut
  {NoStop}%
\bibitem [{\citenamefont {Prüfer}\ \emph {et~al.}(2020)\citenamefont
  {Prüfer}, \citenamefont {Zache}, \citenamefont {Kunkel}, \citenamefont
  {Lannig}, \citenamefont {Bonnin}, \citenamefont {Strobel}, \citenamefont
  {Berges},\ and\ \citenamefont {Oberthaler}}]{Pruefer2019}%
  \BibitemOpen
  \bibfield  {author} {\bibinfo {author} {\bibfnamefont {M.}~\bibnamefont
  {Prüfer}}, \bibinfo {author} {\bibfnamefont {T.~V.}\ \bibnamefont {Zache}},
  \bibinfo {author} {\bibfnamefont {P.}~\bibnamefont {Kunkel}}, \bibinfo
  {author} {\bibfnamefont {S.}~\bibnamefont {Lannig}}, \bibinfo {author}
  {\bibfnamefont {A.}~\bibnamefont {Bonnin}}, \bibinfo {author} {\bibfnamefont
  {H.}~\bibnamefont {Strobel}}, \bibinfo {author} {\bibfnamefont
  {J.}~\bibnamefont {Berges}},\ and\ \bibinfo {author} {\bibfnamefont {M.~K.}\
  \bibnamefont {Oberthaler}},\ }\bibfield  {title} {\bibinfo {title}
  {Experimental extraction of the quantum effective action for a
  non-equilibrium many-body system},\ }\href
  {https://doi.org/10.1038/s41567-020-0933-6} {\bibfield  {journal} {\bibinfo
  {journal} {Nat. Phys.}\ }\textbf {\bibinfo {volume} {16}},\ \bibinfo {pages}
  {1012} (\bibinfo {year} {2020})},\ \Eprint {https://arxiv.org/abs/1909.05120}
  {arXiv:1909.05120 [cond-mat.quant-gas]} \BibitemShut {NoStop}%
\bibitem [{\citenamefont {Gaunt}\ \emph {et~al.}(2013)\citenamefont {Gaunt},
  \citenamefont {Schmidutz}, \citenamefont {Gotlibovych}, \citenamefont
  {Smith},\ and\ \citenamefont {Hadzibabic}}]{Gaunt2013}%
  \BibitemOpen
  \bibfield  {author} {\bibinfo {author} {\bibfnamefont {A.~L.}\ \bibnamefont
  {Gaunt}}, \bibinfo {author} {\bibfnamefont {T.~F.}\ \bibnamefont
  {Schmidutz}}, \bibinfo {author} {\bibfnamefont {I.}~\bibnamefont
  {Gotlibovych}}, \bibinfo {author} {\bibfnamefont {R.~P.}\ \bibnamefont
  {Smith}},\ and\ \bibinfo {author} {\bibfnamefont {Z.}~\bibnamefont
  {Hadzibabic}},\ }\bibfield  {title} {\bibinfo {title} {{Bose--Einstein}
  condensation of atoms in a uniform potential},\ }\href
  {https://doi.org/10.1103/PhysRevLett.110.200406} {\bibfield  {journal}
  {\bibinfo  {journal} {Phys. Rev. Lett.}\ }\textbf {\bibinfo {volume} {110}},\
  \bibinfo {pages} {200406} (\bibinfo {year} {2013})},\ \Eprint
  {https://arxiv.org/abs/1212.4453} {arXiv:1212.4453 [cond-mat.quant-gas]}
  \BibitemShut {NoStop}%
\bibitem [{\citenamefont {Pethick}\ and\ \citenamefont
  {Smith}(2008)}]{Pethick2008}%
  \BibitemOpen
  \bibfield  {author} {\bibinfo {author} {\bibfnamefont {C.~J.}\ \bibnamefont
  {Pethick}}\ and\ \bibinfo {author} {\bibfnamefont {H.}~\bibnamefont
  {Smith}},\ }\href {https://doi.org/10.1017/CBO9780511802850} {\emph {\bibinfo
  {title} {{Bose}--{Einstein} Condensation in Dilute Gases}}},\ \bibinfo
  {edition} {2nd}\ ed.\ (\bibinfo  {publisher} {Cambridge University Press},\
  \bibinfo {address} {Cambridge},\ \bibinfo {year} {2008})\BibitemShut
  {NoStop}%
\bibitem [{\citenamefont {Antoine}\ \emph {et~al.}(2013)\citenamefont
  {Antoine}, \citenamefont {Bao},\ and\ \citenamefont {Besse}}]{Antoine2013}%
  \BibitemOpen
  \bibfield  {author} {\bibinfo {author} {\bibfnamefont {X.}~\bibnamefont
  {Antoine}}, \bibinfo {author} {\bibfnamefont {W.}~\bibnamefont {Bao}},\ and\
  \bibinfo {author} {\bibfnamefont {C.}~\bibnamefont {Besse}},\ }\bibfield
  {title} {\bibinfo {title} {Computational methods for the dynamics of the
  nonlinear {Schrödinger}/{Gross--Pitaevskii} equations},\ }\href
  {https://doi.org/10.1016/j.cpc.2013.07.012} {\bibfield  {journal} {\bibinfo
  {journal} {Comput. Phys. Commun.}\ }\textbf {\bibinfo {volume} {184}},\
  \bibinfo {pages} {2621 } (\bibinfo {year} {2013})},\ \Eprint
  {https://arxiv.org/abs/1305.1093} {arXiv:1305.1093 [math.NA]} \BibitemShut
  {NoStop}%
\bibitem [{\citenamefont {Blanes}\ and\ \citenamefont
  {Moan}(2002)}]{Blanes2002}%
  \BibitemOpen
  \bibfield  {author} {\bibinfo {author} {\bibfnamefont {S.}~\bibnamefont
  {Blanes}}\ and\ \bibinfo {author} {\bibfnamefont {P.}~\bibnamefont {Moan}},\
  }\bibfield  {title} {\bibinfo {title} {Practical symplectic partitioned
  {Runge--Kutta} and {Runge--Kutta--Nyström} methods},\ }\href
  {https://doi.org/10.1016/S0377-0427(01)00492-7} {\bibfield  {journal}
  {\bibinfo  {journal} {J. Comput. Appl. Math.}\ }\textbf {\bibinfo {volume}
  {142}},\ \bibinfo {pages} {313 } (\bibinfo {year} {2002})}\BibitemShut
  {NoStop}%
\bibitem [{\citenamefont {Chin}(2007)}]{Chin2007}%
  \BibitemOpen
  \bibfield  {author} {\bibinfo {author} {\bibfnamefont {S.~A.}\ \bibnamefont
  {Chin}},\ }\bibfield  {title} {\bibinfo {title} {Higher-order splitting
  algorithms for solving the nonlinear {Schrödinger} equation and their
  instabilities},\ }\href {https://doi.org/10.1103/PhysRevE.76.056708}
  {\bibfield  {journal} {\bibinfo  {journal} {Phys. Rev. E}\ }\textbf {\bibinfo
  {volume} {76}},\ \bibinfo {pages} {056708} (\bibinfo {year} {2007})},\
  \Eprint {https://arxiv.org/abs/0710.0396} {arXiv:0710.0396 [physics.comp-ph]}
  \BibitemShut {NoStop}%
\bibitem [{\citenamefont {Gasenzer}\ \emph {et~al.}(2005)\citenamefont
  {Gasenzer}, \citenamefont {Berges}, \citenamefont {Schmidt},\ and\
  \citenamefont {Seco}}]{Gasenzer:2005ze}%
  \BibitemOpen
  \bibfield  {author} {\bibinfo {author} {\bibfnamefont {T.}~\bibnamefont
  {Gasenzer}}, \bibinfo {author} {\bibfnamefont {J.}~\bibnamefont {Berges}},
  \bibinfo {author} {\bibfnamefont {M.~G.}\ \bibnamefont {Schmidt}},\ and\
  \bibinfo {author} {\bibfnamefont {M.}~\bibnamefont {Seco}},\ }\bibfield
  {title} {\bibinfo {title} {Non-perturbative dynamical many-body theory of a
  {Bose-Einstein} condensate},\ }\href
  {https://doi.org/10.1103/PhysRevA.72.063604} {\bibfield  {journal} {\bibinfo
  {journal} {Phys. Rev. A}\ }\textbf {\bibinfo {volume} {72}},\ \bibinfo
  {pages} {063604} (\bibinfo {year} {2005})},\ \Eprint
  {https://arxiv.org/abs/cond-mat/0507480} {arXiv:cond-mat/0507480}
  \BibitemShut {NoStop}%
\bibitem [{\citenamefont {Keldysh}(1964)}]{Keldysh:1964ud}%
  \BibitemOpen
  \bibfield  {author} {\bibinfo {author} {\bibfnamefont {L.}~\bibnamefont
  {Keldysh}},\ }\bibfield  {title} {\bibinfo {title} {Diagram technique for
  nonequilibrium processes},\ }\href@noop {} {\bibfield  {journal} {\bibinfo
  {journal} {Zh. Eksp. Teor. Fiz.}\ }\textbf {\bibinfo {volume} {47}},\
  \bibinfo {pages} {1515} (\bibinfo {year} {1964})},\ \bibinfo {note} {[Sov.
  Phys. JETP \textbf{20}, 1018 (1965)]}\BibitemShut {NoStop}%
\bibitem [{\citenamefont {Berges}(2016)}]{Berges:2015kfa}%
  \BibitemOpen
  \bibfield  {author} {\bibinfo {author} {\bibfnamefont {J.}~\bibnamefont
  {Berges}},\ }\bibinfo {title} {Nonequilibrium quantum fields: From cold atoms
  to cosmology},\ in\ \href
  {https://doi.org/10.1093/acprof:oso/9780198768166.003.0002} {\emph {\bibinfo
  {booktitle} {Strongly Interacting Quantum Systems out of Equilibrium: Lecture
  Notes of the Les Houches Summer School: Volume 99, August 2012}}},\ \bibinfo
  {editor} {edited by\ \bibinfo {editor} {\bibfnamefont {T.}~\bibnamefont
  {Giamarchi}}, \bibinfo {editor} {\bibfnamefont {A.~J.}\ \bibnamefont
  {Millis}}, \bibinfo {editor} {\bibfnamefont {O.}~\bibnamefont {Parcollet}},
  \bibinfo {editor} {\bibfnamefont {H.}~\bibnamefont {Saleur}},\ and\ \bibinfo
  {editor} {\bibfnamefont {L.~F.}\ \bibnamefont {Cugliandolo}}}\ (\bibinfo
  {publisher} {Oxford University Press},\ \bibinfo {address} {Oxford},\
  \bibinfo {year} {2016})\ Chap.~\bibinfo {chapter} {2},\ \Eprint
  {https://arxiv.org/abs/1503.02907} {arXiv:1503.02907 [hep-ph]} \BibitemShut
  {NoStop}%
\bibitem [{\citenamefont {Berges}\ \emph
  {et~al.}(2014{\natexlab{b}})\citenamefont {Berges}, \citenamefont
  {Boguslavski}, \citenamefont {Schlichting},\ and\ \citenamefont
  {Venugopalan}}]{Berges:2013fga}%
  \BibitemOpen
  \bibfield  {author} {\bibinfo {author} {\bibfnamefont {J.}~\bibnamefont
  {Berges}}, \bibinfo {author} {\bibfnamefont {K.}~\bibnamefont {Boguslavski}},
  \bibinfo {author} {\bibfnamefont {S.}~\bibnamefont {Schlichting}},\ and\
  \bibinfo {author} {\bibfnamefont {R.}~\bibnamefont {Venugopalan}},\
  }\bibfield  {title} {\bibinfo {title} {Universal attractor in a highly
  occupied non-{Abelian} plasma},\ }\href
  {https://doi.org/10.1103/PhysRevD.89.114007} {\bibfield  {journal} {\bibinfo
  {journal} {Phys. Rev.}\ }\textbf {\bibinfo {volume} {D89}},\ \bibinfo {pages}
  {114007} (\bibinfo {year} {2014}{\natexlab{b}})},\ \Eprint
  {https://arxiv.org/abs/1311.3005} {arXiv:1311.3005 [hep-ph]} \BibitemShut
  {NoStop}%
\end{thebibliography}%

\end{document}